\documentclass{aa}  

\usepackage{graphicx}
\usepackage{txfonts}
\usepackage{xcolor} 
\usepackage{lscape}

\usepackage{subfig}

\begin{document}

   \title{The GAPS programme at TNG
   \thanks{Based on observations made with the Italian {\it Telescopio Nazionale Galileo} (TNG) operated by the {\it Fundaci\'on Galileo Galilei} (FGG) of the {\it Istituto Nazionale di Astrofisica} (INAF) at the  {\it Observatorio del Roque de los Muchachos} (La Palma, Canary Islands, Spain).}
   \fnmsep \thanks{
   Tables ~\ref{longtable_kinematicis} to ~\ref{longtable_fluxes} 
   are only available in electronic form
   at the CDS via anonymous ftp to cdsarc.u-strasbg.fr (130.79.128.5)
   or via http://cdsweb.u-strasbg.fr/cgi-bin/qcat?J/A+A/
      }
   }
   \subtitle{XXXIV. Activity-rotation, flux-flux relationships, and active region evolution through stellar age}

   \author{J. Maldonado
          \inst{1}
          \and S. Colombo\inst{1}
	  \and A. Petralia\inst{1}
	  \and S. Benatti\inst{1}
	  \and S. Desidera\inst{2}
	  \and L. Malavolta\inst{3}
	  \and A. F. Lanza \inst{4}
	  \and M. Damasso\inst{5}
	  \and G. Micela\inst{1}
	  \and M. Mallonn\inst{6} 
	  \and S. Messina\inst{4}
	  \and A. Sozzetti\inst{5}
	  \and B. Stelzer\inst{7,1}
	  \and K. Biazzo\inst{8}
	  \and R. Gratton\inst{2}
	  \and A. Maggio\inst{1}
	  \and D. Nardiello\inst{2,9}
	  \and G. Scandariato\inst{4}
	  \and L. Affer\inst{1}
	  \and M. Baratella\inst{6}
	  \and R. Claudi\inst{2}
	  \and E. Molinari\inst{10}
	  \and A. Bignamini\inst{11}
	  \and E. Covino\inst{12}
	  \and I. Pagano\inst{4}
	  \and G. Piotto\inst{3}
	  \and E. Poretti\inst{13, 14}
          \and R. Cosentino\inst{14}
	  \and I. Carleo\inst{15, 2}
          }

   \institute{INAF - Osservatorio Astronomico di Palermo,
                 Piazza del Parlamento 1, 90134 Palermo, Italy\\
              \email{jesus.maldonado@inaf.it}
	 \and INAF - Osservatorio Astronomico di Padova, vicolo dell'Osservatorio 5, 35122 Padova, Italy 
	 \and Dipartimento di Fisica e Astronomia Galileo Galilei, Vicolo Osservatorio 3, 35122 Padova, Italy 
	 \and INAF - Osservatorio Astrofisico di Catania, Via S. Sofia 78, 95123, Catania, Italy 
	 \and INAF - Osservatorio Astrofisico di Torino, Via Osservatorio 20, 10025 Pino Torinese, Italy 
         \and Leibniz-Institute for Astrohpysics Potsdam (AIP),  An der Sternwarte 16, D-14482, Potsdam, Germany 
	 \and Institut f\ {u}r Astronomie und Astrophysik, Eberhard-Karls Universit\ {a}t T\ {u}bingen, Sand 1, D-72076, T\ {u}bingen, Germany 
	 \and INAF - Osservatorio Astronomico di Roma, Via Frascati 33, 00078, Monte Porzio Catone (Roma), Italy 
	 \and Aix-Marseille Univ, CNRS, CNES, LAM, Marseille, France 
	 \and INAF - Osservatorio Astronomico di Cagliari and REM, Via della Scienza 5, 09047 Selargius CA, Italy 
         \and INAF - Osservatorio Astronomico di Trieste, Via Tiepolo 11, 34143 Trieste, Italy 
	 \and INAF - Osservatorio Astronomico di Capodimonte, Salita Moiariello 16, 80131 Napoli,Italy 
	 \and INAF - Osservatorio Astronomico di Brera, Via E.Bianchi 46, 23807 Merate, Italy 
	 \and Fundaci\'{o}n Galileo Galilei - INAF, Rambla Jos\'{e} Ana Fernandez P\'{e}rez 7, 38712 - Bre\~{n}a Baja, Spain
	 \and Astronomy Department, 96 Foss Hill Drive, Van Vleck Observatory 101, Wesleyan University, Middletown, CT, 06459, USA
             }

   \date{Received ; accepted }

 
  \abstract
    { Active region evolution plays an important role
    in the generation and variability of magnetic fields on the surface of lower main-sequence stars.
    However, determining the lifetime of active region growth and decay as well as their evolution 
    is a complex task.
    Most previous studies of this phenomenon are based on optical light curves, while little is known about
    the chromosphere and the transition region.} 
   {We aim to test whether the lifetime for active region evolution shows any dependency on the stellar
   parameters, specially on the stellar age. 
   }
   {We identify a sample of stars with well-defined ages via their kinematics and membership to
    young stellar associations and moving groups. 
    We made use  of high-resolution \'echelle spectra 
    from HARPS at La Silla 3.6m-telescope and HARPS-N at TNG
    to compute rotational velocities, activity levels,
    and emission excesses. We use these data to revisit the activity-rotation-age relationship.
    The time-series of the main optical activity indicators, namely Ca~{\sc ii} H \& K, 
    Balmer lines, Na~{\sc i} D$_{\rm 1}$, D$_{\rm 2}$, and He~{\sc i} D$_{\rm 3}$, 
    were analysed together with the available photometry by using state-of-the-art Gaussian  processes to model the stellar
    activity of these stars.  Autocorrelation functions of the available photometry were also analysed.
    We use the derived lifetimes for active region evolution to search for
    correlations with the stellar age, the spectral type, and the level of activity. 
    We also use the pooled variance technique to characterise the activity behaviour of our targets. 
   }
   {Our analysis confirms the decline of activity and rotation as the star ages.
    We also confirm that the rotation rate decays with age more slowly for cooler stars and that, for a given age,
    cooler stars show higher levels of activity. We show that  F- and G-type young stars
    also depart from
    the inactive stars in the flux-flux relationship. 
     The  gaussian process analysis of the different activity indicators does not seem to provide any useful information
    on active region's lifetime and evolution. On the other hand, active region's lifetimes derived from the 
    light-curve analysis might correlate with the stellar age and temperature.
   }
   {
   Although we caution the small number statistics, our results suggest that
   active regions seem to live longer on younger, cooler, and more active stars.

   }


   \maketitle
%
\section{Introduction}\label{intro}

 The relationships among stellar activity, rotation, and stellar age in solar-type
 stars have been widely studied. 
 Chromospheric activity and rotation are linked by the stellar dynamo and
 as the star evolves during the main-sequence phase loosing angular momentum via magnetic braking,
 both rotation and activity diminish { 
 \citep[e.g.][]{1962AnAp...25...18S,1967ApJ...150..551K,1967ApJ...148..217W,1972ApJ...171..565S,1984ApJ...279..763N,1989ApJ...343L..65K,1991ApJ...375..722S,1993MNRAS.261..766J,2001MNRAS.326..877M,2007ApJ...669.1167B,2008ApJ...687.1264M}.
}

 Active region (hereafter AR) growth and decay is another phenomenon related to the surface
 magnetic activity of solar-type stars with convective outer layers. The study of AR is fundamental
 to improve our knowledge about the generation of magnetic fields and their variability.
 However, there are few works dealing with the analysis of AR lifetimes.
 In a series of papers, \cite{1997SoPh..171..191D,1997SoPh..171..211D} use the pooled variance
 technique on calcium data to infer the AR lifetimes of approximately one hundred of lower main-sequence stars. 
 The authors show that AR have rather irregular lifetimes and that different stars might show very
 different pooled variance diagrams depending on their level of activity (age), and colour (mass).

 More recently, several works have developed a methodology based on the decay of the autocorrelation function of light curves
 (in particular using data from the {\sc Kepler} mission) to put constraints on the spot and AR lifetime.
 \cite{2017MNRAS.472.1618G} find that big starspots live longer irrespective of the spectral type of the star and that starspots decay more
 slowly on cooler stars.  
 \cite{2021MNRAS.508..267S} and \cite{2022ApJ...924...31B} discuss the effect of differential rotation
 and how it can destroy the biggest ARs leading to a shorter AR lifetime.  
 It is important to note that these works are based on optical light curves and therefore their conclusions  refer to the  ARs evolution in the stellar photosphere.
 However, it is well known that solar AR in the chromosphere and in the transition region have lifetimes 4-5 times longer than the  ARs
 in the solar photosphere. 
 
 Therefore, a detailed and homogeneous analysis of the chromospheric activity indexes of a 
 large sample of stars with reliable age estimates
 is needed before possible mechanisms for AR growth and decay are invoked.
 This is the goal of this paper, in which we take advantage of the high amount of 
 high-resolution spectra taken within the framework of current radial velocity planet searches
 to derive in an homogeneous way 
 the time-series of the main optical activity indexes for a large sample of stars in open clusters
 and stellar associations with precise age estimates. 
 Some of these stars have also available photometry time series from the TESS mission.
 We also take advantage of state-of-the-art statistical analysis to model stellar activity such as
 Gaussian processing. 

 This paper is organised as follows. We present our stellar sample in Sect.~\ref{sample}.
 Section~\ref{analysis} describes the analysis of the data while in Sect.~\ref{results} we use our
 dataset to revisit the activity, rotation, and age relationships. 
 The dependency of AR lifetimes on spectral type, activity,
 and stellar age are discussed in Section~\ref{tau_evol}.
 Our conclusions follow in Sect.~\ref{conclusions}.

\section{Stellar sample}\label{sample}

\begin{figure}[!htb]
\centering
\includegraphics[scale=0.75]{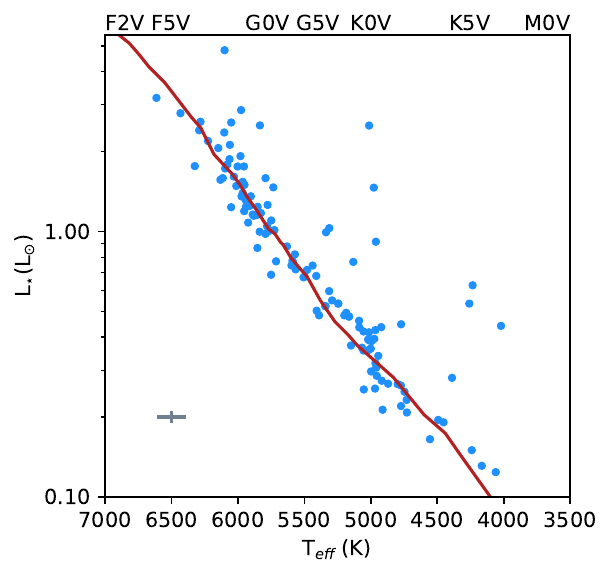}
\caption{
Luminosity versus T$_{\rm eff}$ diagram for the observed stars.
The red line shows the main sequence from \cite{2013ApJS..208....9P}.
}
\label{hrdiagram_plot}
\end{figure}

 The sample analysed in this work is composed of 130 stars in open clusters
 or stellar associations with well known derived ages.
 The bulk of the sample is formed by stars observed within the framework of the Global
 Architecture of Planetary Systems programme \citep[GAPS,][]{2013A&A...554A..28C}.
 In particular, 25 stars were selected from the GAPS Young Objects Project,
 a radial velocity survey aimed to probe the frequency of planets around young stars
 \citep[][]{2020AA...638A...5C}.
 It includes
 young stars in well known star forming regions (e.g. the Taurus complex with 
 an age of $\sim$ 2 Myr)
 as well as bona-fide members of open clusters and moving groups
 (such as Coma Ber or Ursa Major, age $\sim$ 400-600 Myr).
 Additional 48 stars were taken from the GAPS Open Cluster Project,
 a monitoring of selected stars in three open clusters
 (namely the Hyades, M44, and NGC 752)
 aimed  to study the relation between the physical properties of the planets and those of their host stars
 as well as the connection between the physical properties of the cluster environments and those of their planetary systems.
 \citep[][]{2016A&A...588A.118M}.
 Finally, 56 stars members of clusters and moving groups or with well-known ages were selected
 from \citet[][hereafter MH08]{2008ApJ...687.1264M}.
 Table~\ref{list_of_associations} lists the number of stars by open cluster or kinematic group,
 while the corresponding Hertzsprung-Russell (HR) diagram of the observed stars is shown in Fig.~\ref{hrdiagram_plot}.

\begin{table}[!htb]
\centering
\caption{Number of observed stars per cluster or moving group.}
\label{list_of_associations}
\begin{tabular}{lrll}
\hline\noalign{\smallskip}
 Association               &  N stars  & Age     & Ref. \\
                           &           & (Myr)   &           \\ 
\hline
Taurus	                   & 4	       & 1 - 2     &  (a) \\ 
Upper Sco		   & 4         & 10        &  (b) \\ 
Cepheus			   & 2         & 10 - 20   &  (c) \\   
$\beta$ Pic		   & 2         & 24        &  (d) \\ 
Tucana - Horologium        & 4         & 30        &  (e) \\ 
Pleiades                   & 2         & 112       &  (f)  \\ 
AB Dor			   & 2         & 149       &  (g) \\ 
Castor                     & 1         & 200       &  (h) \\ 
Hercules - Lyra            & 1         & 257       &  (i) \\ 
Ursa Major                 & 6         & 414       &  (j) \\ 
Coma Berenices             & 6         & 562       &  (k) \\ 
Praesepe                   & 20        & 578       &  (l) \\ 
Hyades                     & 49        & 750       &  (m) \\ 
Other young  stars         & 2         & 50 - 600  &  (n,o) \\ 
NGC 752                    & 12        & 1340      &  (p)  \\
Old stars                  & 13        & 5300 - 13900 & (q) \\
\hline
Sun			   &           & 4579$^{\dag}$ & (r) \\ 
\hline
\end{tabular}
\tablefoot{(a) \cite{1995ApJS..101..117K}; (b) \cite{2016MNRAS.461..794P}; (c) \cite{2020A&A...637A..43K}; (d) \cite{2015MNRAS.454..593B}; (e) \cite{2008hsf2.book..757T}; (f) \cite{2015ApJ...813..108D}; (g)  \cite{2015MNRAS.454..593B}; (h)  \cite{1998A&A...339..831B}; (i)  \cite{2013A&A...556A..53E}; (j)  \cite{2015ApJ...813...58J}; (k)  \cite{2014A&A...566A.132S}; (l) \cite{2011MNRAS.413.2218D}; (m)  \cite{2015ApJ...807...58B}; (n) \cite{2021AA...645A..71C}; (o) This work;  (p)  \cite{2018ApJ...862...33A}; (q)  \cite{2008ApJ...687.1264M}; (r) \cite{2005Natur.436.1127B}; $^{\dag}$ Minimum age}
\end{table}

 The stars are required to have high-resolution,
 HARPS-N \citep{2012SPIE.8446E..1VC} or HARPS \citep{2003Msngr.114...20M} 
 optical \'echelle spectra. 
 The instrumental setup of HARPS and HARPS-N is almost identical.
 The spectra cover the range 378-691 nm (HARPS) and 383-693 nm (HARPS-N) with a resolving power
 of R $\sim$ 115000. The spectra are provided already reduced using HARPS-N/ESO standard calibration pipelines (Data Reduction Software, DRS version 3.7 and 3.8 respectively )
 and were retrieved from the corresponding ESO\footnote{http://archive.eso.org/wdb/wdb/adp/phase3\_spectral/form?} 
 and TNG\footnote{http://archives.ia2.inaf.it/tng/} archives.
 In addition, several solar spectra taken by the HARPS-N solar telescope
 \citep{2021A&A...648A.103D} were analysed in order to use the Sun as a benchmark. 

\section{Analysis}\label{analysis}
\subsection{Kinematics and age}
 
  Stellar age is one of the most difficult stellar parameter to constrain
  in an accurate way. Solar-type stars evolve too slowly to be dated by their position
  in the Hertzsprung-Russell diagram. 
  Membership to stellar associations and kinematic groups has been proposed
  as a way to overcome this difficulty and used as a methodology
  to identify young stars and to assign ages, specially after the release
  of the {\sc Hipparcos} data.
  Today, the exquisite precision of the recently released {\it Gaia} EDR3 catalogue
  \citep{2020yCat.1350....0G} allows us to compute precise Galactic spatial velocity components
  {\bf $(U, V, W)$} and detailed probabilities of membership to young stellar
  associations. 
  
\begin{figure*}[!htb]
\centering
\begin{minipage}{0.48\linewidth}
\includegraphics[scale=0.45]{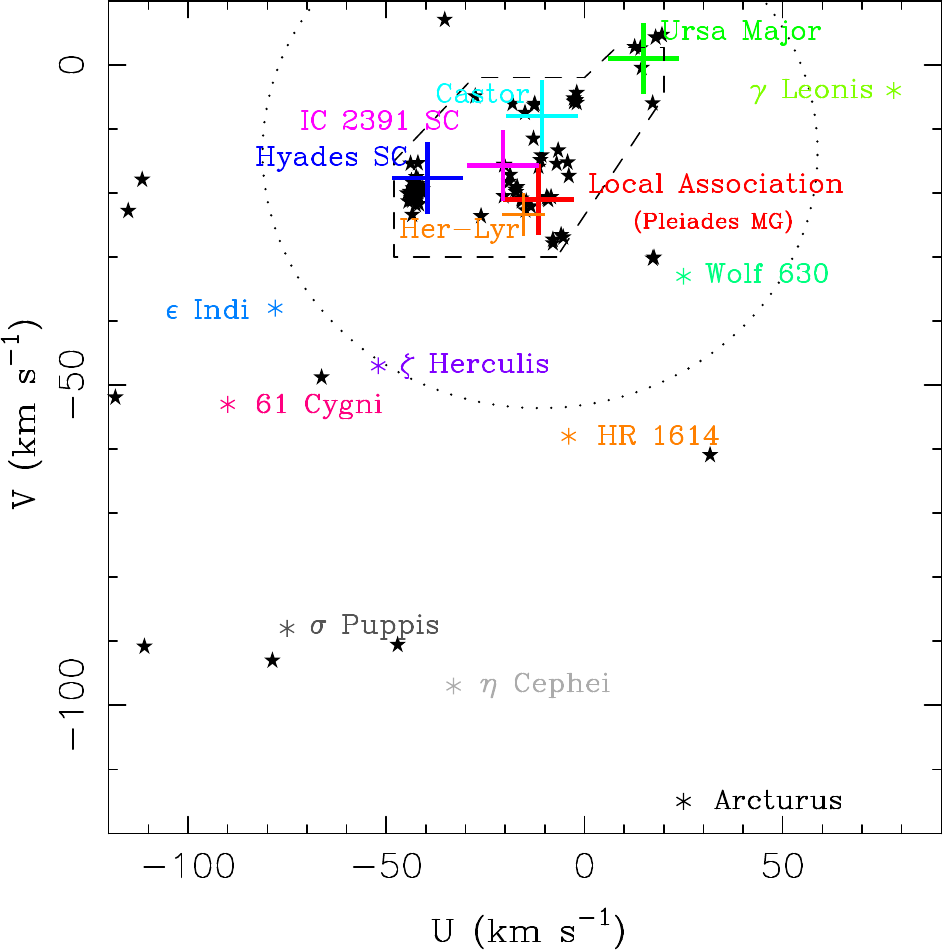}
\end{minipage}
\begin{minipage}{0.48\linewidth}
\includegraphics[scale=0.45]{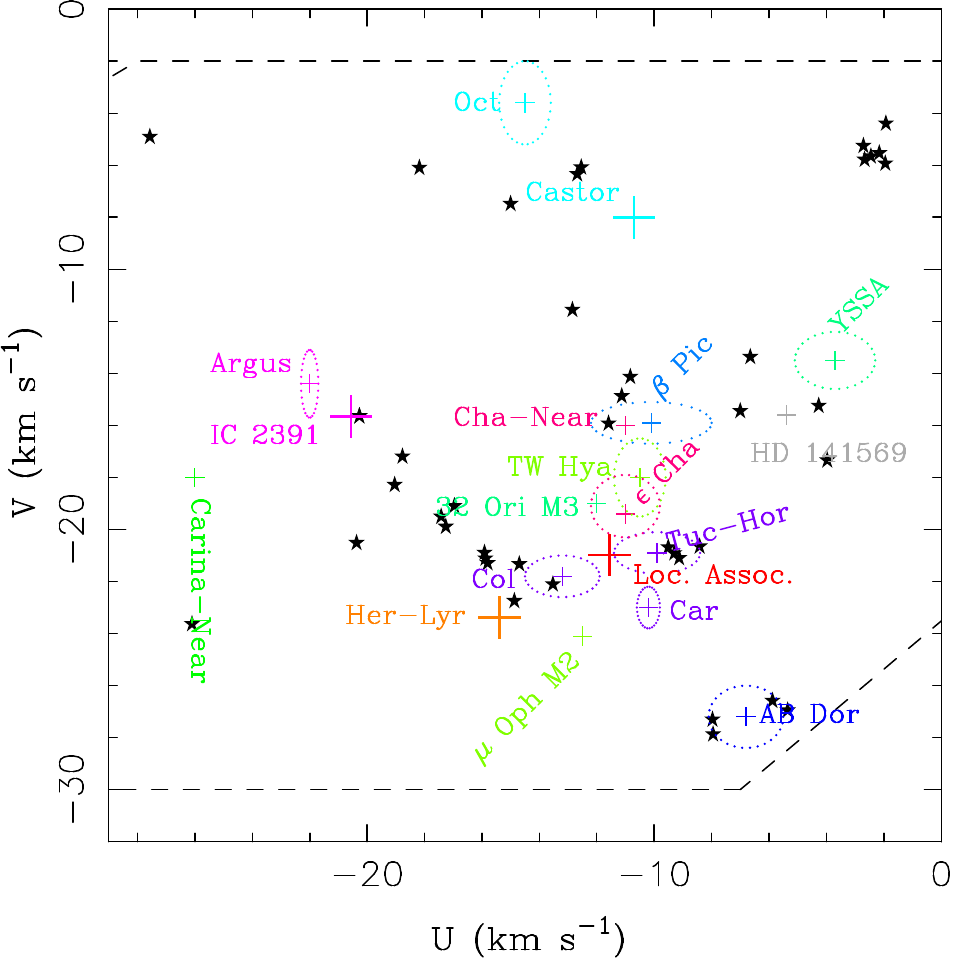}
\end{minipage}
\caption{
Left: {\bf $( U, V)$} plane showing the position of the known kinematic groups
in the solar neighbourhood. Large crosses represent the convergence point of the
``classical'' moving groups.  Coloured asterisks show the position of the so-called ``old'' moving groups.
Our stars are shown by black filled star symbols.
The dashed line represents the boundary of the young disc population as defined by 
\citet{1984AJ.....89.1358E,1989PASP..101..366E}.
The dotted line represents the velocity ellipsoid determined by \cite{2009NewA...14..615F}.
Right: Zoom of the region of the {\bf $( U, V)$} plane around the Local Association.}
\label{uvplane}
\end{figure*}

  Galactic spatial velocity components were computed 
  from the radial velocities, and {\it Gaia} parallaxes
  and proper motions \citep{2020yCat.1350....0G}  
  following the procedure described in \cite{2001MNRAS.328...45M} and
  \cite{2010A&A...521A..12M}. In brief, the original algorithm \citep{1987AJ.....93..864J}
  is adapted to epoch J2000 in the International Celestial Reference System
  (ICRS) as described in Sect. 1.5 of  The Hipparcos and Tycho Catalogues'
  \citep{1997ESASP1200.....E}. To take into account the possible correlation between the
  astrometric parameters, the full covariance matrix was used in computing the
  uncertainties. The corresponding {\bf $(U, V)$} plane is shown in Fig.~\ref{uvplane}.

  It can be seen that most of our targets are in the region of the diagram
  occupied by the young stars (as expected). Only some old stars taken from the literature
  (see above) are outside the boundary of the young star's region. 
  Once young stars are identified we made use of  Bayesian 
  methods to confirm their membership to young stellar associations
  \citep[BANYAN,][]{2018ApJ...856...23G}\footnote{http://www.exoplanetes.umontreal.ca/banyan/banyansigma.php}.


\subsection{Rotational velocity}

 Rotational velocities were computed by means of the Fourier Transform  (FT)
 technique \citep[e.g.][]{2008oasp.book.....G}. 
 In brief, the dominant term in the Fourier transform of the rotational profile
 is a first-order Bessel function that produces a series of relative minima
 at regularly spaced frequencies. 
 The first zero of the Fourier transform is related to $v\sin i$ by:

 \begin{equation}
  v\sin i=\frac{c}{\lambda}\times\frac{k_{1}}{\sigma_{1}}
 \end{equation} 

 \noindent where $c$ is the speed of light, $\lambda$ is the central wavelength of the
 considered line, ${\sigma_{\rm 1}}$ is the position of the first zero of the Fourier
 Transform, and ${k_{\rm 1}}$ is a function of the limb darkening coefficient ($\epsilon$) that can
 be approximated by a fourth-order polynomial degree \citep{1990A&A...237..137D}

 \begin{equation}
 k_{1} = 0.610 + 0.062\epsilon + 0.027\epsilon^{2} + 0.012\epsilon^{3} + 0.004\epsilon^{4}
 \end{equation}

 \noindent where we assume $\epsilon$ = 0.6 \citep[see e.g.][]{2008oasp.book.....G}.  %
  Four spectral lines at  6335.33 \AA, 6378.26 \AA, 6380.75 \AA, and 6393.61 \AA \space were used for the computations.
 An additional line at 6400.11 \AA \space  was used but only for stars with high rotation values,
 since it is a blend of two lines that at low rotation levels are resolved.
 Given that it is not an isolated line, we only considered this blend if the derived
 $v\sin i$ value was compatible with the values obtained from the  other lines.

 In addition to the FT method, we fitted each line profile to a rotational profile following
 the prescriptions of \citet[][]{2008oasp.book.....G}. 
 The fits were performed within a Bayesian framework based on a Monte Carlo Markov Chain (MCMC) sampling of the parameter
 space. Since the rotational profile does not take into account the wings of the line profile 
 (we note that the function is not defined on those points), the profile was convolved with a Lorentzian 
 profile. Therefore, the model contains five parameters namely, the centre of the profile, the depth of the profile, 
 the Lorentzian parameter, the amplitude of the profile, and an additional jitter term. 
 
\begin{figure*}[!htb]
\centering
\begin{minipage}{0.33\linewidth}
\includegraphics[scale=0.45]{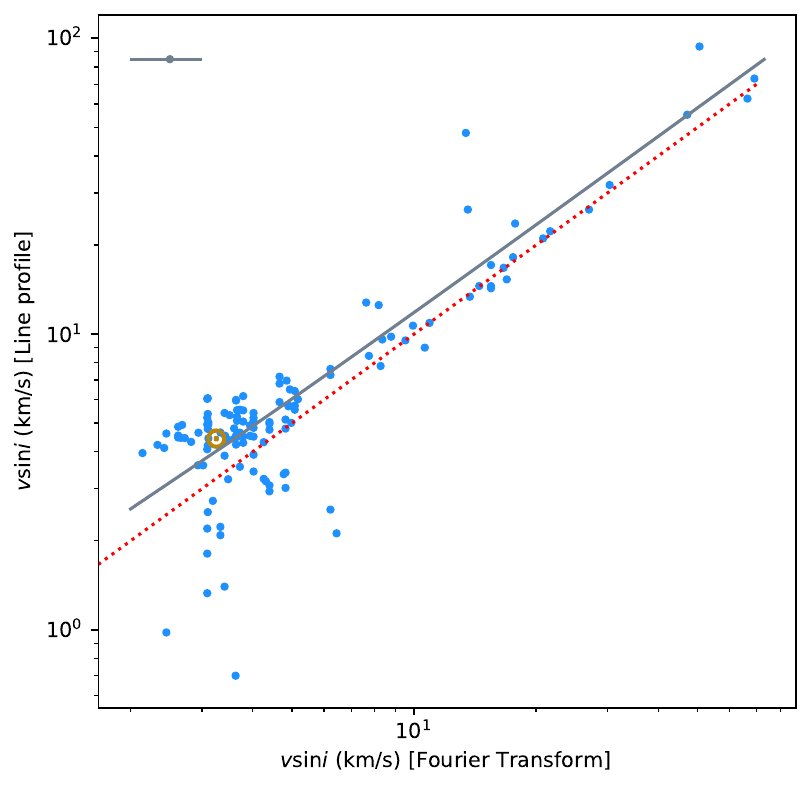}
\end{minipage}
\begin{minipage}{0.33\linewidth}
\includegraphics[scale=0.45]{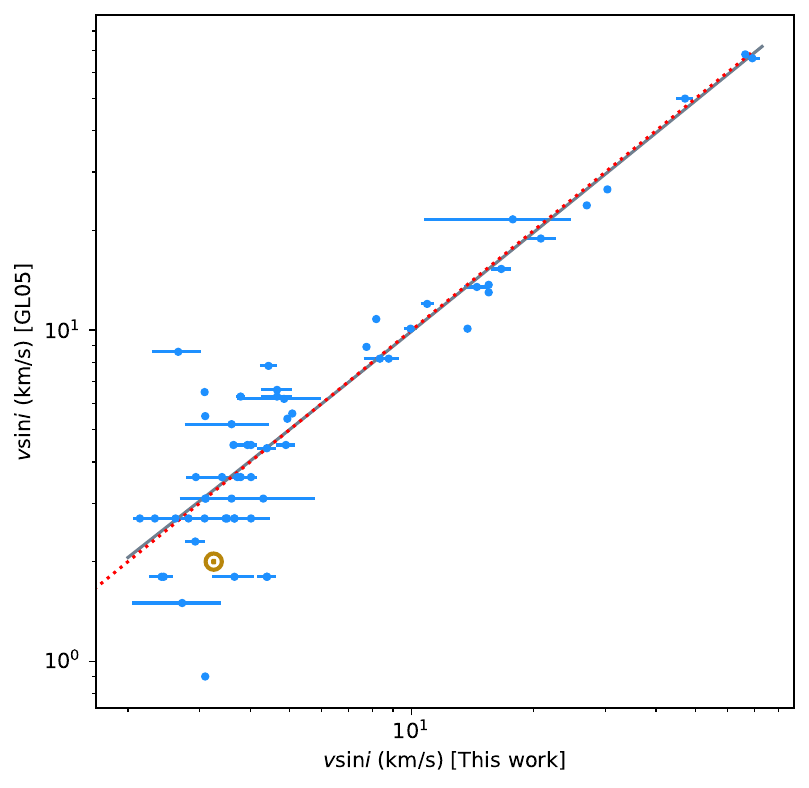}
\end{minipage}
\begin{minipage}{0.33\linewidth}
\includegraphics[scale=0.45]{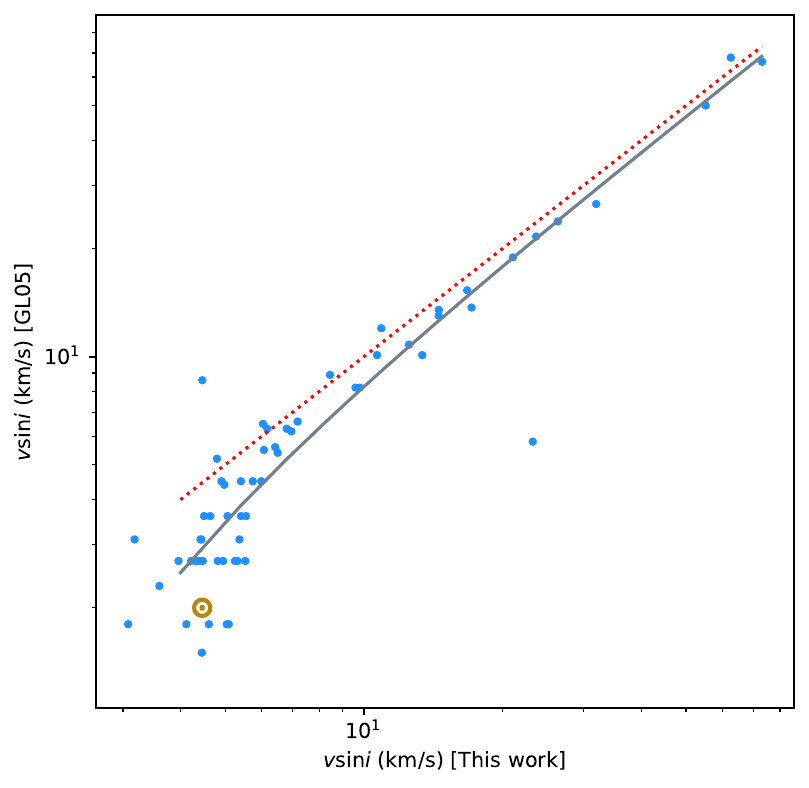}
\end{minipage}
\caption{
Left: Comparison between the $v \sin i$ derived from the fit of line profiles to a rotational profile, and $v \sin i$ derived
by using the FT technique. Centre:
Comparison between the mean $v \sin i $ provided by GL05 and those measured in this work using the FT technique.
Right: Comparison between the mean $v \sin i $ provided by GL05 and those measured in this work by the fitting of line profiles
to a rotational profile.
The red dotted line shows the 1:1 relationship while the grey dashed line shows the best
linear fit. The position of the Sun is shown with the symbol $\odot$. Typical error bars in GL05 data are 
of the order of 1.4 kms$^{\rm -1}$.}
\label{gl05_comp}
\end{figure*}

 A comparison of the derived $v \sin i $ values obtained by using the FT method and those derived using
 the line profile fitting is shown in Fig.~\ref{gl05_comp} (left panel). 
 Although the overall agreement is good,
 it can be seen that the line profile fitting
 method tends to provide slightly larger $v \sin i $ values.
 In particular, we note that for the Sun we obtain a mean value of 3.23 $\pm$ 0.13 kms$^{\rm -1}$ when using the FT method,
 and 4.43 $\pm$ 0.26 kms$^{\rm -1}$ from the fitting profile technique (that can be compared with the
 adopted value of $\sim$ 2.0 kms$^{\rm -1}$).
 Fig.~\ref{gl05_comp} also shows a comparison of our obtained equatorial velocities with those provided in the literature.
 The literature values are taken from the compilation of
 \citet[][hereafter GL05]{2005yCat.3244....0G}. 
 It can be seen that the agreement of our FT values with those from the literature is overall good (centre panel), specially at $v \sin i $ values larger than
 10 kms$^{\rm -1}$, with most stars lying close to the 1:1 relationship. At lower rotation levels, however, the scatter is larger.
 The residual mean square (rms) of the comparison is 1.90 kms$^{\rm -1}$, the root-mean squared
 error (rmse) is 3.6 kms$^{\rm -1}$, and the R$^{\rm 2}$ (coefficient of determination) is $\sim$ 0.98.
 When considering the values derived from the line profile fitting (right panel), we obtain larger values than those found in the literature.
 This effect is more pronounced at the low rotation level. 
 In this case, we obtain an rms value of 3.40 kms$^{\rm -1}$, with an rmse value of 11.6 kms$^{\rm -1}$, and R$^{\rm 2}$ $\sim$ 0.93.

 We conclude that the rotational profile fitting method works better at large rotational velocities.
 At low-rotation levels, however, the width of the line profiles are dominated by
 the intrinsic sources of line broadening such
 as micro and macroturbulence, pressure and magnetic Zeeman
 splitting. As a consequence, in low-rotation stars, the fitting method tends to overestimate the $v \sin i $ values.
 Therefore, in the following, we will consider only the  $v \sin i $ values derived by using the FT method.

\subsection{Activity indexes}

\begin{figure}[!htb]
\centering
\includegraphics[scale=0.60]{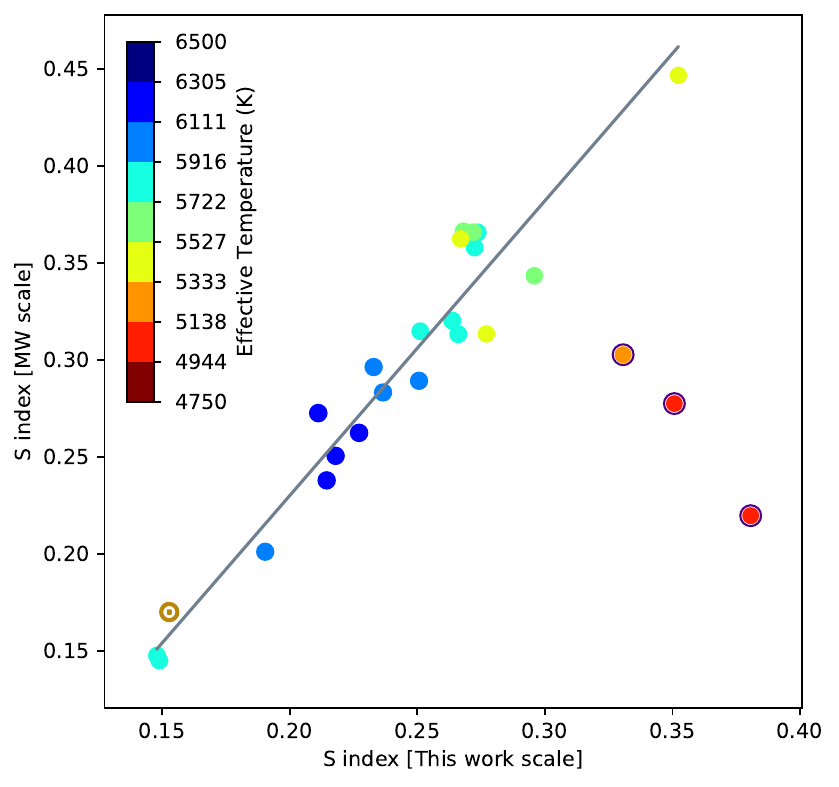}
\caption{
Activity S index in the original Mount Wilson scale as a function of the S index as measured
in this work. Different colours indicate the effective temperature of the stars. 
Outliers are highlighted in purple circles. Errors are within the symbol dimension.  
The best linear fit is shown in grey.}
\label{sindex_comparison}
\end{figure}

 For the examination of activity indexes we use the strong optical lines
 Ca~{\sc ii} H \& K,
 Balmer lines (from H$\alpha$ to H$\epsilon$), Na~{\sc i} D$_{\rm 1}$, D$_{\rm 2}$,
 and He~{\sc i} D$_{\rm 3}$. Our definition of the bandpasses
 for the activity indexes follows \citet[][and references therein]{2019A&A...627A.118M}.
  In order to transform the measured S index into R'$_{\rm HK}$, a mean S index
  was computed for each star and transformed into the Mount Wilson scale by a comparison with
  the stars in common with \cite{1991ApJS...76..383D}. The comparison is shown in Figure~\ref{sindex_comparison}.
  An ordinary least squares fit was performed in order to obtain a relationship between the S index
  measured in this work and the S index in the Mount Wilson scale. A 3$\sigma$ clipping procedure was applied
  to identify outliers to the best linear fit. We note that the outliers correspond to stars for which
  only one measurement is available. Given their rather high S index values
  we speculate that these stars might have a high a level of chromospheric variability.
  We obtain the following relationship

\begin{equation}
S_{\rm MW} =  (1.52 \pm 0.10) \times S_{\rm tw} - (0.074 \pm 0.025)
\end{equation}

 \noindent where S$_{\rm MW}$ is the S index in the Mount Wilson scale and 
 S$_{\rm tw}$ is the S index as measured in this work. 

   The S index contains both the contribution of the photosphere and the chromosphere.   
   Empirical relationships to correct for the photospheric contribution have been calibrated
   using the colour index (B-V). Furthermore, a conversion factor to correct for flux variations
   in the continuum  passbands and normalise to the bolometric luminosity should be applied
   \citep{1984ApJ...279..763N}.
   Unfortunately, for most of our targets there are no reliable 
   B, V magnitudes available in the literature.
   Therefore, we  use the {\it Gaia} DR2 \citep{2018yCat.1345....0G} effective temperature to estimate
   the (B-V) colour of our target stars. In order to do that, we derive a (B-V)-T$_{\rm eff}$
   relationship using the data by \cite{1996ApJ...469..355F}. Details on this calibration
   are given in Appendix~\ref{appendix1}.

   The conversion factor and photospheric corrections most widely used are those provided by
   \citet[][hereafter NO84]{1984ApJ...279..763N}. They are, however, only valid for solar-type stars with 0.44 $<$ (B-V) $<$ 0.82
   (spectral types between F5 and K2). More recently, \cite{2015MNRAS.452.2745S}  derived conversion factors and photospheric
   corrections for cooler stars (up to (B-V) = 1.9).

\subsection{Rotation periods and lifetime of active regions}\label{gp_description} 

  In order to determine the stellar rotation period from the different activity indexes and the available TESS photometric time
  series\footnote{We use the 2-minutes cadence TESS light curves available for the systems. We use the data corrected for time-correlated instrumental signatures, thus the PDCSAP flux column in the FITS file \citep{2016SPIE.9913E..3EJ}.}, 
  we started by using the Generalised Lomb-Scargle periodogram \citep{2009A&A...496..577Z} to identify periodic signals in the data.
  We then model the data using Gaussian Process regression in a Bayesian framework 
  with the following likelihood

  \begin{equation}
  ln p({y_n},{t_n},{\sigma^2_n},\theta) = -\frac{1}{2} \textbf{r}^T K^{-1} \textbf{r} - \frac{1}{2} \ln{\det K} - \frac{N}{2} \ln{2\pi}
  \end{equation}

 \noindent where $y_n$, $t_n$, $\sigma_n$ are, respectively, the data, time of observations and errors, $\theta$ is the array of parameters, $\textbf{r}$ is the residual vector obtained by removing the model from data, $K$ is the covariance matrix and $N$ are the number of observations.

  We selected the widely used Quasi-Periodic function obtained by multiplying a constant term to an exp-sin-squared kernel and to a squared-exponential kernel
  \citep[{\tt george} python package, e.g.][]{2015ITPAM..38..252A,2021A&A...649A.157G,2021A&A...651A..93M} and it is defined as follows

\begin{equation}\label{kernel}
k(i,j) = h^2 \exp{\left(-\frac{(t_i-t_j)^2}{\tau^2} -\frac{\sin^2(\pi(t_i-t_j)/P_{rot})}{2\omega^2}\right)} 
\end{equation}

\noindent where $k(i,j)$ is the $i$-th $j$-th element of the covariance matrix, $t_i$ and $t_j$ are two times of the data set, $h$ is the amplitude of the covariance, $\tau$ is the timescale of the exponential component, $\omega$ is the weight of the periodic component, $P_{rot}$ is the period. 



We do not include an extra error term  ($\sigma_{Jit}$) as the uncertainties of the measured indexes are relative large. However, we added a  linear trend model ($\gamma$ + $\dot{\gamma}t$).
At the beginning of the Gaussian Process regression, both data and errors have been cleaned by a (3) sigma-clip procedure.
The parameter space is sampled with {\tt emcee} \citep{emcee} set with 72 walkers randomly initialised within parameter boundaries, they are reported in Table~\ref{tab:priors}. 
We have imposed uniform priors on $P_{\rm rot}$ with boundaries according
to the False Alarm Probability (FAP) of the maximum power GLS period. 
We used as prior boundaries P$_{\rm GLS}$ $\pm$ 1 d, P$_{\rm GLS}$ $\pm$ 3 d, P$_{\rm GLS}$ $\pm$ 8 d, and  P$_{\rm GLS}$ $\pm$ 15 d for FAP $<$ 0.1\%,  0.1\% $<$ FAP $<$ 1\%, 10\% $<$ FAP $<$ 1\%, and
10\% $>$ FAP, respectively.
We note that for the analysis of the TESS photometric data, better results were obtained in some cases by using gaussian priors
centred around the known values of  P$_{\rm rot}$ (see Sect.~\ref{gp_subsection}).
Finally we set a conservative burn-in phase of 40K, while 10K were used to obtain the posterior distributions. 

\begin{table*}[t]
\centering
\caption{Model priors. Labels $\mathcal{U}$ and $\mathcal{LU}$ represent uniform and log-uniform distribution, respectively, while P$_{GLS}$ is the period of maximum power obtained from the GLS periodogram.}
\label{tab:priors}
\begin{tabular}{l l  l}
\hline
\hline
\noalign{\smallskip}
Parameter & Priors & Description \\
\noalign{\smallskip}
\hline
\noalign{\smallskip}
\multicolumn{3}{c}{\textit{linear trend}} \\
\noalign{\smallskip}
$\gamma$       &     $\mathcal{U}$(min(index), max(index))                             & Minimum and maximum value of the index \\
$\dot{\gamma}$ &     $\mathcal{U}$(min(slope),max(slope))                              & Slopes of the data computed in the first/second \\
               &                                                                      & half-seasons of the observations (d$^{\rm -1}$) \\
\noalign{\smallskip}
\multicolumn{3}{c}{\textit{GP parameters}} \\
\noalign{\smallskip}
$h$             & $\mathcal{L} \mathcal{U}$(10$^{\rm -6}$, 10$^{\rm +6}$)                                     &      \\
$\tau$          &  $\mathcal{L} \mathcal{U}$((minimum $P_{\rm rot}$ prior)/2, 10$^{\rm 4}$)                                                          & (d)  \\
$\omega$        & $\mathcal{L} \mathcal{U}$(10$^{\rm -2}$,10)                 &      \\
$P_{\rm rot}$   & $\mathcal{L} \mathcal{U}$(P$_{\rm GLS}$ $\pm$ $nn$ d)         & $nn$ depends on the FAP, see text (d)  \\
\noalign{\smallskip}
\hline
\noalign{\smallskip}
\end{tabular}
\end{table*}

  

\section{The rotation - age - activity relationships}\label{results}
\subsection{Rotation vs. age and spectral type}

\begin{figure}[!htb]
\centering
\includegraphics[scale=0.45]{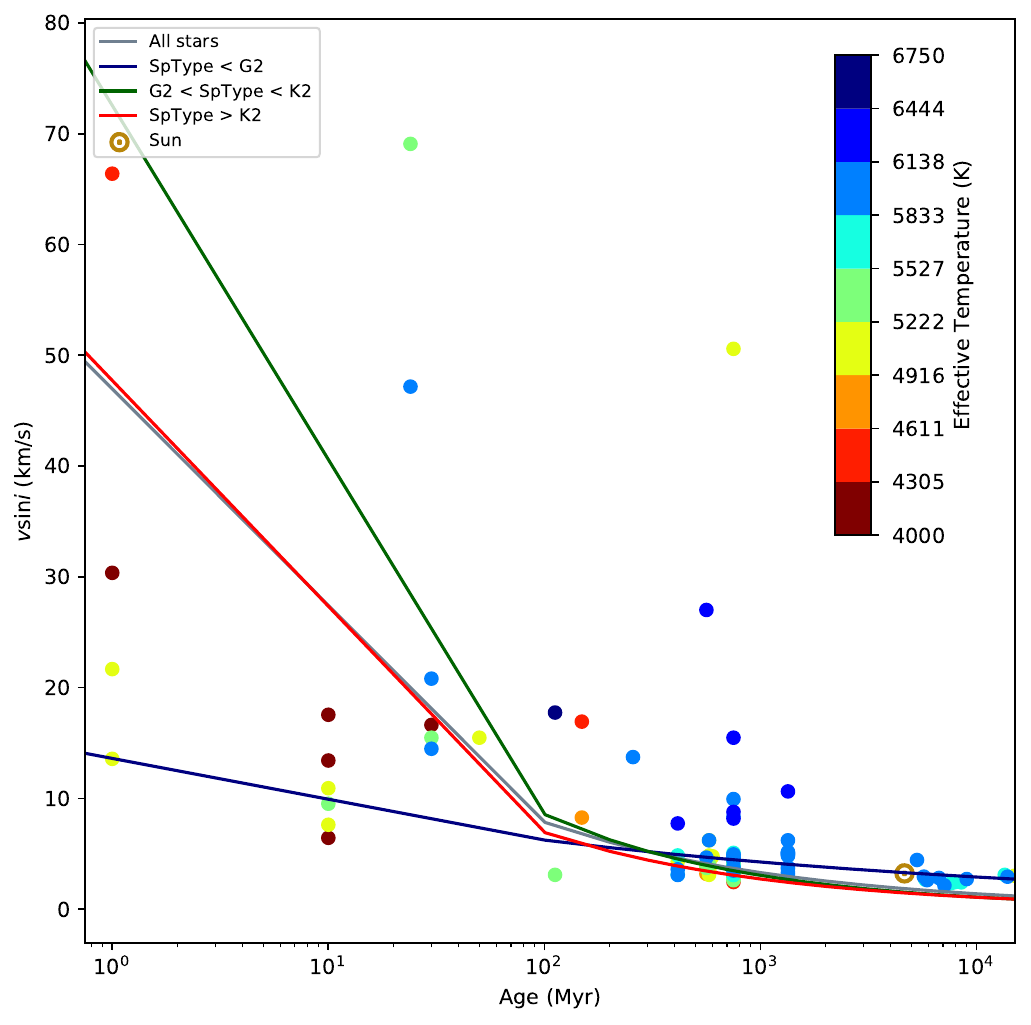}
\caption{ 
Projected rotational velocity, $v \sin i$, as a function of the stellar age. Different colours indicate the effective temperature of the
stars. Solid lines indicate the best fit.}
\label{plot_vsini_age}
\end{figure}

 Figure~\ref{plot_vsini_age} shows the $v \sin i$ values as a function of the stellar age.
 The general tendency of lower rotation rates towards older stellar ages is
 clearly visible. 
 We fit the data to a power law of the form:
 \begin{equation}
 v \sin i \propto \alpha \times t^{\beta}
 \end{equation}

 \noindent where the parameters $\alpha$ and $\beta$ are drawn from
 a bayesian framework using an MCMC simulation. The best-fit parameters
 are given in Table~\ref{tab_vsiniage}. 
 The figure also shows a dependency of the rotation vs. age relationship on the
 stellar spectral type. Rotation in cooler stars shows a lower decay than
 in hotter stars. In order to test that, we divided our target stars into three
 subsamples, namely, stars hotter than 5790 K (that is, a G2-type star),
 stars with effective temperatures between 4800 K and  5790 K
 (spectral type between G2 and K2),
 and stars cooler than K2. The results are given in Table~\ref{tab_vsiniage}. 
 They show that the $\beta$ parameter (the slope)
  is greater
 for stars with spectral type earlier than G2, while the constant of proportionality, $\alpha$,
 does not seem to vary according to the spectral type.

  

\begin{table}
\centering
\caption{ Best derived parameters for the fit $v\sin i = \alpha\times t^{\beta}$. }
\label{tab_vsiniage}
\begin{tabular}{lccc}
\hline
 Sample & $\alpha$ & $\beta$ & $N$ \\
 \hline
 All stars        & 44.34$^{\rm + 0.07}_{\rm -0.07}$  & -0.3760$^{\rm + 0.0003}_{\rm -0.0003}$ & 127 \\
 SpType < G2      & 13.41$^{\rm + 0.15}_{\rm -0.15}$  & -0.166$^{\rm + 0.002}_{\rm -0.002}$ &  45 \\
 G2 < SpType < K2 & 67.29$^{\rm + 0.20}_{\rm -0.20}$  & -0.4478$^{\rm + 0.0005}_{\rm -0.0005}$ &  64 \\
 SpType > K2      & 44.74$^{\rm + 0.08}_{\rm -0.09}$  & -0.4051$^{\rm + 0.0005}_{\rm -0.0005}$ &  18 \\
\hline
\end{tabular}
\end{table}

\subsection{The stellar age - activity relationship}

\begin{figure}[!htb]
\centering
\includegraphics[scale=0.45]{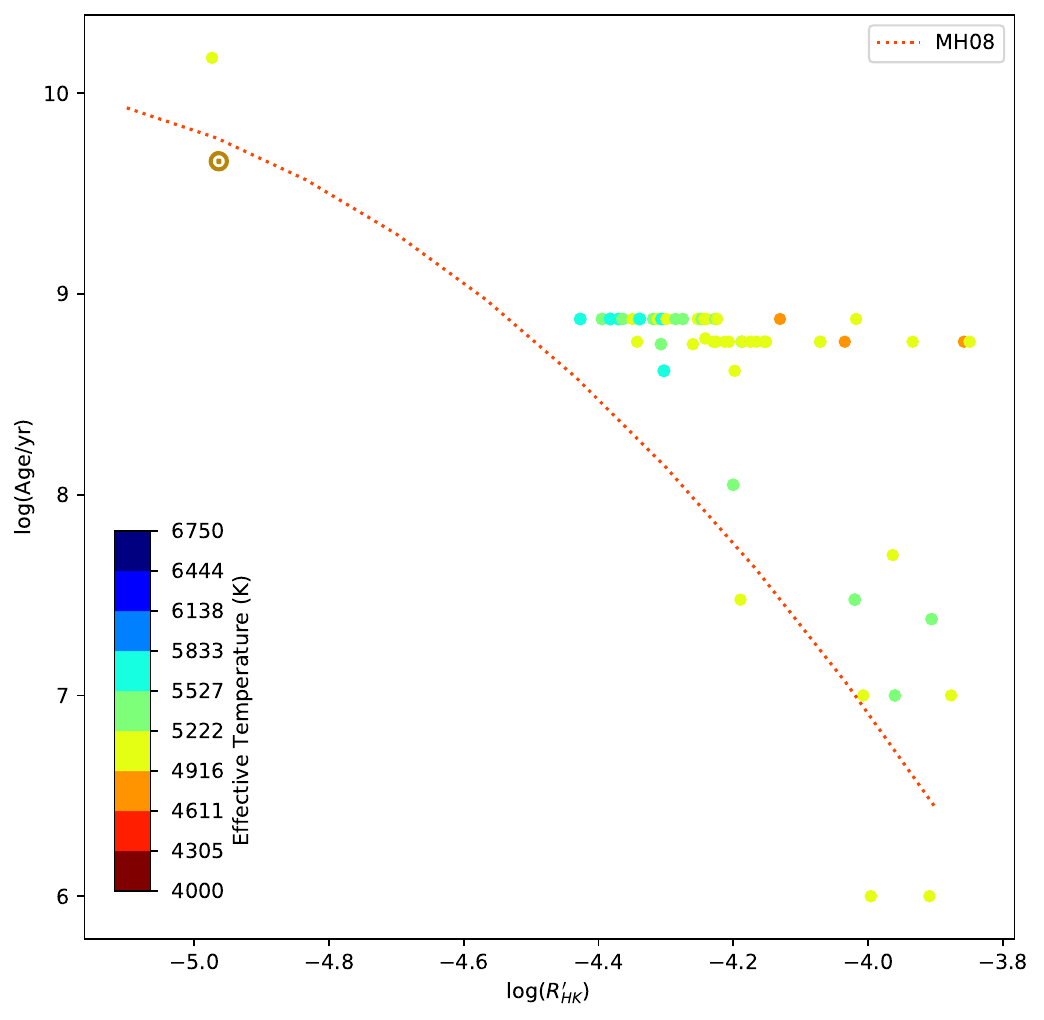}
\caption{
Stellar age as a function of $\log$R$^{\rm '}_{\rm HK}$.
Different colours indicate the effective temperature of the stars.
The dotted line shows the relationship
obtained by MH08. 
The position of the Sun is shown with the symbol $\odot$.}
\label{plot_rhk_age}
\end{figure}

 Figure~\ref{plot_rhk_age} shows the stellar age as a function of the level of stellar
 activity 
 in terms of $\log$R$^{\rm '}_{\rm HK}$.
 For an easy comparison with previous works we used the $\log$R$^{\rm '}_{\rm HK}$
 values derived using the NO84 prescriptions.
 The dotted line shows the empirical
 relationship obtained by MH08.

 It can be seen that the MH08 relationship predicts slightly younger ages
 for stars older than the Hyades.  
 However, our sample is affected by several biases. 
 To start with,
 only 23.3\% of our stars have ages younger than $\sim$ 500 Myr
 (and only one has a colour index within the range of the NO84 calibrations).
 Another bias that might affect our results is the fact that at older ages
 our sample is mainly composed of stars with effective temperatures hotter than
 $\sim$ 5500 K and, therefore, they show lower levels of stellar activity than, otherwise similar, cooler stars.
 The dependency of the age-activity relationship on the spectral type, is quite
 clear when looking at the stars in the Hyades cluster, where it can be seen that cooler stars
 show higher levels of activity.

\subsection{Flux-flux relationships}

 Emission excesses in the Ca~{\sc ii} H \& K and H$\alpha$ lines were
 determined by using the spectral subtraction technique 
 \citep[e.g.][]{1995A&AS..114..287M,2000A&AS..146..103M}. In brief, the basal chromospheric flux
 is removed by using the spectrum of a non-active star of similar
 stellar parameters and chemical composition to the target star
 as reference.
 
 Reference stars were selected from \cite{raquel2010}. Fluxes were derived
 from the measured equivalent width in the subtracted spectra
 by correcting the continuum flux

 \begin{equation}
 \log(F_{\lambda}) = \log(EW) + \log(F^{\rm cont}_{\lambda})
 \end{equation}

 \noindent where the continuum flux, $F^{\rm cont}_{\lambda}$,  was determined by using
 the empirical calibrations with the colour index, $(B-V)$, derived by \cite{1996PASP..108..313H}. 

\begin{figure}[!htb]
\centering
\includegraphics[scale=0.50]{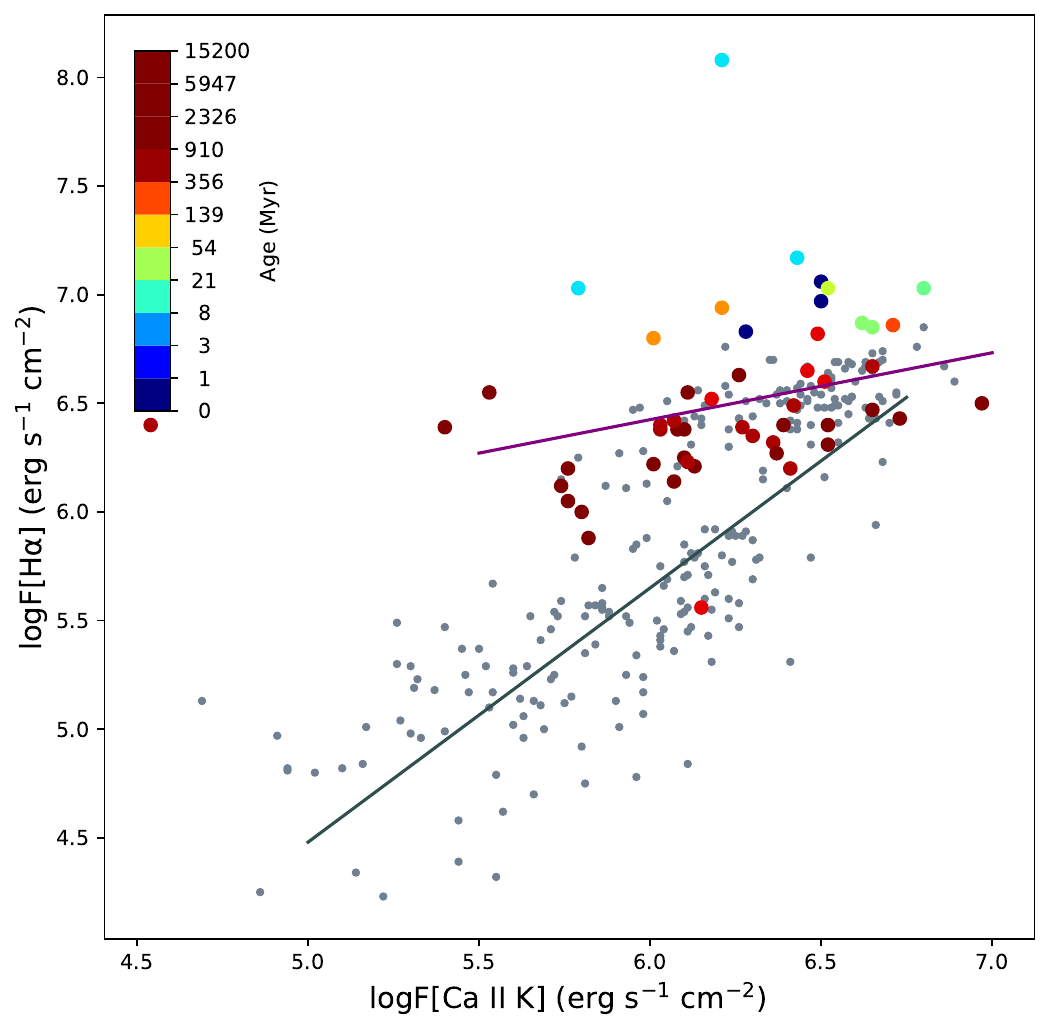} 
\caption{ 
Flux-flux relationship between H$\alpha$ and Ca~{\sc ii} K. Different colours 
indicate the age of the stars. 
For comparison, data for
FGKM stars from the literature \citep{2010A&A...514A..97L,raquel2010} 
are also plotted as grey circles.
Our best linear fit for the upper branch is shown with a purple line, while the
grey line shows the fit for the lower branch derived by \cite{2011MNRAS.417.3100M}.} 
\label{flux_flux}
\end{figure}

 Figure~\ref{flux_flux} shows the comparison between the flux in the H$\alpha$ line
 and the flux in the Ca~{\sc ii} K line. A fit to a power-law function provides
 {
 \begin{equation}
 \log(F_{H\alpha})=(4.58\pm0.67)+(0.31\pm0.11)\times\log(F_{CaK})
 \end{equation}
 }


 \cite{2011MNRAS.414.2629M,2011MNRAS.417.3100M} identified two branches in the H$\alpha$ vs. Ca~{\sc ii} K flux-flux relationship.
 The lower or inactive branch has a slope of 1.17 $\pm$ 0.08 and is composed of field stars. On the other hand,
 the upper or active branch, is composed of young late-K and M stars and has a slope of 0.53 $\pm$ 0.08.
 Our sample provides a slope of 0.33 for the active branch showing that also young F-G stars 
 share the behaviour of cooler young stars.
 However, we note that for most of our inactive stars we were not able to measure any emission excess,
 so we could identify only one star (namely HD 167389) in the inactive branch.
 Figure~\ref{flux_flux} also shows that there seems to be
 a tendency of higher H$\alpha$ fluxes for the youngest stars, that show a rather
 flat H$\alpha$ vs. Ca~{\sc ii} K relationship.
 We note that the different importance of  H$\alpha$ and Ca~{\sc ii} emission might points to
 at a different role of different types of active structure
 \citep[see][for the case of the Sun]{2009A&A...501.1103M}.
 It should also be noted that the formation of the H$\alpha$ line is much more subject to
 non-LTE effects than the Ca~{\sc ii} lines as well as to further complications in cool stars
 (this is because unlike the Ca~{\sc ii} H \& K lines, the H$\alpha$ line is not a resonance transition).
 The star TYC 6779-305-1 shows a very strong emission in the H$\alpha$
 line and departs from the other young stars in the flux-flux relationship.

\section{Temporal evolution of active regions}
\label{tau_evol}

\subsection{Pooled variance analysis}

\begin{figure*}[!htb]
\centering
\begin{minipage}{0.48\linewidth}
\includegraphics[scale=0.45]{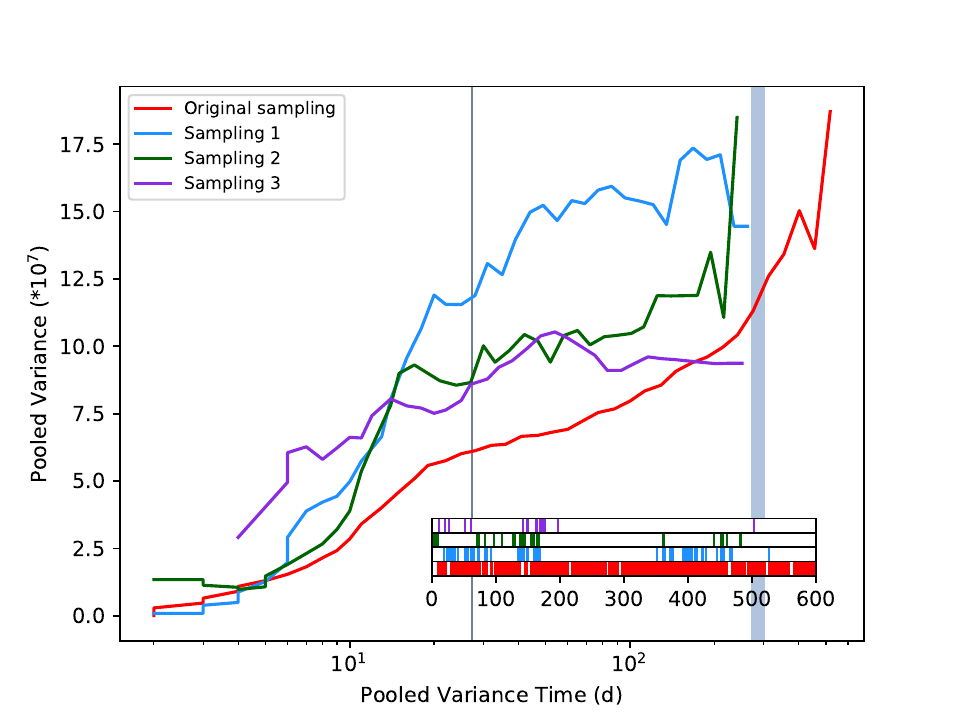} 
\end{minipage}
\begin{minipage}{0.48\linewidth}
\includegraphics[scale=0.45]{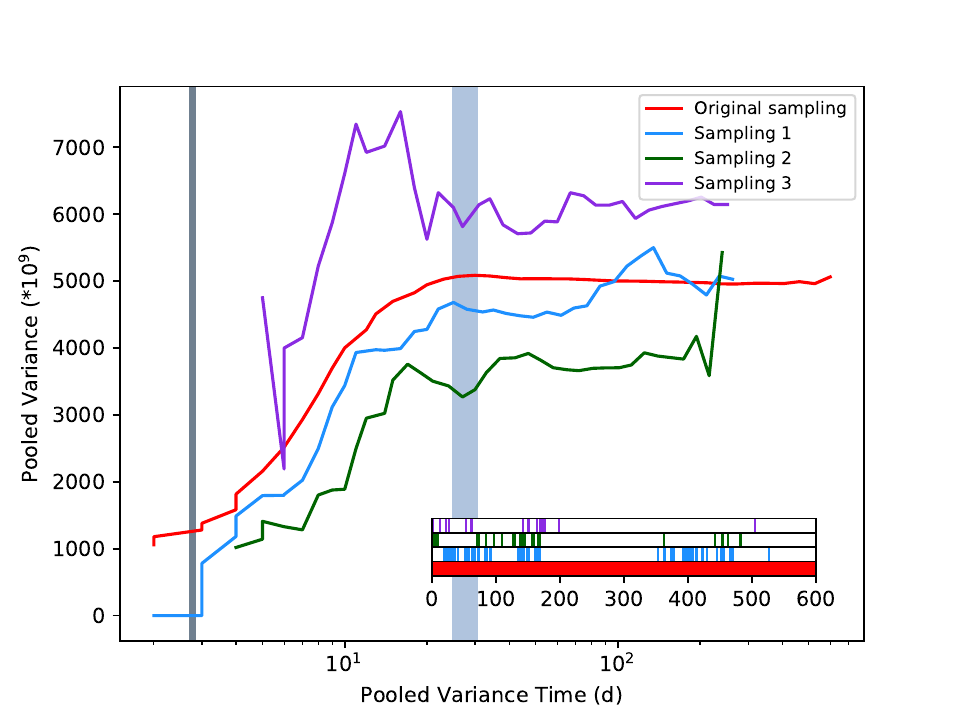}  
\end{minipage}
\begin{minipage}{0.48\linewidth}
\includegraphics[scale=0.45]{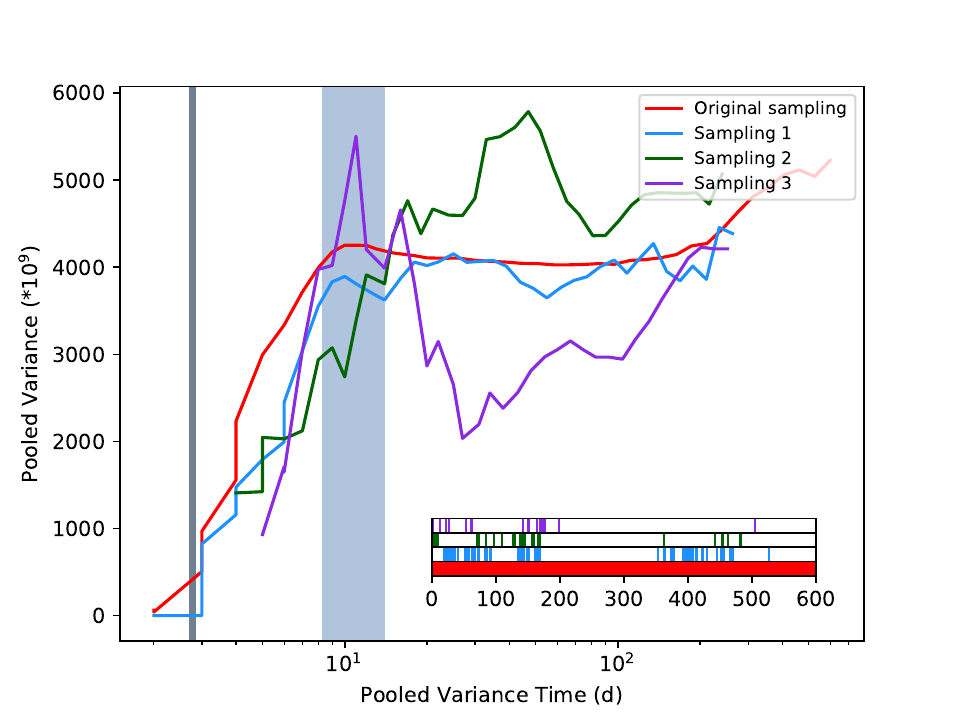}  
\end{minipage}
\begin{minipage}{0.48\linewidth}
\includegraphics[scale=0.45]{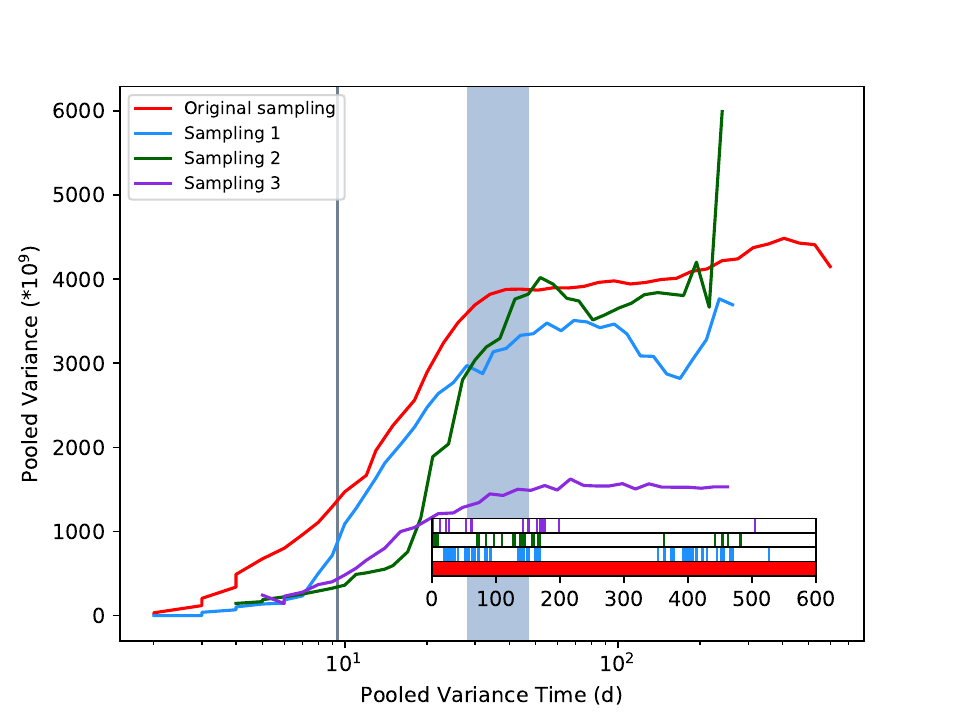}   
\end{minipage}
\caption{ Pooled variance profile for the HARPS-N Solar S-index (up left),
and simulated S-index time series, for different temporal samplings.
The plot shows a smoothed function of the PV for ease reading of the plots.
The vertical lines show the simulated rotation periods and AR lifetimes.
The inset plots show the temporal coverage of the different samplings.}
\label{pv_figure_simu}
\end{figure*}

 We apply the pooled variance (PV) technique \citep[see e.g.][]{1997SoPh..171..191D,1997SoPh..171..211D,2003A&A...409.1017M,2004A&A...425..707L,2017A&A...598A..28S}
 to the time series of the Ca~{\sc ii} H \& K activity index.
 In brief, the data  are binned into time intervals of length $t_{\rm pool}$. 
 Then, first the variance is calculated for each bin, and then the average of these variance values is computed forming
 the so-called pooled variance. This is done for a range of  $t_{\rm pool}$ values across the duration of the monitoring observations.
 The characteristic timescales of the star are manifest as the position where the PV vs $t_{\rm pool}$ changes behaviour.

 Before applying the PV method to our stars we performed a serie of simulations in order to understand the performance of the method
 as well as the effect of the sampling on the derivation of rotation periods and AR lifetimes.
 In our first test we took advantage of the solar spectra taken by the HARPS-N solar telescope and used the three years of S index values
 published in \cite{2019A&A...627A.118M}. We computed the PV diagram using the original dataset and compared it with the PV diagram obtained by sampling the data
 using the observation times of three selected stars. The corresponding plot is shown in Figure~\ref{pv_figure_simu} (top left). 
 In detail, 'sampling 1' contains 83 data points covering a time span of $\sim$ 1.5 yr. The data are divided into two main observing seasons with a gap
 of $\sim$ 200 d between them.
 'Sampling 2' is composed of 41 observation points taken in 1.3 yr. As in 'sampling 1' there is a gap of 200 d between the two main observing seasons, 
 but the second season is less populated than the first one.
 Finally, in 'Sampling 3' we consider only 22 data points, covering $\sim$ 200 d, with a gap of $\sim$ 100 d between the two main observing seasons.  
 The temporal coverage of the different samplings can be seen in the inset of the figure.
 The results show that even in the 'worst' sampling (case 3, in purple) we are able to recover the rotation period with a value in the range
 20 - 30 d. The AR lifetime is, however, only recovered when using the original time-serie (in red), with a value between 200 - 300 d. 

 Since the Sun is clearly not representative of most of our young stars, we performed three additional simulations.
 In simulation 1, (Fig.~\ref{pv_figure_simu}, top right) we consider a short rotation period of 2.74 d and an AR lifetime of 10 rotation periods.
 In simulation 2, (Fig.~\ref{pv_figure_simu}, bottom left) we keep the rotation period in 2.74 d but, consider an AR lifetime of 4 rotation periods.
 Finally, in simulation 3, (Fig.~\ref{pv_figure_simu}, bottom right) we fix the rotation period at 9.4 d, and the AR lifetime to 4 rotation periods.
 The simulations were performed by considering a sinusoidal behaviour, modulated with an exponential decay.
 In order to simulate the effect of spot growth and decay, as the time runs and the amplitude of the variability decays, another sinusoidal signal (with the same period but a different phase) is included.

 It can be seen that in the short period cases, we are not able to recover the injected rotation period (even with the original dataset),
 as the PV steadily increases until the AR lifetime is reached. However, the AR lifetime seems well constrained even for samplings 1 and 2.
 In the case of sampling 3, without an a priory knowledge of the AR lifetime, we would have concluded that the PV diagram is too complex to
 derive any meaningful conclusion. 
 Finally, for simulation 3, we are able to recover the AR lifetime in all cases. The injected rotation period is also 
 recovered, although at a slightly shorter value, $\sim$ 7-8 d.
 
 These simulations show that, with the data at hand, short rotation periods as well as long AR lifetimes may be difficult 
 to identify by means of the PV technique. Therefore, we set a limit of at least 20 observations per star to use
 the method.

\begin{figure*}[!htb]
\centering
\begin{minipage}{0.48\linewidth}
\includegraphics[scale=0.45]{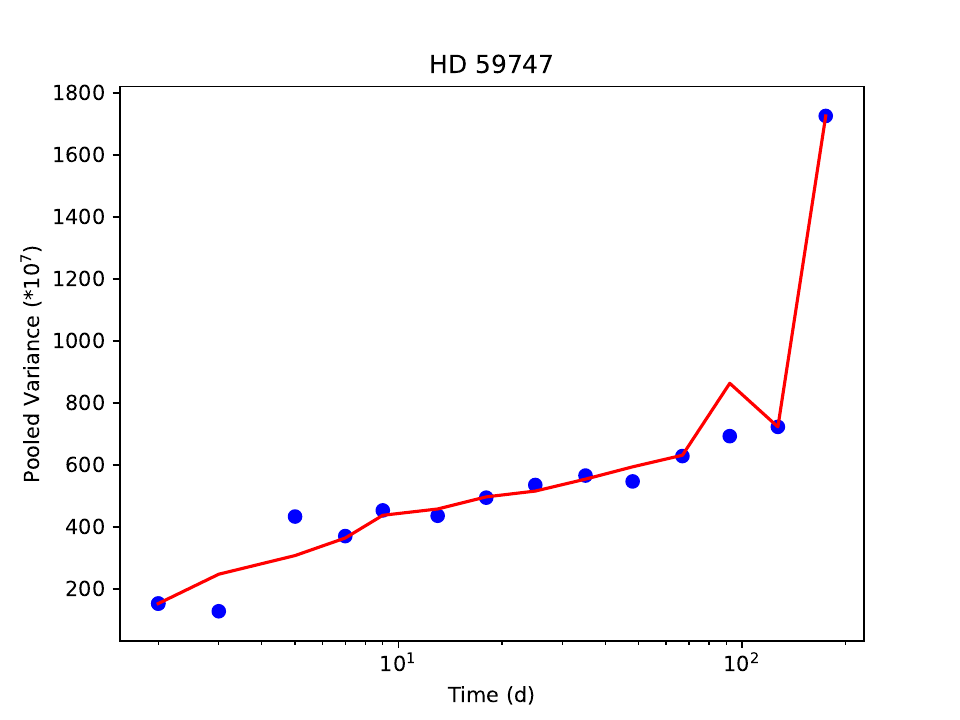}
\end{minipage}
\begin{minipage}{0.48\linewidth}
\includegraphics[scale=0.45]{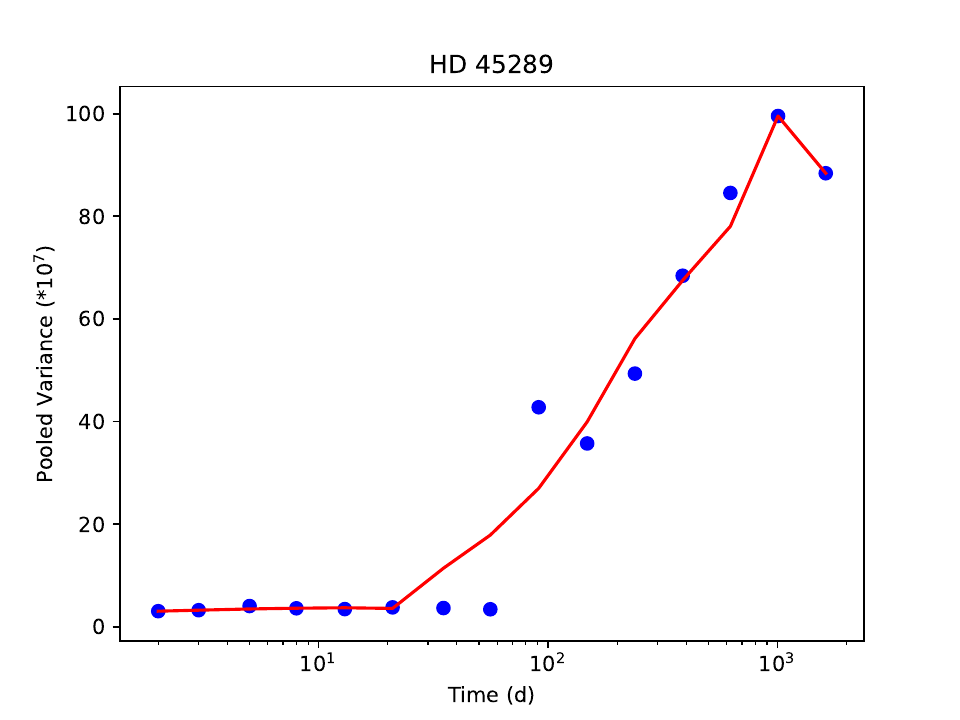}
\end{minipage}
\begin{minipage}{0.48\linewidth}
\includegraphics[scale=0.45]{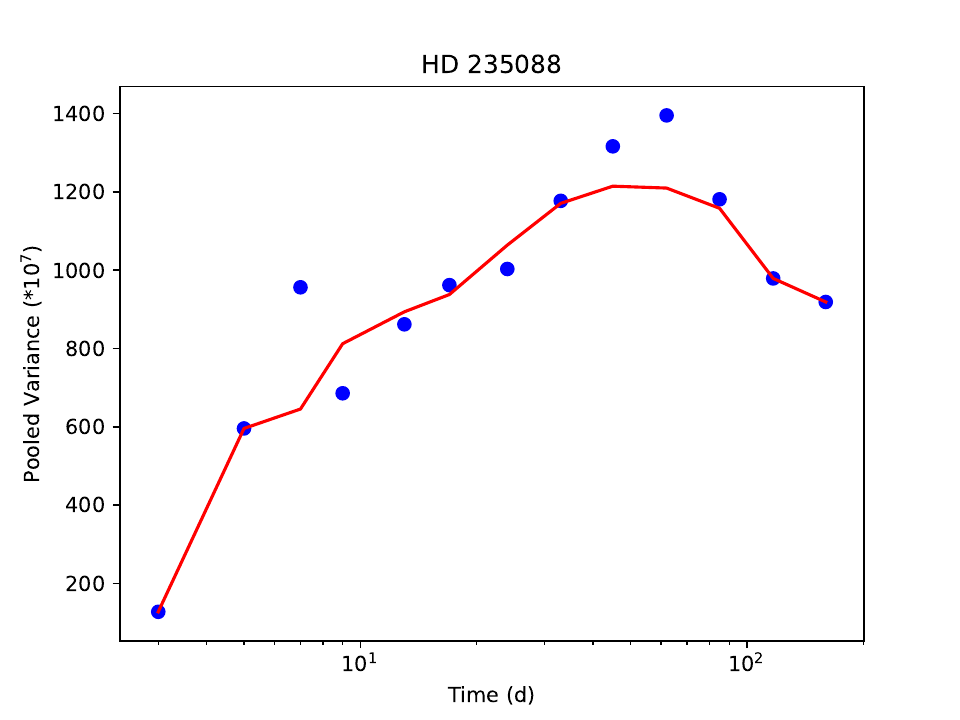}
\end{minipage}
\begin{minipage}{0.48\linewidth}
\includegraphics[scale=0.45]{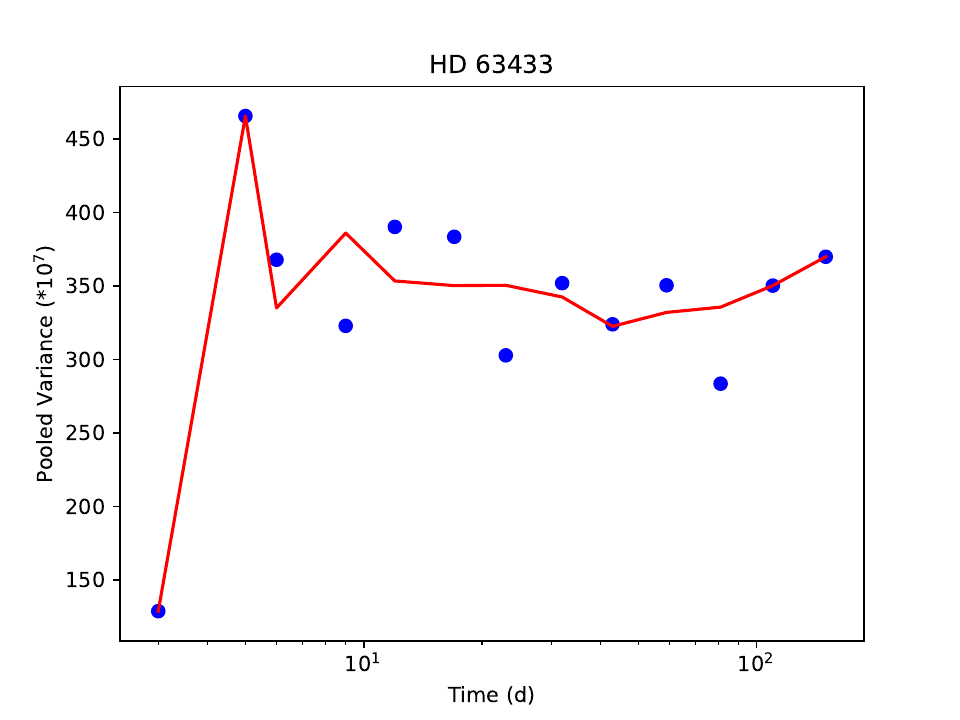}
\end{minipage}
\caption{Pooled variance profile for HD 59747 (up left), HD 45289 (up right),
HD 235088 (bottom left), and HD 63433 (bottom right).
The red line is a smoothed function for easy reading of the plots that highlights the trend of the data.
}
\label{pv_figure}
\end{figure*}

 Figure~\ref{pv_figure} (up left) shows an example of a star (HD 59747)  with a well-defined pattern. It can be seen that for this star the PV 
 steadily increases up to a $t_{\rm pool}$ value of $\sim$ 8 d and then it shows a plateau where the PV remains roughly constant. The PV starts to increase again at $t_{\rm pool}$ $\sim$ 70 - 100  d.
 We conclude that the rotation period of this star is around 8 d, and that active regions have typical time scales for active region evolution  of $\sim$ 10 rotation periods.
 We note that these estimates are in agreement with the results from the GLS and GP analysis. 

 Other stars like HD 45829 (Fig.~\ref{pv_figure}, up right) show a different profile. In this case, the PV shows a small roughly constant value at small $t_{\rm pool}$ values. However, after $\sim$ 20 - 30 d,
 the PV shows a nearly-constant increase of variance with increasing time scale. These stars are dominated by non-periodic variations with substantial
 active region evolution masking the rotational plateau. 

 In stars like HD 235088 (Fig.~\ref{pv_figure}, bottom left) the rotational plateau is not found and the PV increases until the active region evolution time scale is reached at $\sim$ 200 d.
 Other stars, show high PV at short time scale, but then, it diminishes. For example, HD 63433 (Fig.~\ref{pv_figure}, bottom right)  shows a peak at $\sim$ 7 d (in agreement with
 its rotation period) and then the PV steadily decreases (this can be due to statistical 
 fluctuations due to a rather small number of data points in this interval of time or due to the presence of outliers) until it remains constant.
 Finally, some stars have rather complex patterns (e.g. TAP 26), the PV shows a large scatter, and their temporal variation is not well-defined, while stars like
 HIP 21112 show roughly constant patterns. 
 Figure~\ref{pvall1} shows the PV diagram for all stars with more than 20 observations. 

\subsection{AR lifetimes from light curve analysis}
\label{tess_subsection}

  Once we have explored the behaviour of our sample in terms of the activity-age and flux-flux relationships,
 as well as in the pooled variance diagrams,
  we made use of available TESS photometry to 
 study whether the inferred lifetimes of ARs show any dependency with the stellar properties, in particular,
 with the stellar age.      
  Several recent works have made use of light curves to infer the properties
  of active regions. The idea behind these methods is that the decay-time of the autocorrelation function (ACF) is known
  to be related with the characteristic decay time of starspots \citep{nuccio}.
  In particular, \citet[][hereafter GI17]{2017MNRAS.472.1618G} modelled the ACF of light curves
  by using an underdamped harmonic oscillator with an interpulse term.
  However, \citet[][hereafter SA21]{2021MNRAS.508..267S} discussed this choice of the modelling function, 
  and suggested a new modelling with a linear decay.
  On the other hand, \citet[][hereafter BA21]{2022ApJ...924...31B} uses a method based on the strengths
  of the first few normalised autocorrelation peaks.
  
  In order to test whether these methods can be suitable for our stars, we start by computing the ACF
  of the TESS light curves. We note that the ACF can be calculated only for stars for which the
  data have a continuous sampling (or at least that can be interpolated to a continuous sampling). 
  Otherwise, the strength of the ACF peaks might show a complicated dependency on the spectral window.
  The corresponding ACF curves are shown in Fig.~\ref{acf1}, while the properties of the ACF analysis
  is given in Table~\ref{acf_summary}.

  An inspection of the figure reveals that for some stars the peaks of the ACF have always
  the same strength (e.g. V830 Tau, TAP 26). That means that the active regions should be stable
  during the timespan of the observations. In addition, there seems to be no beating in these curves, usually
  due to differential rotation. 
  Since the ACFs of these stars show no sign of time decay, it is unlikely that the methods presented in
  GI17, SA21, or BA21 might work. 
  Indeed, if one try to fit the ACF curve of these stars to one of the functional forms described in GI17 or SA21,
  the result is that the posterior distribution of the AR lifetime is not well constrained, but shifted towards the
  larger prior of the AR lifetime. We illustrate this in the left panel of figure~\ref{posterior}, where we show
  the posterior distribution for the case of the star V830 Tau. It is clear that while all parameters are well constrained,
  the fit is not able to derive any meaningful AR lifetimes.
  We classify these curves are 'sin' (sinusoidal) or 'per' (periodic) to indicate that there is no time-decay
  present in the ACF curve. With the data at hand, the only information that we can extract for these stars is that
  the typical AR lifetime should be much longer than the timespan of the observations.

  Other stars like HIP 92680 and HIP 105388 show a clear time decay. For these stars, we used a bayesian
  framework to model the ACF curve to the functional forms described in GI17 (exponential decay) 
  and SA21 (linear decay).
  As an example, we show the posterior distribution of the fit for the case of HIP 105388 (Fig.~\ref{posterior}, right).
  We used the Bayesian Inference Criterion (BIC) as a measure of the goodness of the two models,
  although in most cases the BIC of the two models are almost identical (that is, there is no significant evidence
  in supporting one model against the other). 
  We note that for some stars, even if the posterior distribution of all the fitted parameters are well constrained,
  the best fit is not able to reproduce all the features seen in the ACF (e.g. V1090 Tau or V1298 Tau).
  This might indicate that these ACFs are not fully regular.
  Indeed, some of our stars show a rather irregular ACF curve that makes difficult its analysis. Some examples
  are HD 285507, or HD 32923.
  Table~\ref{acf_summary} provides the AR lifetimes or lower limits derived, when possible, from the ACF curves.

\begin{table*}[!htb]
\centering
\caption{ Summary of the ACF analysis of the TESS light curves.}
\label{acf_summary}
\begin{tabular}{llll}
\hline\noalign{\smallskip}
Star            &  ACF-fit $\tau$ (d)                      &   ACF-type     & ACF-fit form  \\
\hline
HIP 490         &  232.59$_{\rm - 106.83}^{\rm + 474.93}$      &         sin + decay  &  exp          \\
HIP 1481        &                                          &         sin + decay  &  bad fit      \\
TYC 4500-1478-1 &  132.92$_{\rm - 52.46}^{\rm + 248.64}$       &         sin + decay  &  exp          \\
V1090 Tau       &  135.47$_{\rm - 66.71}^{\rm + 407.07}$       &         sin + decay  &  exp$^{\dag}$ \\
V1298 Tau       &  259.36$_{\rm - 155.05}^{\rm + 687.93}$      &         per + decay  &  lin$^{\dag}$ \\
HD 285507       &                                          &         other        &               \\
HIP 19859       &  $>>$ 26                                 &                     per          &               \\    
TAP 26          &  $>>$ 24                                 &                     per          &               \\
HD 285773       &  $>>$ 25                                 &                     per          &               \\ 
V1202 Tau       &  $>>$ 24                                 &                     sin          &               \\
HIP 21112       &  $>>$ 26                                 &                     sin          &               \\
V830 Tau        &  $>>$ 24                                 &                     sin          &               \\
TYC 5909-319-1  &  387.19$_{\rm - 238.44}^{\rm + 761.66}$      &                   sin + decay? &  exp$^{\dag}$ \\
HD 32923        &                                          &                     other        &               \\
HD 36108        &                                          &                     other        &               \\ 
HD 38283        &  42.19$_{\rm -15.72}^{\rm + 89.43}$          &                   per          & exp$^{\dag}$  \\
HIP 27072       &                                          &                     other        &               \\
HD 45289        &                                          &                     other        &               \\
HD 59747        &  $>>$ 27                                 &                     sin          &               \\
HD 63433        &  62.31$_{\rm -14.29}^{\rm + 37.82}$          &                   sin + decay  & lin$^{\dag}$  \\
HD 70573        &  $ >>$ 25                                &                     sin          &               \\
TYC 1989-0049-1 &                                          &                     other        &               \\
HD 107877       &                                          &                     other        &               \\
BD+26 2342      &  64.07$_{\rm -18.58}^{\rm +55.15}$           &                   sin + decay? &  exp          \\
BD+27 2139      &  $>>$ 27                                 &                     sin          &               \\
HIP 61205       &                                          &                     other        &               \\ 
HD 122862       &                                          &                     other        &               \\
HD 167389       &                                          &                     other          & bad fit       \\
HIP 92680       & 232.93$_{\rm - 155.27}^{\rm +714.18}$        &                   per + decay  & lin           \\
HD 235088       &                                          &                     other        &               \\
HD 191408       &                                          &                     other        &               \\
HD 196378       &                                          &                     other        &               \\
HIP 105388      & 82.95$_{\rm -18.16}^{\rm +33.26}$           &                    per + decay  & exp           \\
\hline
\end{tabular}
\tablefoot{'sin': the ACF is clearly sinusoidal without an apparent time-decay; 'per': the ACF is periodic without an apparent time-decay;
'decay': a time-decay is seen in the ACF; 'other': the ACF does not fit in the other categories; 'exp': fit to an exponential decay
following GI17; 'lin': fit to a linear decay following SA21; $^{\dag}$ the best fit does not properly model the curve}
\end{table*}



\subsection{AR lifetime as a function of the stellar parameters}

\begin{figure*}[!htb]
\centering
\begin{minipage}{0.48\linewidth}
\includegraphics[scale=0.80]{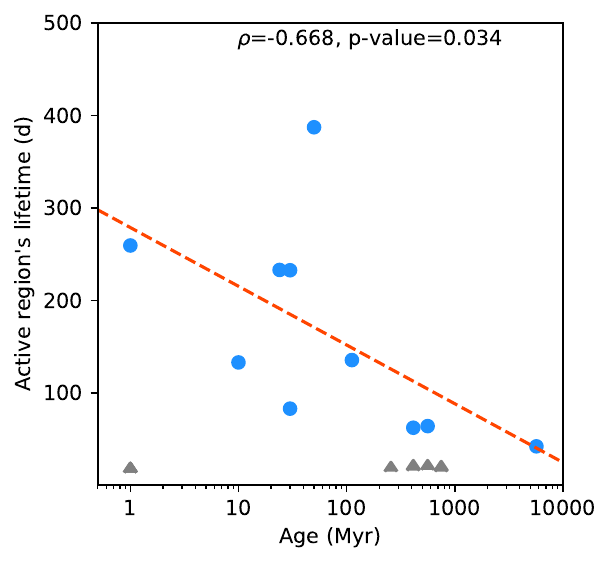}
\end{minipage}
\begin{minipage}{0.48\linewidth}
\includegraphics[scale=0.80]{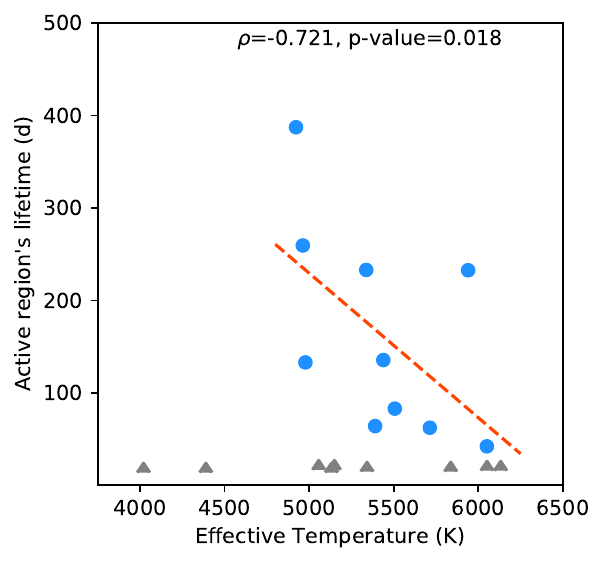}
\end{minipage}
\begin{minipage}{0.48\linewidth}
\includegraphics[scale=0.80]{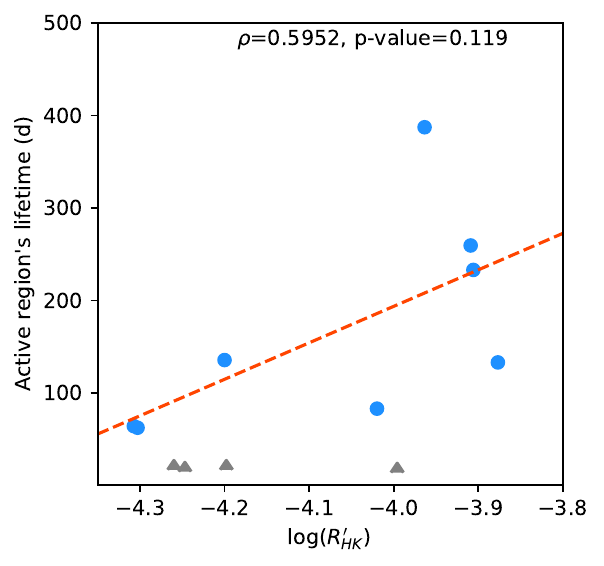}
\end{minipage}
\begin{minipage}{0.48\linewidth}
\includegraphics[scale=0.80]{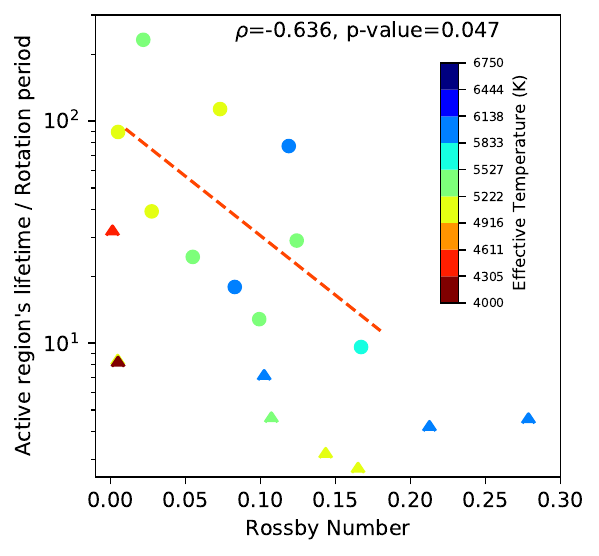}
\end{minipage}
\caption{
Timescale of AR evolution derived from the ACF analysis of the TESS light curves as a function of the stellar age (top left),
the effective temperature (top right), the $\log$R$^{'}_{\rm HK}$ value (bottom left), and the Rossby number (bottom right).
Stars with lower limits on AR timescale are shown with triangles.
A linear fit is (dashed orange-red line) is shown for  for guiding the eye.}  
\label{ar_evolution_acf}
\end{figure*}

 Figure~\ref{ar_evolution_acf}, top left, shows the timescale of AR evolution
 derived from the ACF analysis of the TESS light curves
 as a function
 of the stellar age. 
 Given that the number of points is rather low and also the uncertainties involved in age and AR lifetime, any conclusion from this figure should be taken
 with caution. Nevertheless, the figure reveals a tendency of younger stars to show
 longer AR lifetimes. The Spearman's rank test, $\rho$, is -0.67 with a p-value of 0.03 (the lower limits on AR lifetimes were not considered).
  
 We also checked for correlations between AR lifetimes, the effective temperature of the star (Fig.~\ref{ar_evolution_acf}, top right),
 and the level of activity (as measured by the $\log$R$^{'}_{\rm HK}$ value), Fig.~\ref{ar_evolution_acf}, bottom left.
 A general tendency of increasing AR lifetime with cooler temperatures and higher activity
 levels seems to be present in the data. Whilst the AR lifetime correlation with T$_{\rm eff}$ might be statistically significant
 (with a p-value lower than 0.02), the one with the level of activity does not (p-value $\sim$ 0.12).
 These results are in agreement with GI17, SA21 or BA22 who analysing a large dataset of {\sc Kepler} light curves concluded that
 ARs decay more slowly in cooler stars.

\begin{figure}[!htb]
\centering
\includegraphics[scale=0.55]{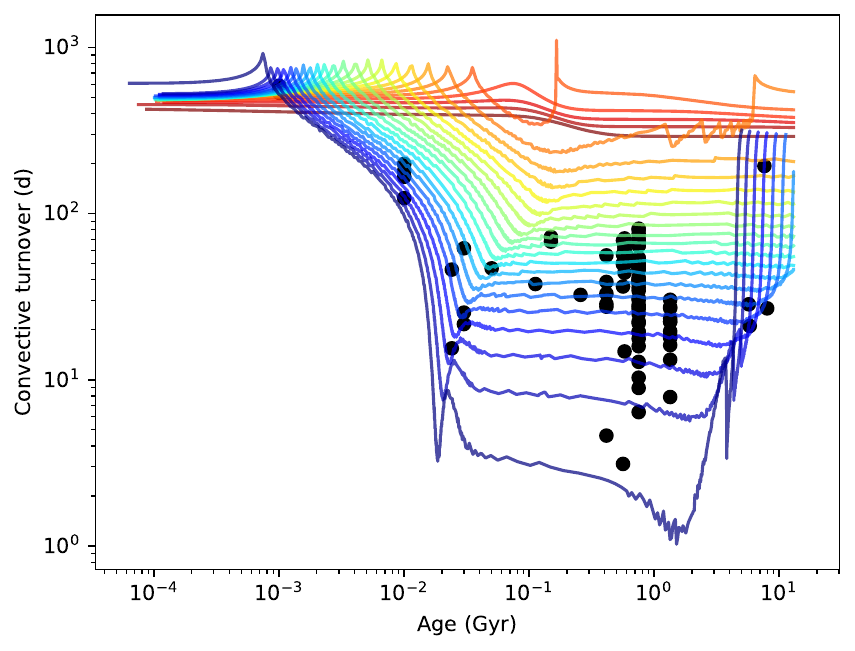}
\caption{
Convective turnover timescale as a function of the stellar age. The black circles show the stars studied in this work while
the continuous lines  represents the tracks provided by  \cite{2013ApJ...776...87S}. The models have solar metallicities and
stellar masses ranging from 0.10 M $_{\odot}$ to 1.25 M$_{\odot}$ (from red to violet) with a mass step of 0.05 M$_{\odot}$.}
\label{tau_conv_plot}
\end{figure}

 It is worth noticing that when comparing stars with different stellar parameters, other properties like
 the convective turnover time might be different as well. The use of the
 Rossby number has been shown to improve substantially the observed activity-rotation relations for main sequence, solar-type
 stars \citep[e.g.][]{1984ApJ...279..763N}.
 To compute the Rossby number, we first estimate the mass of our stars by using the {\it Gaia} DR2 luminosities \citep{2018yCat.1345....0G}
 and the mass-luminosity relationship provided by \cite{2018A&A...619L...1W}. We then derive 
 the convective turnover timescales by interpolating (in stellar mass and age) the theoretical tracks provided by
 \cite{2013ApJ...776...87S}. Figure~\ref{tau_conv_plot} shows the position of our target stars in the stellar mass-age diagram where it can
 be seen that most of our targets have masses in the 0.8 - 1.1 M$_{\odot}$ range.
 Finally, the Rossby number is computed as

 \begin{equation}
  R_{\rm 0} =  \frac{P_{\rm rot}}{\tau_{\rm conv}} 
 \end{equation}
 
  Figure~\ref{ar_evolution_acf}, bottom right, shows the timescale of AR evolution
  as a function
 of the Rossby number. For a better comparison between stars with different properties we show
 the AR lifetime in units of the corresponding rotational period. 
 The figure shows a clear tendency of decreasing AR lifetimes with increasing Rossby number,
 which would imply that ARs survive longer
 in stars with larger convective turnover timescales and shorter rotation period.
 A  Spearman's correlation test returns the values $\rho$ = -0.64 and $p$-value = 0.05.

\subsection{Gaussian process analysis of the spectroscopic indexes}
\label{gp_subsection}


 In this section we explore whether our spectroscopic time series can be used to infer the AR lifetimes.
   That would be of the utmost interest as it will provide a complementary approach to the use of light curve ACFs.
   Through this analysis we use the results of the GP analysis (see Sect.~\ref{gp_description}) and
    make the assumption that the GP hyperparameter $\tau$ (that is, the timescale
 of the exponential decay, see Eqn.~\ref{kernel}) corresponds to the AR growth and decay lifetime.
 Whether this assumption is well founded or not will be discussed below.

 For this analysis we focus only on stars with more than twenty observations and with available TESS photometry.
 Since most of our targets are young, they should have short rotation periods and therefore, the rotation periods
 derived from the TESS data should be reliable.
 We note that the use of GP analysis to derive rotation period  has already been used in the literature
 \citep[e.g.][]{2018MNRAS.474.2094A}.
 However, given that the rotation period is a key parameter of the analysis,
 we performed a comparison with other photometric surveys like ASAS \citep{2014AAS...22323603S,2019MNRAS.485..961J}, 
 SWAPS \citep{wasp},
 STELLA \citep{2004AN....325..527S}, as well as other literature sources.
 Table~\ref{tab_prot_comparison} provides a summary of the derived rotation
 periods. The TESS rotation periods are derived from our GP analysis.

\begin{figure}[!htb]
\centering
\includegraphics[scale=0.45]{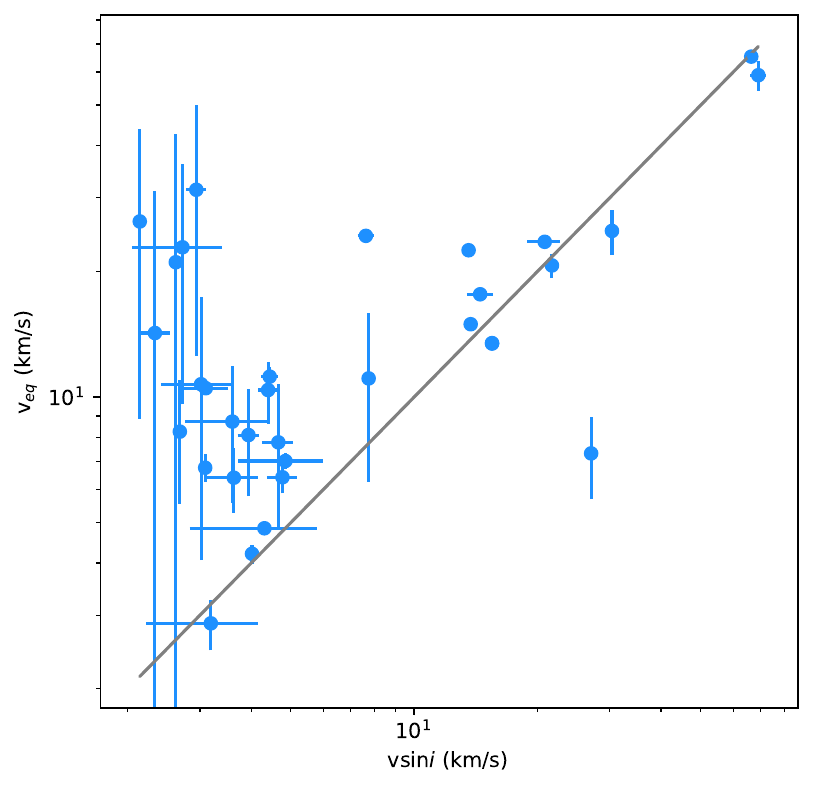}
\caption{ Equatorial velocities (derived from the TESS P$_{\rm rot}$ and R$_{\star}$) versus projected rotational velocities, v$\sin i $.}
\label{equ_vs_vsini}
\end{figure}

 In addition, the TESS-derived periods can be translated into equatorial velocities, v$_{\rm eq}$, and compared
 with the corresponding v$\sin i$ values.
 In order to perform this conversion, stellar radii are taken from {\it Gaia} DR2 \citep{2018yCat.1345....0G}. 
 The corresponding plot is shown in Fig.~\ref{equ_vs_vsini}. As expected,
 most of the targets lie in the region v$_{\rm eq}$ larger than v$\sin i$, while 14 stars are close to the line
 v$_{\rm eq}$ $\sim$ v$\sin i$ and should have inclination angles $\sim$ 90 degrees. 
  We note that one star, namely HD 107877, have P$_{\rm rot}$ a value that translates into non-physical v$_{\rm eq}$ values (i.e., shorter than v$\sin i$).
  For this star no clear P$_{\rm rot}$ was found in the analysis of the ASAS or SWAPS photometry, while the STELLA data show
 two peaks at $\sim$ 7.3 d and $\sim$ 1.6 d. We note that the 7.3 d signal is still too large, to be compatible with the v$\sin i$ value. 

\begin{table*}
\centering
\caption{Rotational periods derived from TESS, ASAS, SWAPS, PV analysis and the literature.}
\label{tab_prot_comparison}
\begin{tabular}{lccclc}
\hline
Star            &	TESS	&	ASAS	&	SWAPS	&  Other        & Pooled Variance	\\
                &       (d)     &        (d)    &       (d)     &   (d)         &     (d)               \\ 
\hline
HIP 490	        &	3.01	&		&		&		&	        	\\
HIP 1481	&	2.41	&		&		&		&	$\sim$ 3	\\
TYC 4500-1478-1$^{\dag}$	&	3.39	&	3.42	&		& 3.5 (STELLA)  &	$\sim$ 6	\\
V1090 Tau$^{\dag}$	&	4.68	&	4.76	&	4.71	&		&	        	\\
V1298 Tau$^{\dag}$	&	2.91	&		& 2.89/1.44	& 2.91 \citep{2021NatAs.tmp..252S} &	\\
HD 285507	&	5.76	&	10.57	&		& 2.24 \citep{2020AA...638A...5C} & 	\\
HIP 19859	&	5.85	&	  	&		&		&	  	        \\
TAP 26$^{\dag}$	        &	0.71	&	0.71	&		& 0.71 \citep{2013AstL...39..251G} &	\\
HD 285773	&	5.14	& 1.9/4.6	&		& 10.7? \citep{2019ApJ...879..100D} &	$\sim$ 6	\\
V1202 Tau$^{\dag}$	&	2.72	&	2.7	&	1.59 	& 2.68 (STELLA)	&		        \\
HIP 21112	&	5.40	&		&		&		&		        \\
V830 Tau$^{\dag}$	&	2.77	&	2.74	&	1.37    & 2.74 \citep{2020AA...642A.133D} &    \\ 
TYC 5909-319-1$^{\dag}$	&	3.42	&	3.37	&	1.41/3.4& 3.43 \citep{2021AA...645A..71C} &	\\
HD 32923	&	3.43	&		&		& 32 \citep{2020AN....341..497S} & $\sim$ 3/4	        \\
HD 36108	&	2.48	&		&  2.99?	&		& $\sim$ 2/3	        \\
HD 38283	&	2.36	&		&		&		&		\\
HIP 27072$^{\dag}$	&	6.21	&		&		& 5.9 \citep{benja}		&		\\
HD 45289	&	4.37	&		&		&		&		\\
HD 59747$^{\dag}$	&	8.04	&		&		&		& $\sim$ 8		\\
HD 63433$^{\dag}$	&	6.48	&		&	7.98	&6.45 \citep{2020AJ....160..179M} & $\sim$ 4/5?	\\
HD 70573$^{\dag}$	&	3.32	&		&		&	3.28 (STELLA)	&		\\
TYC 1989-0049-1	&	12.16	&	8.27	&	10.86	& 5.5/11 (STELLA)	&		\\
HD 107877	&	9.25	&		&		& 7.3?/1.16? (STELLA)     &    \\
BD+26 2342$^{\dag}$	&	4.99	&		&		&	4.6 (GAPS data)	&		\\
BD+27 2139$^{\dag}$	&	9.37	&		&	9.29	&	9.28 (STELLA)	& $\sim$ 4/5?	\\
HIP 61205	&	5.91	&		&	7.58	&	7.39 (STELLA)	&		\\
HD 122862	&	3.80	&		&		&		&	$\sim$ 3	\\
HD 167389$^{\dag}$	&	7.70	&		&		& 8.85 (GAPS data)	& $\sim$ 7/8	\\
HIP 92680	&	1.00	&		&		&		&		\\
HD 235088	&	6.14	&		&		& 14.1 (REM), 12.8-13.5 (STELLA)	&		\\
HD 191408	&	3.44	&		&		&		&		\\
HD 196378	&	8.86	&		&		&		&		\\
HIP 105388	&	3.39	&		&		&		&		\\
\hline
\end{tabular}
\tablefoot
{
   $^{\dag}$ Star selected for the detailed comparison of the different activity indexes. 
 {  They are shown in Fig.~\ref{tess_prot} and Fig.~\ref{tess_tau}. }
   } 
\end{table*}

  In the following we will retain for study only those star for which
  the number of spectroscopic observations is larger than twenty, they have TESS photometry, and
  we have at least one independent confirmation that the TESS-derived period is correct. 
  These stars are highlighted in Table~\ref{tab_prot_comparison}.

\begin{figure}[!htb]
\centering
\includegraphics[scale=0.45]{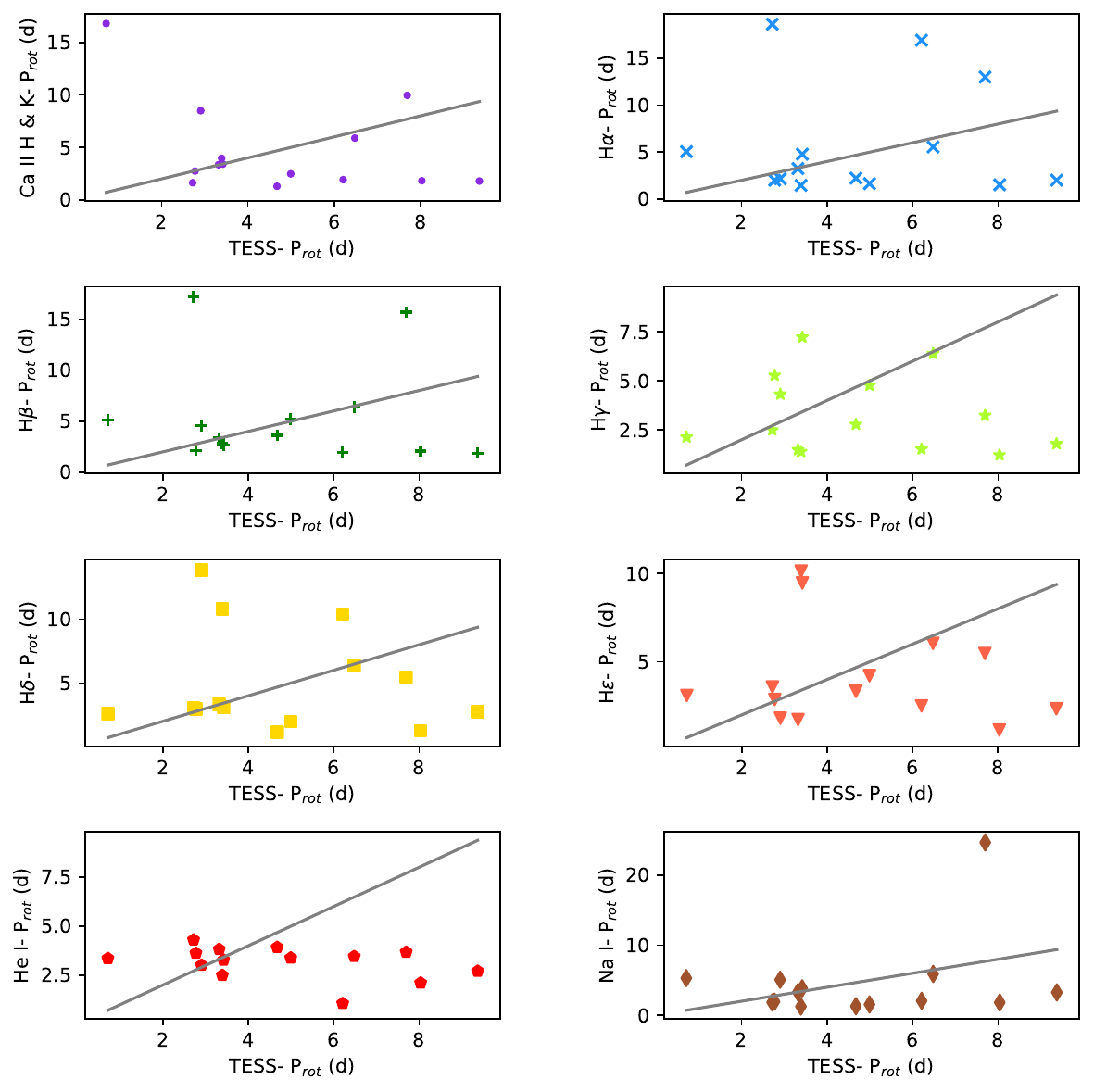}
\caption{ Derived rotation periods  
from the GP analysis
of the different spectroscopic indexes as a function of the values derived from the GP analysis
of the TESS photometry.}
\label{tess_prot}
\end{figure}
\begin{figure}[!htb]
\includegraphics[scale=0.45]{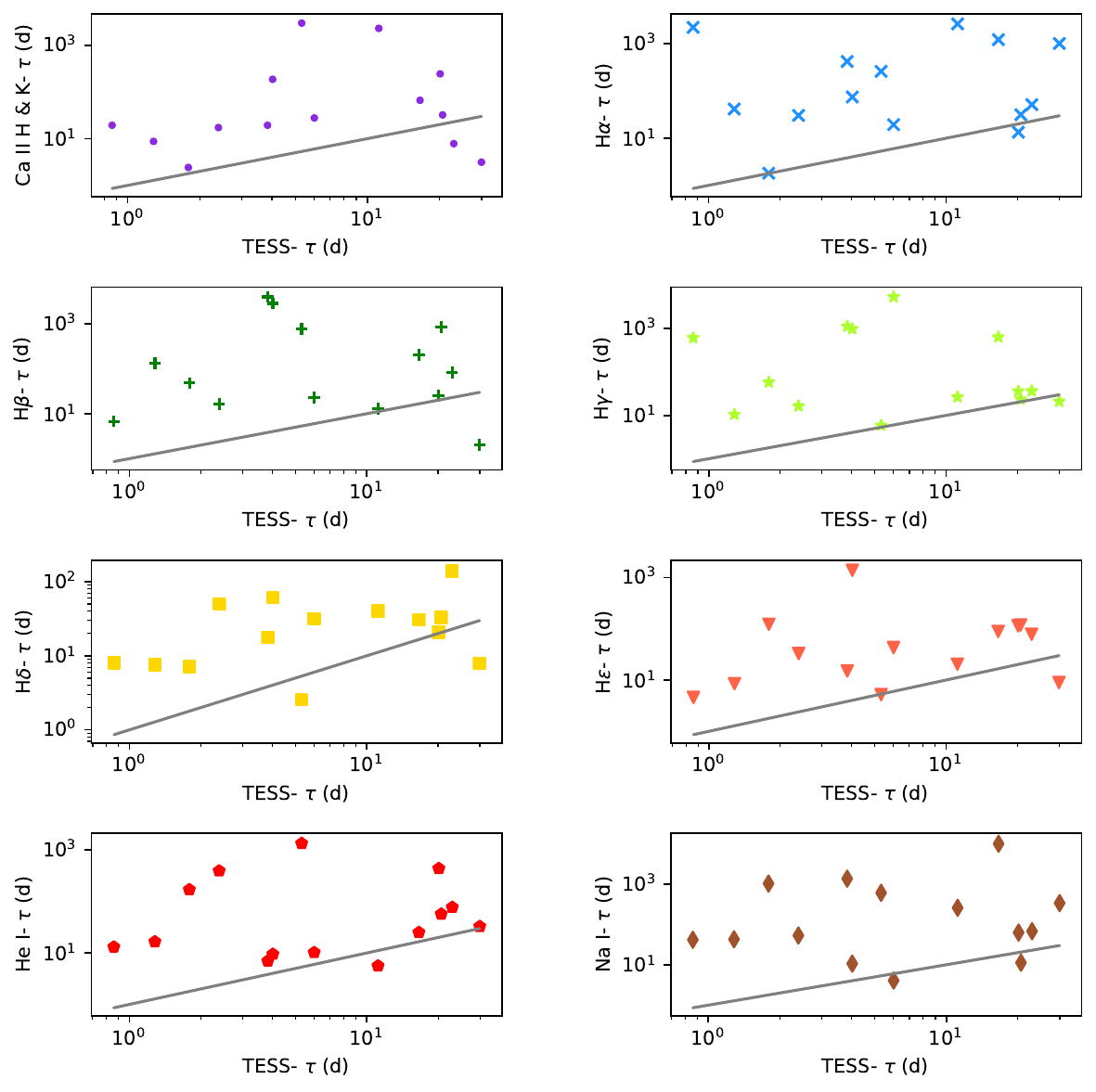}
\caption{ Derived AR lifetimes from the GP analysis
of the different spectroscopic indexes as a function of the values derived from the GP analysis
of the TESS photometry.}
\label{tess_tau}
\end{figure}

  Figure~\ref{tess_prot} shows the rotation periods 
 derived from the GP analysis of the Ca~{\sc ii},
 the Balmer lines, He~{\sc i} D$_{\rm 3}$, and Na~{\sc i} D$_{\rm 1}$, D$_{\rm 2}$ activity indexes
 as a function of the reference TESS-derived rotatin periods.
 The corresponding comparison for the derived AR lifetimes is shown in Fig.~\ref{tess_tau}.

 We note that for most of the stars the rotation periods derived from the spectroscopic indexes are
 significantly shorter than the TESS-derived periods. 
 For example, for BD+27 2139 (which has a TESS-derived period of 9.37 d)
 the analysis of the different spectroscopic indexes provide values in the range $\sim$ 2 -3 d.  
 On the other hand, AR lifetimes derived from
 spectroscopic indexes are much longer than those derived from the TESS data.

 Furthermore, even if the periods from different indexes agree, AR lifetimes can be very different
 from one index to another.
 An example is the star V830 Tau, for which we recover a rotation period of $\sim$ 2.77 d in TESS, Ca~{\sc ii},
 H$\delta$, and H$\epsilon$ data. However, the AR lifetime varies from $\sim$ 5.32 d in the TESS data to  $\sim$ 2970 d in
 the Ca~{\sc ii} H \& K data  (we note that for this star, from the ACF light curve analysis we only concluded that
 its AR lifetime should be longer than 24 d.)
 

  We conclude that the GP analysis of the spectroscopic indexes does not allow us
  to measure AR lifetimes to a useful accuracy.
  Several explanations can be put forward to account for this result.
   The first one deals with the data and the assumptions used in this work.
    We note that the bulk of the stars analysed in this work comes from a radial velocity exoplanet program,
     and therefore, the number of observations, temporal baseline, and sampling
      vary considerably from one star to another and in some cases it  may not be optimal.
       Although stating the obvious, we recall that only stars in which potential planetary signals
        are identified are observed with a high cadence.
	 Furthermore, it is important to keep in mind
	   that GPs are simplified {\it ad hoc} models of stellar activity and that the correspondence
	    between the GP hyperparameters and the physical properties of AR should be further analysed.

\section{Conclusions}\label{conclusions}
 In this work a detailed analysis of a large sample of young stars with
 well known derived ages determined from their membership to 
 kinematic associations and moving groups is performed.
 Projected rotational velocities and activity indexes are determined in an homogeneous
 way from high-resolution optical spectra. The temporal series of the different activity
 indexes are used together with a gaussian process regression analysis to
 infer rotational periods and the lifetime of AR growth and decay.

 We characterise our sample in terms of  activity-rotation-age and
 flux-flux relationships and confirm the well known trend of decreasing
 activity and rotation with stellar age. We also show that cooler stars show
 higher levels of activity, and that their rotation rate shows a lower
 age-decay than their hotter counterparts. 
 We also find that young F, G stars depart from the inactive stars in the flux-flux relationships.

 We search for correlations between the ARs evolution lifetime and the stellar properties,
 namely age, effective temperature, and level of activity.
  AR lifetimes derived from the ACF analysis of light curves show a 
 tendency to decrease with the stellar age.
 ARs lifetimes are also found to be lower in hotter and inactive stars.
 A global tendency of larger ARs lifetimes versus lower Rossby number
 is also found. However, we caution that these relationships are affected by the low number
 of stars for which a reliable AR lifetime could be obtained.
 Finally, one cannot forget the assumptions linked to the models used to determine
               stellar ages or the convective turnover timescale.
 We also tried to derive AR lifetimes from a GP modelling of the spectroscopic time-series, but
 the results were largely unsatisfactory, even if we restricted the analysis to stars with well
 known rotation periods from photometric data.

 Further observations of stars covering a wide range of stellar ages, together with  a better understanding of
 how to model stellar activity, as well as an accurate determination of the stellar properties will help us to understand 
 whether ARs have rather irregular lifetimes 
 or if there is some unknown relationship between ARs lifetimes and stellar properties.

\begin{acknowledgements}

  J.M., S.C., A.P., G.M acknowledge support from the \emph{Accordo Attuativo ASI-INAF n. 2021-5-HH.0,
  Partecipazione alla fase B2/C della missione Ariel (ref. G. Micela).}
  S.C acknowledge financial contribution from the agreement \emph{ASI-INAF n.2018-16-HH.0 (THE StellaR PAth project)}.
  We sincerely appreciate the careful reading of the manuscript and the constructive comments of the referee Gibor Basri.

\end{acknowledgements}

%
\bibliographystyle{aa}
\bibliography{yo_activity_rotation.bib}


\begin{appendix} 
\section{The (B-V) colour - T$_{\rm eff}$ relationship}
\label{appendix1}

 In order to derive a relationship between the effective temperature and
 the (B-V) colour we use the data from \citet[][Table~3]{1996ApJ...469..355F}.
 The data was fitted to a seven order polynomial fit of the form:
 $(B-V)$ = a$_{\rm 0}$ + a$_{\rm 1}$$\times$$(\log T_{\rm eff})$ + a$_{\rm 2}$$\times$$(\log T_{\rm eff})^2$ + ... +
 a$_{\rm 7}$$\times$$(\log T_{\rm eff})^7$.
 Table~\ref{table_fgk_parameters} gives the coefficients of the polynomial fit.

\begin{figure}[!htb]
\centering
\includegraphics[scale=0.60]{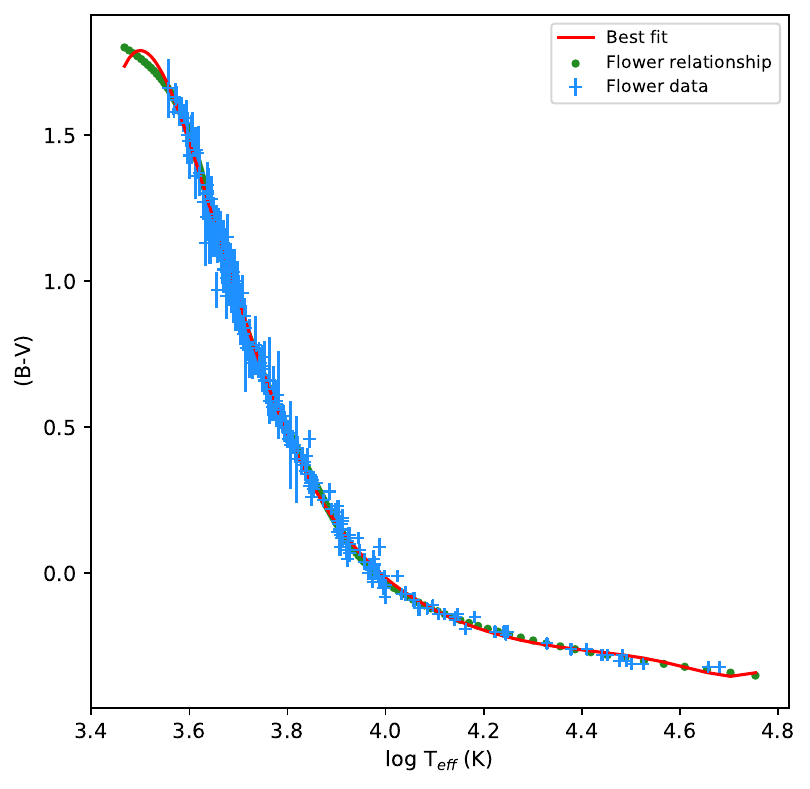}
\caption{(B-V) vs. $\log$T$_{\rm eff}$ relationship.
The blue crosses show the data for the individual main-sequence stars
while the green dots corresponds to the colour-T$_{\rm eff}$ scale presented in  \cite{1996ApJ...469..355F}.
Our best fit is shown in red.}
\label{bvflower}
\end{figure}

\begin{table}[!htb]
\centering
\caption{Coefficients of the (B-V) vs. $\log$T$_{\rm eff}$ relationship. }
\label{table_fgk_parameters}
\begin{tabular}{lc}
  \hline\noalign{\smallskip}
 Coefficient & Value \\ 
\hline 
a$_{\rm 0}$  & -6.5459$\times$10$^{\rm 5}$ \\ 
a$_{\rm 1}$  &  1.0991$\times$10$^{\rm 6}$ \\ 
a$_{\rm 2}$  & -7.8965$\times$10$^{\rm 5}$ \\ 
a$_{\rm 3}$  &  3.1471$\times$10$^{\rm 5}$ \\ 
a$_{\rm 4}$  & -7.5147$\times$10$^{\rm 4}$ \\ 
a$_{\rm 5}$  &  1.0752$\times$10$^{\rm 4}$ \\ 
a$_{\rm 6}$  & -8.5347$\times$10$^{\rm 2}$ \\ 
a$_{\rm 7}$  &  2.8998$\times$10$^{\rm 1}$ \\ 
\hline
\end{tabular}
\end{table}

\section{Online Figures}\label{pv_plots}

 Figure~\ref{pvall1} shows the pooled variance profile for all the stars with more than 20 observations.

\begin{figure*}[!htb]
\centering
\subfloat{
\includegraphics[scale=0.45]{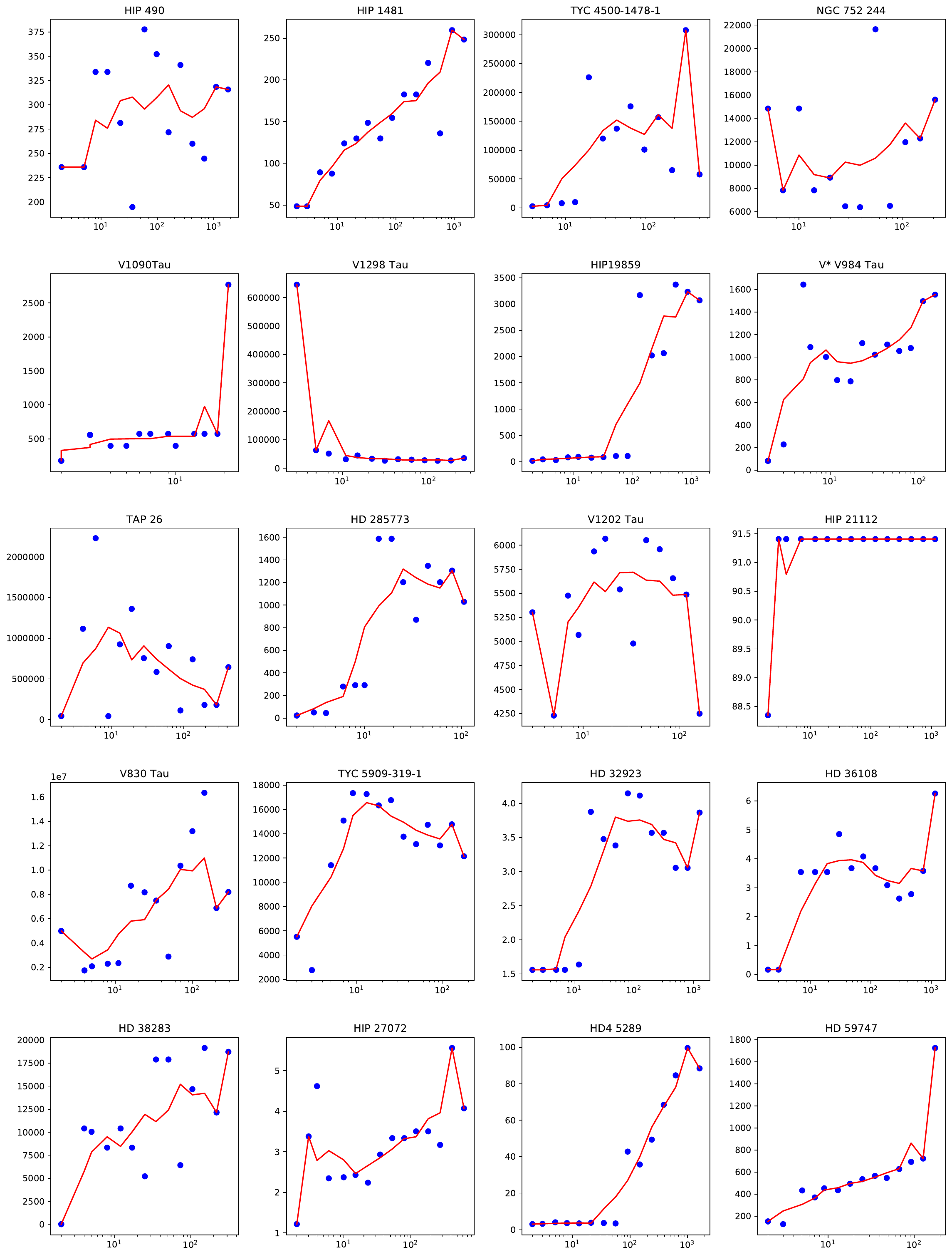}}
\caption{ Pooled variance profile for stars with more than 20 observations.  The red line is a smoothed function for ease reading of the plots.}
\label{pvall1}
\end{figure*}

\begin{figure*}[!htb]
\centering
\ContinuedFloat
\subfloat{
\includegraphics[scale=0.45]{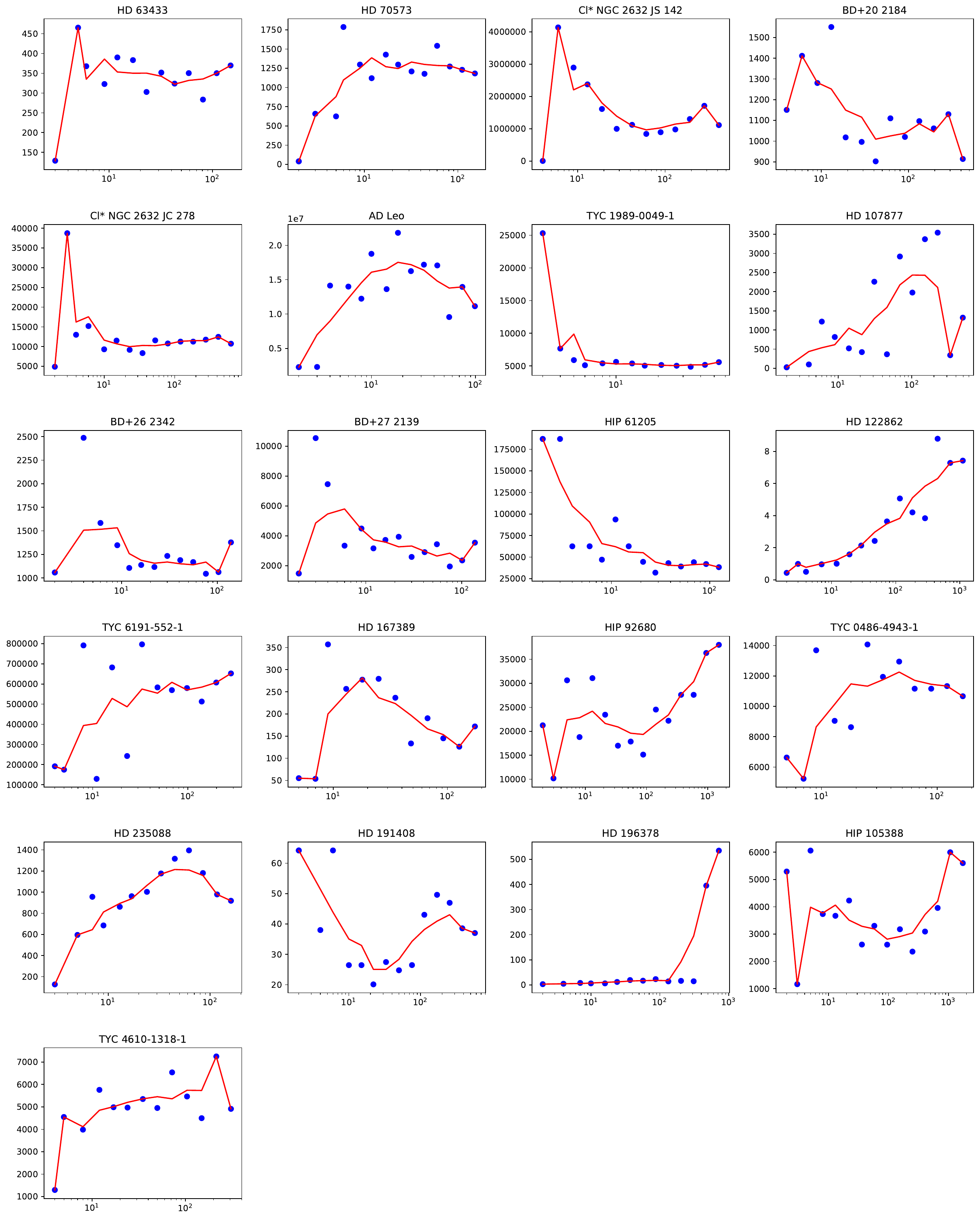}}
\caption{ Continued.}
\label{pvall1}
\end{figure*}

\begin{figure*} 
\centering
\subfloat{
\includegraphics[scale=0.45]{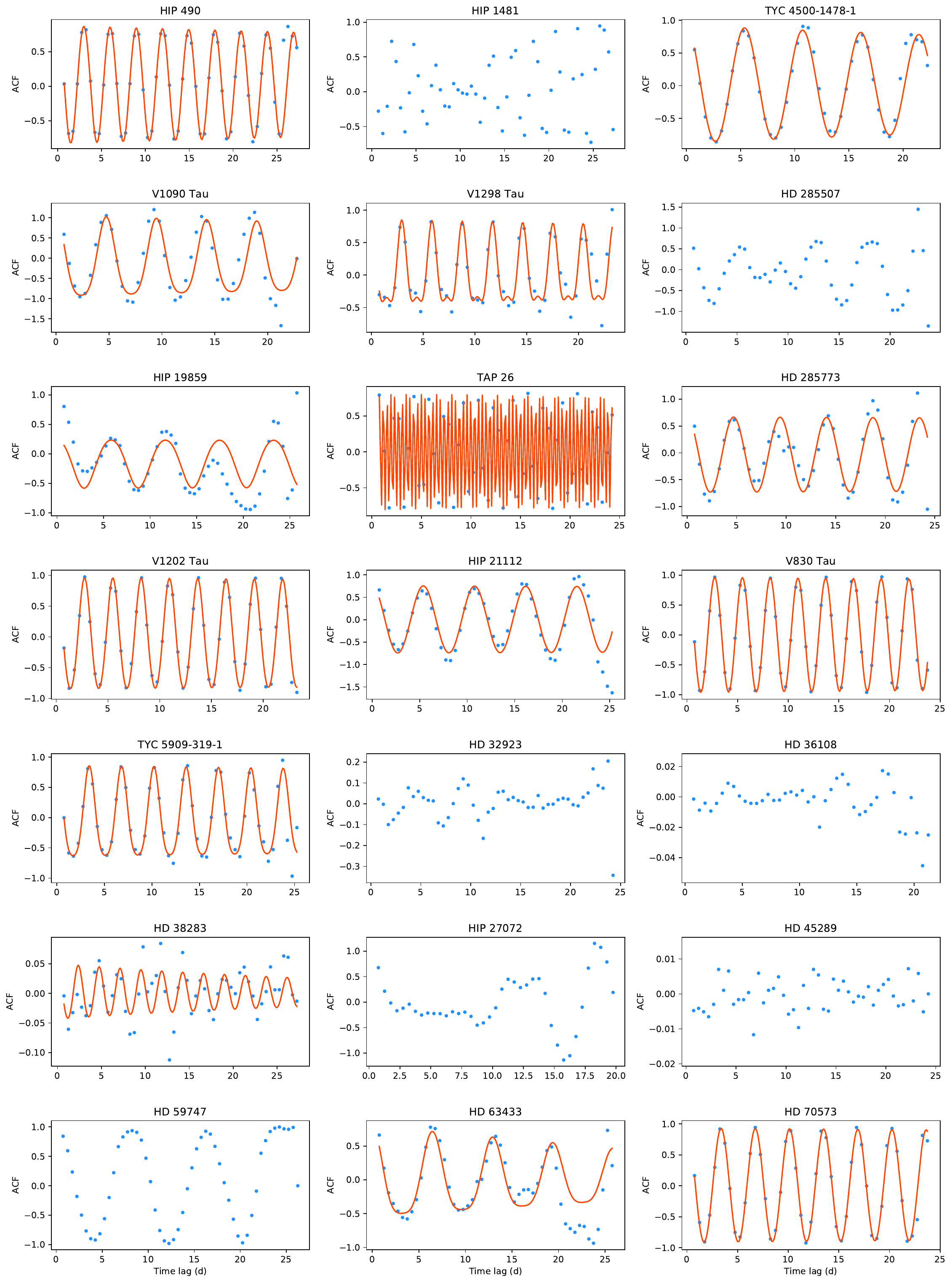}}
\caption{ ACF curves of the TESS data. The red line shows the best fit.} 
\label{acf1}
\end{figure*}

\begin{figure*} 
\centering
\ContinuedFloat
\subfloat{
\includegraphics[scale=0.45]{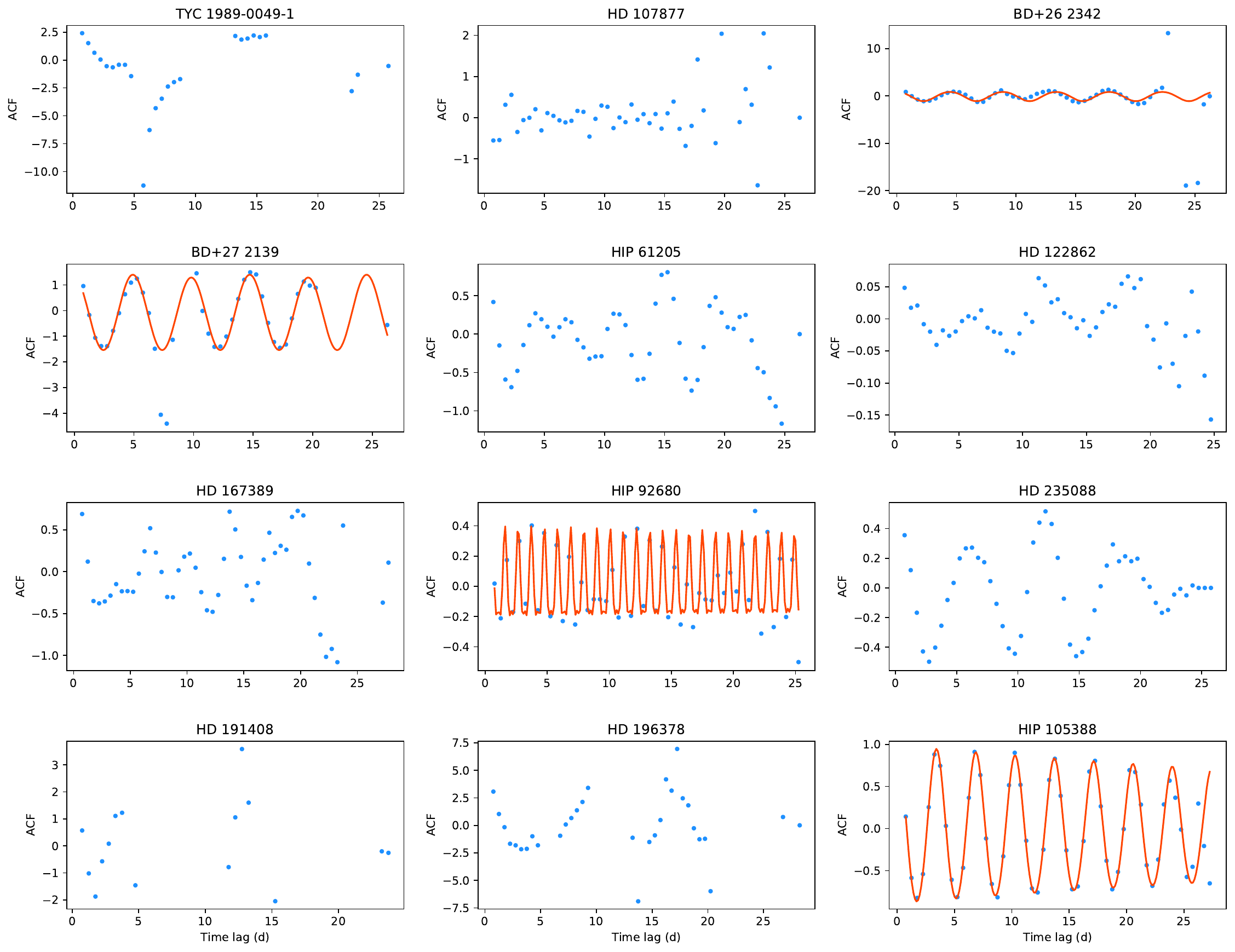}}
\caption{ Continued.}
\label{acf1}
\end{figure*}

\begin{figure*}[!htb]
\centering
\begin{minipage}{0.48\linewidth}
\includegraphics[scale=0.80]{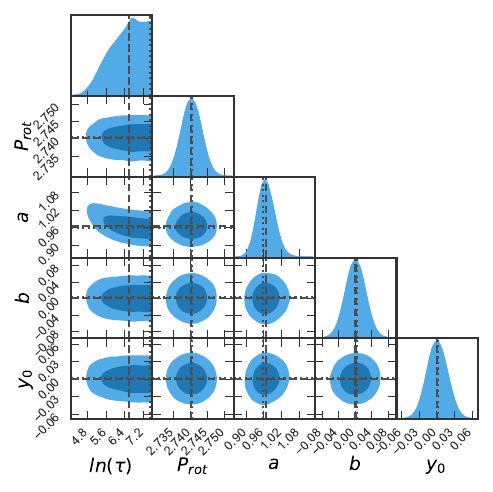}
\end{minipage}
\begin{minipage}{0.48\linewidth}
\includegraphics[scale=0.80]{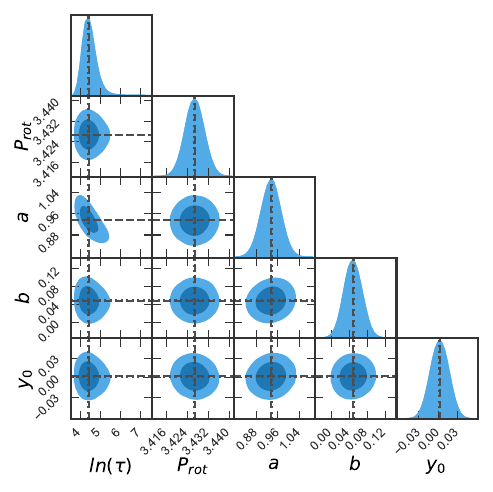}
\end{minipage}
\caption{ Posterior distribution of the ACF modelling for the star V830 Tau (left)
and for the star HIP 105388 (right).}
\label{posterior}
\end{figure*}


\section{Online Tables}

Table~\ref{longtable_kinematicis} provides the kinematic data of the stars analysed in this work.
Namely, star identifier (column 1), galactic-spatial velocity components (columns 2, 3, and 4),
preliminary young stellar group or association (column 5), best hypothesis and probability
(columns 6 and 7)
for stellar group or association membership
obtained using the BANYAN tool \citep{2018ApJ...856...23G}.

\longtab[1]{
\begin{longtable}{lccccccc}
\caption{
 {Galactic-spatial velocity components and membership to stellar kinematic groups and associations.}
  }\label{longtable_kinematicis}\\
\hline
 Star & $U$                  & $V$              & $W$              & Association & Best hypothesis &  Probability \\
      & (kms$^{\rm -1}$)     & (kms$^{\rm -1}$) & (kms$^{\rm -1}$) &             &                 &      \\
\hline
\endfirsthead
\caption{Continued.} \\
\hline
 Star & $U$                  & $V$              & $W$              & Association & Best hypothesis &  Probability \\
      & (kms$^{\rm -1}$)     & (kms$^{\rm -1}$) & (kms$^{\rm -1}$) &             &                 &      \\
\hline
\endhead
\hline
\endfoot
\hline
\endlastfoot
HIP 490	&	-9.30	$\pm$	0.06	&	-20.92	$\pm$	0.04	&	-1.91	$\pm$	0.23	&	TUC	&	TUC	&	0.9998	\\
HIP 1481	&	-9.51	$\pm$	0.12	&	-20.69	$\pm$	0.15	&	-0.38	$\pm$	0.26	&	TUC	&	TUC	&	0.9999	\\
TYC 4500-1478-1	 	&	-6.65	$\pm$	0.68	&	-13.37	$\pm$	1.07	&	-4.67	$\pm$	0.37	&	Cepheus	&	Field	&	0.0000	\\
HD 3823	&	-111.57	$\pm$	0.13	&	-17.94	$\pm$	0.06	&	-35.52	$\pm$	0.12	&	Field	&	Field	&	0.0000	\\
HIP 8486	&	19.56	$\pm$	0.14	&	4.66	$\pm$	0.05	&	-2.45	$\pm$	0.21	&	Ursa Major	&	Field	&	0.0000	\\
NGC 752 48	&	-17.26	$\pm$	0.89	&	-19.90	$\pm$	1.15	&	-21.24	$\pm$	0.83	&	NGC 752	&	Field	&	0.0000	\\
Cl* NGC 752 RV 144	&	-15.88	$\pm$	0.22	&	-21.12	$\pm$	0.27	&	-18.57	$\pm$	0.20	&	NGC 752	&	Field	&	0.0000	\\
NGC 752 80 	&	-18.77	$\pm$	1.63	&	-17.20	$\pm$	1.53	&	-20.35	$\pm$	0.98	&	NGC 752	&	Field	&	0.0000	\\
NGC 752 151	&	-15.91	$\pm$	0.36	&	-20.90	$\pm$	0.36	&	-18.43	$\pm$	0.23	&	NGC 752	&	Field	&	0.0000	\\
NGC 752 185	&	-20.37	$\pm$	1.87	&	-20.52	$\pm$	1.78	&	-19.87	$\pm$	1.12	&	NGC 752	&	Field	&	0.0000	\\
NGC 752 184	&	-14.87	$\pm$	0.41	&	-22.75	$\pm$	0.41	&	-17.34	$\pm$	0.26	&	NGC 752	&	Field	&	0.0000	\\
NGC 752 211	&	-17.42	$\pm$	0.87	&	-19.51	$\pm$	0.82	&	-18.48	$\pm$	0.51	&	NGC 752	&	Field	&	0.0000	\\
NGC 752 229	&	-20.27	$\pm$	2.37	&	-15.65	$\pm$	2.19	&	-22.02	$\pm$	1.41	&	NGC 752	&	Field	&	0.0000	\\
NGC 752 229	&	-20.27	$\pm$	2.37	&	-15.65	$\pm$	2.19	&	-22.02	$\pm$	1.41	&	NGC 752	&	Field	&	0.0000	\\
NGC 752 236	&	-19.04	$\pm$	2.71	&	-18.29	$\pm$	2.50	&	-20.33	$\pm$	1.61	&	NGC 752	&	Field	&	0.0000	\\
NGC 752 244	&	-13.53	$\pm$	0.24	&	-22.11	$\pm$	0.27	&	-18.65	$\pm$	0.19	&	NGC 752	&	Field	&	0.0000	\\
NGC 752 268	&	-16.96	$\pm$	2.14	&	-19.10	$\pm$	1.99	&	-18.92	$\pm$	1.24	&	NGC 752	&	Field	&	0.0000	\\
HIP 14976	&	-40.85	$\pm$	0.20	&	-19.41	$\pm$	0.10	&	-0.88	$\pm$	0.09	&	HYA	&	HYA	&	0.8410	\\
HIP 15310	&	-43.47	$\pm$	0.12	&	-23.43	$\pm$	0.03	&	-1.47	$\pm$	0.10	&	HYA	&	Field	&	0.0000	\\
HD 20794	&	-78.71	$\pm$	0.08	&	-93.03	$\pm$	0.23	&	-29.41	$\pm$	0.36	&	Field	&	Field	&	0.0000	\\
HIP 16529	&	-41.71	$\pm$	0.11	&	-19.17	$\pm$	0.04	&	-0.69	$\pm$	0.06	&	HYA	&	HYA	&	0.9956	\\
HD 22879	&	-111.00	$\pm$	0.15	&	-90.86	$\pm$	0.10	&	-43.08	$\pm$	0.15	&	Field	&	Field	&	0.0000	\\
TYC 1804-1924-1	&	-7.97	$\pm$	0.66	&	-27.31	$\pm$	0.21	&	-15.60	$\pm$	0.29	&	PLE	&	PLE	&	0.9994	\\
V1090 Tau	&	-7.95	$\pm$	0.29	&	-27.88	$\pm$	0.10	&	-13.88	$\pm$	0.13	&	PLE	&	PLE	&	0.9998	\\
  V* V471 Tau         	&	-43.68	$\pm$	0.44	&	-18.26	$\pm$	0.06	&	-2.67	$\pm$	0.23	&	HYA	&	HYA	&	0.9739	\\
  HD 283066           	&	-41.82	$\pm$	0.35	&	-19.23	$\pm$	0.08	&	-1.12	$\pm$	0.15	&	HYA	&	HYA	&	0.9990	\\
  HD 283044           	&	-43.23	$\pm$	0.33	&	-19.31	$\pm$	0.08	&	-1.71	$\pm$	0.14	&	HYA	&	HYA	&	0.9972	\\
  HD 286363           	&	-42.21	$\pm$	0.17	&	-18.87	$\pm$	0.02	&	-1.43	$\pm$	0.10	&	HYA	&	HYA	&	0.9911	\\
HIP 18327	&	-42.59	$\pm$	0.16	&	-18.55	$\pm$	0.03	&	-1.40	$\pm$	0.08	&	HYA	&	HYA	&	0.9995	\\
  HD 285348           	&	-41.31	$\pm$	0.33	&	-19.70	$\pm$	0.05	&	-1.07	$\pm$	0.15	&	HYA	&	HYA	&	0.9975	\\
 V1298 Tau		 	&	-12.68	$\pm$	0.01	&	-6.35	$\pm$	0.02	&	-9.07	$\pm$	0.02	&	Taurus	&	Taurus	&	0.9976	\\
  HG 7-88             	&	-42.85	$\pm$	0.24	&	-19.26	$\pm$	0.04	&	-1.57	$\pm$	0.11	&	HYA	&	HYA	&	0.9991	\\
HIP 19098	&	-42.02	$\pm$	0.19	&	-19.35	$\pm$	0.03	&	-0.50	$\pm$	0.09	&	HYA	&	HYA	&	0.9993	\\
HIP 19148	&	-42.31	$\pm$	0.13	&	-18.62	$\pm$	0.02	&	-1.43	$\pm$	0.07	&	HYA	&	HYA	&	0.9982	\\
 HD 285507		 	&	-42.74	$\pm$	0.38	&	-18.66	$\pm$	0.03	&	-1.60	$\pm$	0.19	&	HYA	&	HYA	&	0.9992	\\
HIP 19781	&	-42.01	$\pm$	0.16	&	-17.93	$\pm$	0.02	&	-0.92	$\pm$	0.08	&	HYA	&	HYA	&	0.9908	\\
HIP 19786	&	-42.06	$\pm$	0.20	&	-19.34	$\pm$	0.03	&	-1.17	$\pm$	0.10	&	HYA	&	HYA	&	0.9983	\\
HIP 19793	&	-42.46	$\pm$	0.14	&	-19.34	$\pm$	0.03	&	-1.37	$\pm$	0.05	&	HYA	&	HYA	&	0.9992	\\
HIP 19796	&	-42.40	$\pm$	0.27	&	-18.74	$\pm$	0.02	&	-1.15	$\pm$	0.14	&	HYA	&	HYA	&	0.9984	\\
HIP 19859	&	14.44	$\pm$	0.13	&	-0.50	$\pm$	0.02	&	-10.04	$\pm$	0.08	&	Ursa Major	&	Field	&	0.0000	\\
  HD 285590           	&	-42.46	$\pm$	0.58	&	-19.42	$\pm$	0.03	&	-0.98	$\pm$	0.27	&	HYA	&	HYA	&	0.9996	\\
  V* V984 Tau         	&	-42.54	$\pm$	0.16	&	-19.06	$\pm$	0.03	&	-1.51	$\pm$	0.06	&	HYA	&	HYA	&	0.9984	\\
 TAP 26			 	&	-18.18	$\pm$	5.27	&	-6.10	$\pm$	0.23	&	-11.53	$\pm$	2.22	&	Taurus	&	Taurus	&	0.9965	\\
HIP 20130	&	-42.40	$\pm$	0.24	&	-19.22	$\pm$	0.03	&	-1.41	$\pm$	0.09	&	HYA	&	HYA	&	0.9996	\\
HIP 20146	&	-41.63	$\pm$	0.27	&	-19.13	$\pm$	0.03	&	-0.75	$\pm$	0.11	&	HYA	&	HYA	&	0.9996	\\
HIP 20237	&	-42.07	$\pm$	0.15	&	-19.13	$\pm$	0.04	&	-1.27	$\pm$	0.06	&	HYA	&	HYA	&	0.9997	\\
HIP 20480	&	-42.21	$\pm$	0.23	&	-18.68	$\pm$	0.03	&	-1.66	$\pm$	0.08	&	HYA	&	HYA	&	0.9989	\\
  V* V988 Tau         	&	-42.42	$\pm$	1.86	&	-17.54	$\pm$	0.58	&	-3.91	$\pm$	0.77	&	HYA	&	HYA	&	0.9739	\\
  HD 27771             	&	-42.36	$\pm$	0.24	&	-18.65	$\pm$	0.02	&	-0.96	$\pm$	0.11	&	HYA	&	HYA	&	0.9995	\\
HIP 20557	&	-42.03	$\pm$	0.15	&	-19.09	$\pm$	0.03	&	-1.22	$\pm$	0.05	&	HYA	&	HYA	&	0.9996	\\
  V* V990 Tau           	&	-42.72	$\pm$	0.17	&	-19.32	$\pm$	0.02	&	-1.05	$\pm$	0.07	&	HYA	&	HYA	&	0.9997	\\
  HD 285742           	&	-42.01	$\pm$	0.82	&	-19.06	$\pm$	0.03	&	-1.63	$\pm$	0.33	&	HYA	&	HYA	&	0.9942	\\
HIP 20741	&	-40.99	$\pm$	0.14	&	-19.56	$\pm$	0.03	&	-1.56	$\pm$	0.06	&	HYA	&	HYA	&	0.9991	\\
  HD 286820           	&	-43.87	$\pm$	0.70	&	-15.43	$\pm$	0.28	&	-5.16	$\pm$	0.43	&	HYA	&	Field	&	0.0046	\\
HIP 20815	&	-41.96	$\pm$	0.15	&	-19.14	$\pm$	0.02	&	-1.22	$\pm$	0.06	&	HYA	&	HYA	&	0.9997	\\
HIP 20826	&	-41.48	$\pm$	0.16	&	-19.19	$\pm$	0.02	&	-0.85	$\pm$	0.08	&	HYA	&	HYA	&	0.9994	\\
HIP 20850	&	-43.63	$\pm$	0.65	&	-18.17	$\pm$	0.08	&	-1.96	$\pm$	0.29	&	HYA	&	HYA	&	0.9990	\\
  HD 28258            	&	-43.63	$\pm$	0.65	&	-18.17	$\pm$	0.08	&	-1.96	$\pm$	0.29	&	HYA	&	HYA	&	0.9990	\\
HIP 20899	&	-41.94	$\pm$	0.14	&	-19.37	$\pm$	0.02	&	-0.45	$\pm$	0.06	&	HYA	&	HYA	&	0.9996	\\
HIP 20951	&	-42.10	$\pm$	0.10	&	-19.30	$\pm$	0.02	&	-1.28	$\pm$	0.04	&	HYA	&	HYA	&	0.9997	\\
  HD 285773           	&	-42.10	$\pm$	0.10	&	-19.30	$\pm$	0.02	&	-1.28	$\pm$	0.04	&	HYA	&	HYA	&	0.9997	\\
HIP 20978	&	-42.57	$\pm$	0.21	&	-19.17	$\pm$	0.02	&	-1.38	$\pm$	0.08	&	HYA	&	HYA	&	0.9998	\\
  HD 28462             	&	-42.57	$\pm$	0.21	&	-19.17	$\pm$	0.02	&	-1.38	$\pm$	0.08	&	HYA	&	HYA	&	0.9998	\\
HIP 21099	&	-42.63	$\pm$	0.14	&	-19.42	$\pm$	0.02	&	-1.07	$\pm$	0.05	&	HYA	&	HYA	&	0.9996	\\
 V1202 Tau		 	&	-12.54	$\pm$	0.02	&	-6.09	$\pm$	0.02	&	-9.59	$\pm$	0.02	&	Taurus	&	Taurus	&	0.9991	\\
HIP 21112	&	-42.01	$\pm$	0.21	&	-15.36	$\pm$	0.12	&	-1.43	$\pm$	0.15	&	HYA	&	Field	&	0.0116	\\
 V830 Tau		 	&	-12.85	$\pm$	2.48	&	-11.56	$\pm$	0.26	&	-8.56	$\pm$	0.70	&	Taurus	&	Taurus	&	0.9981	\\
HIP 21317	&	-42.01	$\pm$	0.13	&	-19.39	$\pm$	0.02	&	-1.61	$\pm$	0.05	&	HYA	&	HYA	&	0.9996	\\
HIP 21654	&	-43.59	$\pm$	0.21	&	-21.08	$\pm$	0.33	&	-0.49	$\pm$	0.28	&	HYA	&	HYA	&	0.9939	\\
HIP 22203	&	-42.81	$\pm$	0.28	&	-19.35	$\pm$	0.08	&	-1.51	$\pm$	0.11	&	HYA	&	HYA	&	0.9995	\\
  HD 284787            	&	-44.69	$\pm$	1.71	&	-20.07	$\pm$	0.21	&	-3.08	$\pm$	0.47	&	HYA	&	HYA	&	0.9607	\\
HIP 22422	&	-42.13	$\pm$	0.13	&	-19.19	$\pm$	0.02	&	-0.80	$\pm$	0.05	&	HYA	&	HYA	&	0.9995	\\
HIP 23069	&	-42.24	$\pm$	0.27	&	-18.92	$\pm$	0.04	&	-1.20	$\pm$	0.09	&	HYA	&	HYA	&	0.9983	\\
HIP 23498	&	-41.78	$\pm$	0.32	&	-18.88	$\pm$	0.05	&	-1.15	$\pm$	0.10	&	HYA	&	HYA	&	0.9977	\\
HIP 23750	&	-42.15	$\pm$	0.26	&	-19.13	$\pm$	0.03	&	-1.38	$\pm$	0.06	&	HYA	&	HYA	&	0.9969	\\
 TYC 5909-319-1		 	&	-14.70	$\pm$	0.71	&	-21.34	$\pm$	0.61	&	-4.45	$\pm$	0.58	&	Field	&	Field	&	0.0000	\\
HD 32923	&	-26.10	$\pm$	0.15	&	-23.64	$\pm$	0.10	&	28.35	$\pm$	0.14	&	Field	&	Field	&	0.0000	\\
HIP 25486	&	-11.59	$\pm$	0.38	&	-15.93	$\pm$	0.26	&	-8.99	$\pm$	0.21	&	Beta Pic	&	Beta Pic	&	0.9993	\\
HD 36108	&	-35.23	$\pm$	0.09	&	7.03	$\pm$	0.10	&	-24.36	$\pm$	0.07	&	Field	&	Field	&	0.0000	\\
HD 38283	&	31.70	$\pm$	0.04	&	-60.94	$\pm$	0.15	&	-7.82	$\pm$	0.09	&	Field	&	Field	&	0.0000	\\
HIP 27072	&	17.95	$\pm$	0.02	&	4.25	$\pm$	0.00	&	-11.75	$\pm$	0.02	&	Ursa Major	&	Field	&	0.0000	\\
HD 45289	&	-115.09	$\pm$	0.06	&	-22.82	$\pm$	0.13	&	-3.76	$\pm$	0.06	&	Field	&	Field	&	0.0000	\\
 HD 59747		 	&	12.65	$\pm$	0.15	&	2.76	$\pm$	0.01	&	-10.42	$\pm$	0.06	&	Ursa Major	&	Field	&	0.0000	\\
 HD 63433		 	&	14.10	$\pm$	0.18	&	2.54	$\pm$	0.04	&	-7.97	$\pm$	0.08	&	Ursa Major	&	Field	&	0.0000	\\
 HD 70573		 	&	-15.80	$\pm$	0.19	&	-21.30	$\pm$	0.18	&	-10.53	$\pm$	0.10	&	Hercules-Lyra	&	Field	&	0.0000	\\
  Cl* NGC 2632 JC 61  	&	-43.54	$\pm$	1.18	&	-20.31	$\pm$	0.58	&	-8.25	$\pm$	0.81	&	PRAE	&	PRAE	&	0.9998	\\
  Cl* NGC 2632 JS 133 	&	-42.17	$\pm$	0.67	&	-19.78	$\pm$	0.31	&	-10.15	$\pm$	0.47	&	PRAE	&	PRAE	&	0.9999	\\
Cl* NGC 2632 JS 142	&	-41.98	$\pm$	1.71	&	-19.88	$\pm$	0.78	&	-9.71	$\pm$	1.18	&	PRAE	&	PRAE	&	0.9998	\\
  Cl* NGC 2632 JS 143 	&	-42.62	$\pm$	0.99	&	-20.19	$\pm$	0.46	&	-9.54	$\pm$	0.69	&	PRAE	&	PRAE	&	0.9999	\\
Cl* NGC 2632 JS 247	&	-43.76	$\pm$	1.15	&	-20.77	$\pm$	0.53	&	-9.02	$\pm$	0.81	&	PRAE	&	PRAE	&	0.9999	\\
Cl* NGC 2632 JC 141	&	-43.40	$\pm$	0.81	&	-20.37	$\pm$	0.40	&	-9.32	$\pm$	0.57	&	PRAE	&	PRAE	&	0.9999	\\
Cl* NGC 2632 JC 158	&	-42.91	$\pm$	1.00	&	-20.52	$\pm$	0.48	&	-9.04	$\pm$	0.71	&	PRAE	&	PRAE	&	1.0000	\\
Cl* NGC 2632 JC 172 	&	-44.45	$\pm$	1.78	&	-21.35	$\pm$	0.86	&	-8.31	$\pm$	1.26	&	PRAE	&	PRAE	&	0.9999	\\
Cl* NGC 2632 JC 208	&	-43.24	$\pm$	1.30	&	-19.34	$\pm$	0.64	&	-9.41	$\pm$	0.93	&	PRAE	&	PRAE	&	0.9999	\\
  Cl* NGC 2632 JC 221	&	-42.10	$\pm$	1.20	&	-20.44	$\pm$	0.61	&	-8.49	$\pm$	0.85	&	PRAE	&	PRAE	&	0.9998	\\
Cl* NGC 2632 JS 404 	&	-43.44	$\pm$	0.66	&	-21.18	$\pm$	0.31	&	-9.97	$\pm$	0.48	&	PRAE	&	PRAE	&	0.9999	\\
Cl* NGC 2632 JC 234 	&	-44.04	$\pm$	0.60	&	-19.86	$\pm$	0.30	&	-9.84	$\pm$	0.44	&	PRAE	&	PRAE	&	0.9998	\\
K2 101	&	-42.89	$\pm$	0.67	&	-20.44	$\pm$	0.34	&	-10.01	$\pm$	0.49	&	PRAE	&	PRAE	&	1.0000	\\
BD+20 2184	&	-41.81	$\pm$	0.37	&	-20.45	$\pm$	0.18	&	-9.36	$\pm$	0.29	&	PRAE	&	PRAE	&	0.9999	\\
Cl* NGC 2632 WJJP 792	&	-42.52	$\pm$	0.81	&	-21.88	$\pm$	0.39	&	-9.72	$\pm$	0.59	&	PRAE	&	PRAE	&	0.9997	\\
Cl* NGC 2632 JC 278 	&	-42.67	$\pm$	0.36	&	-20.07	$\pm$	0.18	&	-9.92	$\pm$	0.27	&	PRAE	&	PRAE	&	1.0000	\\
  Cl* NGC 2632 KW 551 	&	-42.01	$\pm$	0.64	&	-20.70	$\pm$	0.33	&	-9.77	$\pm$	0.47	&	PRAE	&	PRAE	&	0.9999	\\
Cl* NGC 2632 JS 563	&	-41.78	$\pm$	0.53	&	-19.78	$\pm$	0.28	&	-9.31	$\pm$	0.40	&	PRAE	&	PRAE	&	0.9999	\\
  Cl* NGC 2632 JS 576 	&	-42.15	$\pm$	1.23	&	-20.22	$\pm$	0.59	&	-9.27	$\pm$	0.91	&	PRAE	&	PRAE	&	0.9999	\\
 AD Leo			 	&	-15.00	$\pm$	0.00	&	-7.49	$\pm$	0.00	&	3.52	$\pm$	0.00	&	 Castor MG	&	Field	&	0.0000	\\
 TYC 1989-0049-1	 	&	-2.16	$\pm$	0.04	&	-5.54	$\pm$	0.03	&	-0.90	$\pm$	0.37	&	CBER	&	CBER	&	1.0000	\\
 HD 107877		 	&	-2.71	$\pm$	0.03	&	-5.26	$\pm$	0.03	&	-0.29	$\pm$	0.35	&	CBER	&	CBER	&	0.9999	\\
 BD+26 2342		 	&	-2.45	$\pm$	0.05	&	-5.66	$\pm$	0.04	&	-0.64	$\pm$	0.61	&	CBER	&	CBER	&	1.0000	\\
 BD+27 2139		 	&	-2.67	$\pm$	0.02	&	-5.79	$\pm$	0.02	&	0.80	$\pm$	0.21	&	CBER	&	CBER	&	0.9999	\\
 HIP 61205		 	&	-1.95	$\pm$	0.04	&	-5.94	$\pm$	0.03	&	-0.68	$\pm$	0.32	&	CBER	&	Field	&	0.3090	\\
 BD+23 2472		 	&	-1.93	$\pm$	0.02	&	-4.40	$\pm$	0.05	&	-2.33	$\pm$	0.39	&	CBER	&	CBER	&	0.9027	\\
HD 122862	&	-27.57	$\pm$	0.10	&	-4.91	$\pm$	0.12	&	37.76	$\pm$	0.04	&	Field	&	Field	&	0.0000	\\
TYC 6779-305-1	&	-3.97	$\pm$	0.62	&	-17.35	$\pm$	0.13	&	-6.68	$\pm$	0.26	&	Upper Sco	&	Upper Sco	&	0.9959	\\
 TYC 6191-552-1		 	&	-7.00	$\pm$	0.62	&	-15.45	$\pm$	0.09	&	-6.29	$\pm$	0.30	&	Upper Sco	&	Upper Sco	&	0.9983	\\
 GSC 06204-00812	 	&	-4.27	$\pm$	1.15	&	-15.25	$\pm$	0.13	&	-7.37	$\pm$	0.55	&	Upper Sco	&	Upper Sco	&	0.9987	\\
 V866	Sco 	&				&				&				&	Upper Sco	&	Upper Sco	&	0.9711	\\
 HD 167389		 	&	17.20	$\pm$	0.05	&	-6.03	$\pm$	0.13	&	-14.79	$\pm$	0.06	&	Ursa Major	&	Field	&	0.0000	\\
HIP 92680	&	-11.13	$\pm$	0.18	&	-14.88	$\pm$	0.05	&	-7.25	$\pm$	0.07	&	Beta Pic	&	Beta Pic	&	0.9739	\\
 TYC 0486-4943-1	 	&	-5.35	$\pm$	0.08	&	-26.96	$\pm$	0.07	&	-12.22	$\pm$	0.02	&	AB Dor	&	Field	&	0.1888	\\
HD 186408	&	17.54	$\pm$	0.02	&	-30.11	$\pm$	0.15	&	-0.34	$\pm$	0.03	&	Field	&	Field	&	0.0000	\\
HD 186427	&	17.23	$\pm$	0.02	&	-30.26	$\pm$	0.15	&	-1.80	$\pm$	0.04	&	Field	&	Field	&	0.0000	\\
 HD 235088		 	&	-41.73	$\pm$	0.02	&	-21.77	$\pm$	0.26	&	-18.95	$\pm$	0.06	&	Field	&	Field	&	0.0000	\\
HD 191408	&	-118.33	$\pm$	0.12	&	-51.95	$\pm$	0.03	&	46.99	$\pm$	0.07	&	Field	&	Field	&	0.0000	\\
HD 196378	&	-66.30	$\pm$	0.16	&	-48.81	$\pm$	0.14	&	-1.35	$\pm$	0.12	&	Field	&	Field	&	0.0000	\\
 TYC 1090-543-1		 	&	-5.88	$\pm$	1.74	&	-26.59	$\pm$	2.63	&	-12.31	$\pm$	1.28	&	AB Dor	&	Field	&	0.0321	\\
HIP 105388	&	-8.42	$\pm$	0.01	&	-20.66	$\pm$	0.02	&	-0.69	$\pm$	0.00	&	TUC	&	TUC	&	0.9997	\\
HD 210918	&	-47.12	$\pm$	0.09	&	-90.59	$\pm$	0.08	&	-8.20	$\pm$	0.12	&	Field	&	Field	&	0.0000	\\
HIP 116748	&	-9.13	$\pm$	0.15	&	-21.10	$\pm$	0.16	&	-0.90	$\pm$	0.23	&	TUC	&	TUC	&	0.9999	\\
 TYC 4610-1318-1	 	&	-10.83	$\pm$	0.22	&	-14.14	$\pm$	0.37	&	-5.48	$\pm$	0.13	&	Cepheus	&	Field	&	0.0000	\\
\hline\noalign{\smallskip}
\end{longtable}
\tablefoot{TUC: Tucana-Horologium; PLE: Pleiades; CBER: Coma Berenices; PRAE: Praesepe (M44); HYA: Hyades. }
}%

 Table~\ref{longtable_other_properties} lists for each star (column 1) its corresponding
 age (column 2), effective temperature (column 3), (B-V) colour (column 4), 
 $\log(R^{\rm '}_{\rm HK})$ computed using the prescriptions given in \citet[][]{1984ApJ...279..763N} (column 5),
 projected rotational velocity, v$\sin i$, (column 6), stellar mass, radius, and luminosity (columns 7, 8, and 9; for simplicity, 
 asymmetric uncertainties were averaged into a single error estimate),  turnover convective timescale (column 10),
 number of observations (column 11), time span (column 12), and mean signal-to-noise ratio measured at $\sim$ 550 nm. 
 Table~\ref{longtable_fluxes} provides the emission excess in the  Ca~{\sc ii} H (column 2), Ca~{\sc ii} K (column 3), and H$\alpha$ (column 4) lines.



\longtab[2]{
\begin{landscape}
\begin{longtable}{lcccccccccccc}
\caption{
 { Derived properties for the observed stars.}
  }\label{longtable_other_properties}\\
\hline
 Star & Age   &  T$_{\rm eff}$  & (B-V) &  $\log$R$^{\rm '}_{\rm HK}$ &  v$\sin i$         & M$_{\star}$   &  R$_{\star}$   & L$_{\star}$   & $\tau_{\rm conv}$  & n$_{\rm obs}$ & T$_{\rm span}$ & SNR\\
      & (Myr) &  (K)            &       &                             &   (kms$^{\rm -1}$) & (M$_{\odot}$) &  (R$_{\odot}$) & (L$_{\odot}$) &       (d) & & (yr) & \\ 
\hline
\endfirsthead
\caption{Continued.} \\
\hline
Star & Age   &  T$_{\rm eff}$  & (B-V) &  $\log$R$^{\rm '}_{\rm HK}$ &  v$\sin i$         & M$_{\star}$   &  R$_{\star}$   & L$_{\star}$   & $\tau_{\rm conv}$  & n$_{\rm obs}$ & T$_{\rm span}$ & SNR\\
      & (Myr) &  (K)            &       &                             &   (kms$^{\rm -1}$) & (M$_{\odot}$) &  (R$_{\odot}$) & (L$_{\odot}$) &       (d) & & (yr) & \\
\hline
\endhead
\hline
\endfoot
\hline
\endlastfoot
\hline\noalign{\smallskip}
HIP490	&	30	&	5940	$\pm$	77	&	0.35	&				&	14.48	$\pm$	1.05	&	1.044	$\pm$	0.000	&	1.05	$\pm$	0.03	&	1.240	$\pm$	0.003	&	25.37	&	50	&	14.711	&	123	\\
HIP1481	&	30	&	6115	$\pm$	67	&	0.31	&				&	20.81	$\pm$	1.92	&	1.099	$\pm$	0.001	&	1.12	$\pm$	0.02	&	1.589	$\pm$	0.003	&	21.63	&	24	&	12.046	&	106	\\
 TYC 4500-1478-1	 	&	10	&	4978	$\pm$	72	&	0.74	&	-3.877	$\pm$	0.059	&	7.62	$\pm$	0.34	&	1.160	$\pm$	0.002	&	1.63	$\pm$	0.04	&	1.463	$\pm$	0.014	&	123.92	&	45	&	3.318	&	59	\\
HD3823	&	6700	&	6103	$\pm$	57	&	0.32	&				&	2.82	$\pm$	0.21	&	1.214	$\pm$	0.001	&	1.37	$\pm$	0.02	&	2.360	$\pm$	0.005	&		&	9	&	1.938	&	459	\\
HIP8486	&	414	&	5793	$\pm$	52	&	0.41	&				&	3.11	$\pm$	0.01	&	0.993	$\pm$	0.001	&	0.98	$\pm$	0.01	&	0.979	$\pm$	0.003	&	32.99	&	1	&		&	148	\\
NGC 752 48	&	1340	&	6061	$\pm$	162	&	0.32	&				&	4.98	$\pm$	1.27	&	1.185	$\pm$	0.021	&	1.32	$\pm$	0.07	&	2.120	$\pm$	0.148	&	7.88	&	1	&		&	24	\\
Cl* NGC 752 RV 144	&	1340	&	5924	$\pm$	224	&	0.37	&				&	3.33	$\pm$	0.24	&	1.010	$\pm$	0.009	&	0.99	$\pm$	0.05	&	1.080	$\pm$	0.037	&	30.30	&	9	&	0.090	&	14	\\
NGC 752 80 	&	1340	&	5940	$\pm$	202	&	0.35	&				&	4.90	$\pm$	0.27	&	1.058	$\pm$	0.007	&	1.08	$\pm$	0.05	&	1.308	$\pm$	0.036	&	23.40	&	3	&	0.011	&	20	\\
NGC 752 151	&	1340	&	6325	$\pm$	334	&	0.24	&				&	10.63	$\pm$	6.11	&	1.114	$\pm$	0.008	&	1.11	$\pm$	0.10	&	1.763	$\pm$	0.050	&	16.16	&	6	&	0.013	&	19	\\
NGC 752 185	&	1340	&	5955	$\pm$	205	&	0.36	&				&	6.22	$\pm$	0.03	&	1.138	$\pm$	0.011	&	1.25	$\pm$	0.06	&	1.758	$\pm$	0.066	&	13.21	&	7	&	0.038	&	22	\\
NGC 752 184	&	1340	&	5964	$\pm$	159	&	0.35	&				&	5.08	$\pm$	0.09	&	1.100	$\pm$	0.008	&	1.16	$\pm$	0.04	&	1.538	$\pm$	0.046	&	17.87	&	13	&	0.156	&	18	\\
NGC 752 211	&	1340	&	5889	$\pm$	50	&	0.39	&				&	4.82	$\pm$	0.18	&	1.030	$\pm$	0.007	&	1.03	$\pm$	0.02	&	1.158	$\pm$	0.030	&	27.38	&	9	&	0.156	&	19	\\
NGC 752 229	&	1340	&	5972	$\pm$	83	&	0.35	&				&	3.11	$\pm$	0.01	&	1.066	$\pm$	0.008	&	1.09	$\pm$	0.03	&	1.356	$\pm$	0.041	&	22.35	&	11	&	0.241	&	19	\\
NGC 752 229	&	1340	&	5972	$\pm$	83	&	0.35	&				&	5.08	$\pm$	0.09	&	1.066	$\pm$	0.008	&	1.09	$\pm$	0.03	&	1.356	$\pm$	0.041	&	22.35	&	10	&	0.090	&	18	\\
NGC 752 236	&	1340	&	6013	$\pm$	484	&	0.34	&				&	5.17	$\pm$	0.83	&	1.087	$\pm$	0.008	&	1.12	$\pm$	0.10	&	1.484	$\pm$	0.044	&	19.59	&	11	&	0.090	&	19	\\
NGC 752 244	&	1340	&	5954	$\pm$	149	&	0.36	&				&	3.64	$\pm$	0.07	&	1.033	$\pm$	0.008	&	1.03	$\pm$	0.06	&	1.193	$\pm$	0.037	&	26.94	&	27	&	1.709	&	16	\\
NGC 752 268	&	1340	&	5968	$\pm$	62	&	0.35	&				&	3.79	$\pm$	0.09	&	1.095	$\pm$	0.008	&	1.15	$\pm$	0.03	&	1.508	$\pm$	0.047	&	18.53	&	15	&	0.156	&	26	\\
HIP14976	&	750	&	5586	$\pm$	23	&	0.48	&	-4.427			&	3.60	$\pm$	0.84	&	0.946	$\pm$	0.001	&	0.93	$\pm$	0.00	&	0.765	$\pm$	0.004	&	38.03	&	1	&		&	171	\\
HIP15310	&	750	&	5951	$\pm$	95	&	0.36	&				&	4.94	$\pm$	0.07	&	1.094	$\pm$	0.001	&	1.15	$\pm$	0.03	&	1.499	$\pm$	0.004	&	19.40	&	15	&	4.776	&	136	\\
HD20794	&	13500	&	5751	$\pm$	656	&	0.42	&				&	3.10	$\pm$	0.02	&	0.912	$\pm$	0.001	&	0.83	$\pm$	0.17	&	0.687	$\pm$	0.002	&		&	6	&	6.417	&	596	\\
HIP16529	&	750	&	5186	$\pm$	64	&	0.65	&	-4.312			&	3.10	$\pm$	0.49	&	0.871	$\pm$	0.001	&	0.87	$\pm$	0.02	&	0.494	$\pm$	0.002	&	48.59	&	1	&		&	107	\\
HD22879	&	13900	&	5912	$\pm$	38	&	0.37	&				&	2.93	$\pm$	0.16	&	1.048	$\pm$	0.001	&	1.06	$\pm$	0.02	&	1.248	$\pm$	0.003	&		&	4	&	11.080	&	338	\\
TYC 1804-1924-1	&	112	&	6614	$\pm$	430	&	0.17	&				&	17.75	$\pm$	7.00	&	1.270	$\pm$	0.005	&	1.36	$\pm$	0.16	&	3.186	$\pm$	0.042	&		&	8	&	0.008	&	85	\\
V1090Tau	&	112	&	5438	$\pm$	107	&	0.54	&	-4.200	$\pm$	0.025	&	3.10	$\pm$	0.42	&	0.949	$\pm$	0.002	&	0.97	$\pm$	0.03	&	0.744	$\pm$	0.007	&	37.68	&	24	&	0.178	&	39	\\
  V* V471 Tau         	&	750	&	4994	$\pm$	123	&	0.73	&	-4.018	$\pm$	0.050	&	50.58	$\pm$	6.63	&	0.833	$\pm$	0.001	&	0.84	$\pm$	0.03	&	0.391	$\pm$	0.002	&	54.12	&	3	&	0.005	&	36	\\
  HD 283066           	&	750	&	4493	$\pm$	130	&	0.98	&				&	2.45	$\pm$	0.13	&	0.726	$\pm$	0.001	&	0.73	$\pm$	0.03	&	0.195	$\pm$	0.001	&	70.07	&	16	&	0.821	&	48	\\
  HD 283044           	&	750	&	4165	$\pm$	179	&	1.15	&				&	3.09	$\pm$	0.01	&	0.674	$\pm$	0.001	&	0.70	$\pm$	0.05	&	0.131	$\pm$	0.001	&	80.14	&	9	&	0.175	&	29	\\
  HD 286363           	&	750	&	4730	$\pm$	188	&	0.85	&				&	3.09	$\pm$	0.01	&	0.745	$\pm$	0.001	&	0.72	$\pm$	0.04	&	0.232	$\pm$	0.001	&	67.36	&	11	&	0.175	&	43	\\
HIP18327	&	750	&	5087	$\pm$	60	&	0.69	&	-4.253	$\pm$	0.020	&	4.02	$\pm$	0.14	&	0.850	$\pm$	0.001	&	0.85	$\pm$	0.02	&	0.435	$\pm$	0.001	&	51.74	&	4	&	9.892	&	55	\\
  HD 285348           	&	750	&	4747	$\pm$	208	&	0.85	&				&	3.33	$\pm$	0.24	&	0.757	$\pm$	0.001	&	0.74	$\pm$	0.04	&	0.249	$\pm$	0.001	&	65.54	&	11	&	0.279	&	45	\\
 V1298 Tau		 	&	1	&	4962	$\pm$	88	&	0.74	&	-3.909	$\pm$	0.046	&	13.57	$\pm$	0.48	&	1.032	$\pm$	0.002	&	1.29	$\pm$	0.04	&	0.915	$\pm$	0.008	&	586.56	&	116	&	2.081	&	56	\\
  HG 7-88             	&	750	&	4060	$\pm$	133	&	1.22	&				&	3.41	$\pm$	0.31	&	0.670	$\pm$	0.001	&	0.71	$\pm$	0.05	&	0.124	$\pm$	0.001	&	81.12	&	8	&	0.169	&	27	\\
HIP19098	&	750	&	4966	$\pm$	179	&	0.74	&	-4.350	$\pm$	0.002	&	4.02	$\pm$	0.14	&	0.852	$\pm$	0.001	&	0.88	$\pm$	0.10	&	0.425	$\pm$	0.002	&	51.44	&	2	&	0.027	&	90	\\
HIP19148	&	750	&	5965	$\pm$	31	&	0.35	&				&	4.66	$\pm$	0.41	&	1.074	$\pm$	0.001	&	1.11	$\pm$	0.02	&	1.396	$\pm$	0.004	&	21.89	&	8	&	9.919	&	66	\\
 HD 285507		 	&	750	&	4453	$\pm$	127	&	1.00	&				&	3.63	$\pm$	0.53	&	0.724	$\pm$	0.001	&	0.73	$\pm$	0.04	&	0.191	$\pm$	0.001	&	70.36	&	18	&	0.451	&	40	\\
HIP19781	&	750	&	5631	$\pm$	47	&	0.46	&	-4.306			&	3.10	$\pm$	0.02	&	0.977	$\pm$	0.001	&	0.99	$\pm$	0.02	&	0.879	$\pm$	0.003	&	34.39	&	1	&		&	90	\\
HIP19786	&	750	&	5828	$\pm$	77	&	0.40	&				&	3.64	$\pm$	0.07	&	1.037	$\pm$	0.001	&	1.06	$\pm$	0.03	&	1.173	$\pm$	0.006	&	26.67	&	1	&		&	115	\\
HIP19793	&	750	&	5750	$\pm$	100	&	0.43	&				&	5.08	$\pm$	0.09	&	1.026	$\pm$	0.001	&	1.06	$\pm$	0.02	&	1.100	$\pm$	0.004	&	28.18	&	1	&		&	129	\\
HIP19796	&	750	&	6293	$\pm$	17	&	0.25	&				&	15.48	$\pm$	0.03	&	1.206	$\pm$	0.001	&	1.30	$\pm$	0.02	&	2.406	$\pm$	0.006	&	6.39	&	1	&		&	123	\\
HIP19859	&	414	&	6052	$\pm$	70	&	0.33	&				&	3.60	$\pm$	0.84	&	1.036	$\pm$	0.001	&	1.01	$\pm$	0.02	&	1.234	$\pm$	0.002	&	27.53	&	63	&	11.121	&	127	\\
  HD 285590           	&	750	&	4240	$\pm$	87	&	1.12	&				&	3.33	$\pm$	0.24	&	0.693	$\pm$	0.001	&	0.72	$\pm$	0.03	&	0.150	$\pm$	0.001	&	75.49	&	7	&	0.169	&	32	\\
  V* V984 Tau         	&	750	&	5292	$\pm$	155	&	0.61	&	-4.319	$\pm$	0.030	&	2.62	$\pm$	0.04	&	0.889	$\pm$	0.001	&	0.88	$\pm$	0.07	&	0.550	$\pm$	0.002	&	45.90	&	49	&	1.251	&	65	\\
 TAP 26			 	&	1	&	4389	$\pm$	107	&	1.03	&				&	66.39	$\pm$	0.37	&	0.801	$\pm$	0.002	&	0.92	$\pm$	0.03	&	0.281	$\pm$	0.002	&	532.54	&	45	&	3.432	&	29	\\
HIP20130	&	750	&	5566	$\pm$	32	&	0.50	&	-4.383			&	3.51	$\pm$	0.32	&	0.933	$\pm$	0.001	&	0.91	$\pm$	0.03	&	0.720	$\pm$	0.003	&	39.79	&	1	&		&	150	\\
HIP20146	&	750	&	5572	$\pm$	41	&	0.49	&	-4.371	$\pm$	0.040	&	3.09	$\pm$	0.01	&	0.964	$\pm$	0.001	&	0.97	$\pm$	0.02	&	0.820	$\pm$	0.003	&	35.88	&	3	&		&	62	\\
HIP20237	&	750	&	6064	$\pm$	136	&	0.31	&				&	9.94	$\pm$	0.37	&	1.149	$\pm$	0.001	&	1.24	$\pm$	0.05	&	1.873	$\pm$	0.011	&	12.82	&	1	&		&	159	\\
HIP20480	&	750	&	5500	$\pm$	70	&	0.53	&	-4.365			&	3.66	$\pm$	0.43	&	0.922	$\pm$	0.001	&	0.90	$\pm$	0.04	&	0.675	$\pm$	0.003	&	41.28	&	1	&		&	98	\\
  V* V988 Tau         	&	750	&	4974	$\pm$	86	&	0.75	&	-4.299	$\pm$	0.018	&	3.09	$\pm$	0.01	&	0.836	$\pm$	0.002	&	0.85	$\pm$	0.02	&	0.394	$\pm$	0.004	&	53.70	&	5	&	0.112	&	43	\\
  HD27771             	&	750	&	5200	$\pm$	147	&	0.65	&	-4.243	$\pm$	0.015	&	3.79	$\pm$	0.09	&	0.866	$\pm$	0.001	&	0.86	$\pm$	0.04	&	0.483	$\pm$	0.002	&	49.34	&	8	&	0.087	&	61	\\
HIP20557	&	750	&	6225	$\pm$	87	&	0.26	&				&	8.18	$\pm$	0.07	&	1.183	$\pm$	0.001	&	1.27	$\pm$	0.05	&	2.192	$\pm$	0.006	&	8.93	&	1	&		&	95	\\
  V*V990Tau           	&	750	&	4774	$\pm$	53	&	0.84	&				&	4.40	$\pm$	0.24	&	0.766	$\pm$	0.001	&	0.75	$\pm$	0.02	&	0.263	$\pm$	0.001	&	64.12	&	12	&	0.284	&	41	\\
  HD 285742           	&	750	&	4798	$\pm$	141	&	0.82	&	-4.131	$\pm$	0.061	&	3.09	$\pm$	0.01	&	0.767	$\pm$	0.001	&	0.75	$\pm$	0.03	&	0.266	$\pm$	0.002	&	63.96	&	7	&	0.085	&	45	\\
HIP20741	&	750	&	5781	$\pm$	8	&	0.42	&				&	4.40	$\pm$	0.24	&	1.011	$\pm$	0.001	&	1.02	$\pm$	0.01	&	1.047	$\pm$	0.004	&	30.24	&	1	&		&	84	\\
  HD 286820           	&	750	&	4773	$\pm$	131	&	0.83	&				&				&	0.875	$\pm$	0.006	&	0.98	$\pm$	0.04	&	0.447	$\pm$	0.011	&	48.00	&	1	&		&	9	\\
HIP20815	&	750	&	6148	$\pm$	45	&	0.30	&				&	8.78	$\pm$	0.53	&	1.171	$\pm$	0.001	&	1.27	$\pm$	0.03	&	2.062	$\pm$	0.008	&	10.30	&	1	&		&	101	\\
HIP20826	&	750	&	6097	$\pm$	26	&	0.30	&				&	8.35	$\pm$	0.74	&	1.123	$\pm$	0.001	&	1.18	$\pm$	0.02	&	1.727	$\pm$	0.004	&	15.91	&	1	&		&	120	\\
HIP20850	&	750	&	5246	$\pm$	44	&	0.62	&	-4.286			&	3.66	$\pm$	0.43	&	0.885	$\pm$	0.001	&	0.89	$\pm$	0.02	&	0.535	$\pm$	0.002	&	46.50	&	1	&		&	92	\\
  HD 28258            	&	750	&	5246	$\pm$	44	&	0.62	&	-4.275	$\pm$	0.025	&	4.02	$\pm$	0.14	&	0.885	$\pm$	0.001	&	0.89	$\pm$	0.02	&	0.535	$\pm$	0.002	&	46.50	&	6	&	0.085	&	68	\\
HIP20899	&	750	&	5904	$\pm$	52	&	0.37	&				&	4.66	$\pm$	0.41	&	1.070	$\pm$	0.001	&	1.11	$\pm$	0.02	&	1.357	$\pm$	0.004	&	22.39	&	1	&		&	128	\\
HIP20951	&	750	&	5342	$\pm$	42	&	0.58	&	-4.227			&	3.09	$\pm$	0.01	&	0.875	$\pm$	0.001	&	0.84	$\pm$	0.02	&	0.523	$\pm$	0.002	&	48.00	&	1	&		&	75	\\
  HD 285773           	&	750	&	5342	$\pm$	42	&	0.58	&	-4.247	$\pm$	0.017	&	2.68	$\pm$	0.10	&	0.875	$\pm$	0.001	&	0.84	$\pm$	0.02	&	0.523	$\pm$	0.002	&	48.00	&	25	&	0.874	&	75	\\
HIP20978	&	750	&	5164	$\pm$	148	&	0.66	&	-4.240			&	3.41	$\pm$	0.31	&	0.865	$\pm$	0.001	&	0.86	$\pm$	0.04	&	0.478	$\pm$	0.002	&	49.49	&	1	&		&	55	\\
  HD28462             	&	750	&	5164	$\pm$	148	&	0.66	&	-4.249	$\pm$	0.015	&	3.11	$\pm$	0.38	&	0.865	$\pm$	0.001	&	0.86	$\pm$	0.04	&	0.478	$\pm$	0.002	&	49.49	&	9	&	0.090	&	61	\\
HIP21099	&	750	&	5481	$\pm$	144	&	0.53	&	-4.395			&	3.72	$\pm$	0.16	&	0.937	$\pm$	0.000	&	0.94	$\pm$	0.06	&	0.716	$\pm$	0.003	&	39.25	&	1	&		&	101	\\
 V1202 Tau		 	&	1	&	5132	$\pm$	218	&	0.68	&	-3.996	$\pm$	0.021	&	21.67	$\pm$	0.85	&	0.976	$\pm$	0.002	&	1.11	$\pm$	0.08	&	0.768	$\pm$	0.006	&	564.86	&	41	&	1.322	&	59	\\
HIP21112	&	750	&	6133	$\pm$	44	&	0.31	&				&	4.40	$\pm$	0.24	&	1.094	$\pm$	0.001	&	1.11	$\pm$	0.02	&	1.566	$\pm$	0.008	&	19.40	&	35	&	9.387	&	58	\\
 V830 Tau		 	&	1	&	4020	$\pm$	414	&	1.24	&				&	30.36	$\pm$	0.57	&	0.923	$\pm$	0.002	&	1.37	$\pm$	0.17	&	0.441	$\pm$	0.004	&	550.83	&	146	&	2.409	&	29	\\
HIP21317	&	750	&	5852	$\pm$	19	&	0.38	&				&	3.79	$\pm$	0.09	&	1.050	$\pm$	0.001	&	1.08	$\pm$	0.01	&	1.239	$\pm$	0.004	&	24.88	&	1	&		&	113	\\
HIP21654	&	750	&	5858	$\pm$	55	&	0.39	&				&	3.48	$\pm$	0.99	&	1.031	$\pm$	0.002	&	1.04	$\pm$	0.04	&	1.154	$\pm$	0.012	&	27.49	&	1	&		&	97	\\
HIP22203	&	750	&	5726	$\pm$	45	&	0.44	&				&	3.66	$\pm$	0.43	&	1.006	$\pm$	0.001	&	1.02	$\pm$	0.01	&	1.012	$\pm$	0.006	&	30.92	&	1	&		&	101	\\
  HD284787            	&	750	&	5190	$\pm$	112	&	0.64	&	-4.224	$\pm$	0.027	&	4.02	$\pm$	0.14	&	0.867	$\pm$	0.001	&	0.86	$\pm$	0.03	&	0.485	$\pm$	0.003	&	49.19	&	16	&	0.290	&	54	\\
HIP22422	&	750	&	6031	$\pm$	21	&	0.33	&				&	4.90	$\pm$	0.27	&	1.108	$\pm$	0.001	&	1.16	$\pm$	0.02	&	1.609	$\pm$	0.006	&	17.70	&	1	&		&	91	\\
HIP23069	&	750	&	5597	$\pm$	61	&	0.49	&	-4.383			&	3.94	$\pm$	0.23	&	0.940	$\pm$	0.001	&	0.92	$\pm$	0.02	&	0.745	$\pm$	0.003	&	38.84	&	1	&		&	94	\\
HIP23498	&	750	&	5411	$\pm$	95	&	0.56	&	-4.340			&	4.02	$\pm$	0.14	&	0.930	$\pm$	0.001	&	0.94	$\pm$	0.03	&	0.680	$\pm$	0.003	&	40.20	&	1	&		&	86	\\
HIP23750	&	750	&	5590	$\pm$	23	&	0.49	&	-4.339	$\pm$	0.028	&	3.64	$\pm$	0.07	&	0.950	$\pm$	0.001	&	0.94	$\pm$	0.01	&	0.777	$\pm$	0.003	&		&	8	&	10.015	&	42	\\
 TYC 5909-319-1		 	&	50	&	4922	$\pm$	53	&	0.76	&	-3.964	$\pm$	0.026	&	15.48	$\pm$	0.03	&	0.860	$\pm$	0.002	&	0.91	$\pm$	0.02	&	0.436	$\pm$	0.003	&	46.83	&	58	&	1.455	&	30	\\
HD32923	&	9000	&	5835	$\pm$	187	&	0.39	&				&	2.72	$\pm$	0.67	&	1.254	$\pm$	0.001	&	1.55	$\pm$	0.08	&	2.510	$\pm$	0.014	&		&	48	&	10.266	&	180	\\
HIP25486	&	24	&	6080	$\pm$	43	&	0.32	&				&	47.17	$\pm$	2.30	&	1.134	$\pm$	0.001	&	1.21	$\pm$	0.02	&	1.788	$\pm$	0.003	&	15.47	&	5	&	0.169	&	110	\\
HD36108	&	7100	&	5982	$\pm$	51	&	0.35	&				&	2.14	$\pm$	0.08	&	1.162	$\pm$	0.001	&	1.29	$\pm$	0.03	&	1.922	$\pm$	0.003	&		&	30	&	9.494	&	217	\\
HD38283	&	5700	&	6051	$\pm$	72	&	0.33	&				&	2.94	$\pm$	0.16	&	1.245	$\pm$	0.001	&	1.46	$\pm$	0.03	&	2.574	$\pm$	0.005	&	28.46	&	71	&	2.570	&	140	\\
HIP27072	&	414	&	6283	$\pm$	231	&	0.26	&				&	7.74	$\pm$	0.01	&	1.229	$\pm$	0.002	&	1.36	$\pm$	0.18	&	2.590	$\pm$	0.013	&	4.62	&	50	&	5.117	&	269	\\
HD45289	&	7600	&	5734	$\pm$	104	&	0.43	&				&	2.33	$\pm$	0.20	&	1.103	$\pm$	0.000	&	1.23	$\pm$	0.10	&	1.466	$\pm$	0.002	&	193.17	&	112	&	13.330	&	178	\\
 HD 59747		 	&	414	&	5150	$\pm$	91	&	0.66	&	-4.198	$\pm$	0.024	&	4.31	$\pm$	1.48	&	0.814	$\pm$	0.001	&	0.77	$\pm$	0.02	&	0.372	$\pm$	0.001	&	56.04	&	83	&	1.445	&	129	\\
 HD 63433		 	&	414	&	5714	$\pm$	212	&	0.44	&	-4.303	$\pm$	0.013	&	4.85	$\pm$	1.13	&	0.941	$\pm$	0.000	&	0.90	$\pm$	0.04	&	0.772	$\pm$	0.001	&	38.80	&	72	&	1.238	&	201	\\
 HD 70573		 	&	257	&	5837	$\pm$	80	&	0.40	&				&	13.73	$\pm$	0.16	&	0.995	$\pm$	0.001	&	0.98	$\pm$	0.02	&	0.998	$\pm$	0.004	&	32.43	&	47	&	1.253	&	96	\\
  Cl* NGC 2632 JC 61  	&	578	&	4996	$\pm$	69	&	0.73	&	-4.153	$\pm$	0.033	&	4.82	$\pm$	0.18	&	0.830	$\pm$	0.003	&	0.83	$\pm$	0.02	&	0.386	$\pm$	0.005	&	53.96	&	5	&	0.123	&	26	\\
  Cl* NGC 2632 JS 133 	&	578	&	4944	$\pm$	86	&	0.76	&	-4.187	$\pm$	0.041	&	3.72	$\pm$	0.16	&	0.807	$\pm$	0.003	&	0.79	$\pm$	0.03	&	0.340	$\pm$	0.005	&	57.29	&	7	&	0.983	&	21	\\
Cl* NGC 2632 JS 142	&	578	&	5314	$\pm$	71	&	0.60	&	-4.188	$\pm$	0.558	&	4.02	$\pm$	0.14	&	0.905	$\pm$	0.005	&	0.91	$\pm$	0.03	&	0.596	$\pm$	0.013	&	44.07	&	74	&	3.386	&	27	\\
  Cl* NGC 2632 JS 143 	&	578	&	4955	$\pm$	126	&	0.75	&	-3.934	$\pm$	0.230	&	3.72	$\pm$	0.16	&	0.772	$\pm$	0.004	&	0.73	$\pm$	0.04	&	0.286	$\pm$	0.005	&	62.94	&	3	&	0.014	&	11	\\
Cl* NGC 2632 JS 247	&	578	&	5021	$\pm$	783	&	0.72	&	-4.230	$\pm$	0.063	&	3.66	$\pm$	0.43	&	0.832	$\pm$	0.003	&	0.83	$\pm$	0.11	&	0.392	$\pm$	0.005	&	53.68	&	4	&	0.134	&	21	\\
Cl* NGC 2632 JC 141	&	578	&	5012	$\pm$	26	&	0.73	&	-4.155	$\pm$	0.047	&	4.40	$\pm$	0.24	&	0.831	$\pm$	0.004	&	0.83	$\pm$	0.01	&	0.389	$\pm$	0.006	&	53.82	&	11	&	1.166	&	23	\\
Cl* NGC 2632 JC 158	&	578	&	5015	$\pm$	244	&	0.72	&	-4.226	$\pm$	0.054	&	4.40	$\pm$	0.24	&	0.813	$\pm$	0.003	&	0.79	$\pm$	0.05	&	0.358	$\pm$	0.006	&	56.43	&	19	&	1.283	&	22	\\
Cl* NGC 2632 JC 172 	&	578	&	4960	$\pm$	116	&	0.75	&	-4.166	$\pm$	0.029	&	3.10	$\pm$	0.02	&	0.786	$\pm$	0.004	&	0.75	$\pm$	0.03	&	0.307	$\pm$	0.006	&	60.62	&	5	&	0.120	&	24	\\
Cl* NGC 2632 JC 208	&	578	&	4918	$\pm$	151	&	0.77	&	-4.071	$\pm$	0.063	&	4.82	$\pm$	0.18	&	0.766	$\pm$	0.003	&	0.72	$\pm$	0.04	&	0.274	$\pm$	0.004	&	63.94	&	5	&	0.109	&	18	\\
  Cl* NGC 2632 JC 221	&	578	&	5053	$\pm$	96	&	0.71	&	-3.849	$\pm$	0.223	&	4.26	$\pm$	0.38	&	0.745	$\pm$	0.003	&	0.66	$\pm$	0.05	&	0.254	$\pm$	0.004	&	67.30	&	5	&	0.087	&	12	\\
Cl* NGC 2632 JS 404 	&	578	&	5014	$\pm$	105	&	0.73	&	-4.213	$\pm$	0.057	&	3.79	$\pm$	0.09	&	0.845	$\pm$	0.003	&	0.86	$\pm$	0.04	&	0.417	$\pm$	0.005	&	51.79	&	16	&	0.374	&	22	\\
Cl* NGC 2632 JC 234 	&	578	&	5001	$\pm$	57	&	0.73	&	-4.207	$\pm$	0.027	&	3.79	$\pm$	0.09	&	0.816	$\pm$	0.003	&	0.80	$\pm$	0.02	&	0.362	$\pm$	0.005	&	55.99	&	7	&	0.123	&	27	\\
K2 101	&	578	&	4969	$\pm$	122	&	0.74	&	-4.151	$\pm$	0.071	&	6.22	$\pm$	0.03	&	0.750	$\pm$	0.003	&	0.68	$\pm$	0.04	&	0.256	$\pm$	0.004	&	66.58	&	2	&	0.008	&	17	\\
BD+20 2184	&	578	&	6003	$\pm$	181	&	0.35	&				&	6.22	$\pm$	0.03	&	1.135	$\pm$	0.005	&	1.23	$\pm$	0.12	&	1.757	$\pm$	0.030	&	14.81	&	37	&	3.454	&	53	\\
Cl* NGC 2632 WJJP 792	&	578	&	4912	$\pm$	108	&	0.77	&	-3.858	$\pm$	0.236	&	4.26	$\pm$	0.38	&	0.720	$\pm$	0.002	&	0.64	$\pm$	0.04	&	0.213	$\pm$	0.003	&	70.89	&	5	&	0.290	&	13	\\
Cl* NGC 2632 JC 278 	&	578	&	5089	$\pm$	73	&	0.68	&	-4.175	$\pm$	0.039	&	4.82	$\pm$	0.18	&	0.862	$\pm$	0.002	&	0.87	$\pm$	0.03	&	0.461	$\pm$	0.004	&	49.55	&	56	&	5.164	&	24	\\
  Cl* NGC 2632 KW 551 	&	578	&	4965	$\pm$	445	&	0.74	&	-4.071	$\pm$	0.023	&	4.02	$\pm$	0.14	&	0.793	$\pm$	0.003	&	0.76	$\pm$	0.08	&	0.319	$\pm$	0.004	&	59.47	&	4	&	0.134	&	19	\\
Cl* NGC 2632 JS 563	&	578	&	5055	$\pm$	160	&	0.70	&	-4.343			&	3.41	$\pm$	0.31	&	0.844	$\pm$	0.003	&	0.84	$\pm$	0.04	&	0.420	$\pm$	0.005	&	51.94	&	1	&		&	16	\\
  Cl* NGC 2632 JS 576 	&	578	&	4870	$\pm$	143	&	0.79	&	-4.035	$\pm$	0.070	&	3.48	$\pm$	0.40	&	0.764	$\pm$	0.003	&	0.73	$\pm$	0.05	&	0.267	$\pm$	0.003	&	64.27	&	4	&	0.959	&	11	\\
 AD Leo			 	&	200	&	3859	$\pm$	131	&	1.32	&				&				&				&				&				&		&	66	&	0.816	&	74	\\
 TYC 1989-0049-1	 	&	562	&	4773	$\pm$	151	&	0.84	&				&	3.19	$\pm$	0.97	&	0.733	$\pm$	0.001	&	0.69	$\pm$	0.04	&	0.220	$\pm$	0.001	&	68.86	&	2	&	0.238	&	58	\\
 HD 107877		 	&	562	&	6433	$\pm$	99	&	0.21	&				&	27.01	$\pm$	0.19	&	1.241	$\pm$	0.002	&	1.34	$\pm$	0.06	&	2.789	$\pm$	0.022	&	3.12	&	84	&	4.063	&	100	\\
 BD+26 2342		 	&	562	&	5390	$\pm$	190	&	0.57	&	-4.307	$\pm$	0.035	&	3.94	$\pm$	0.23	&	0.855	$\pm$	0.001	&	0.80	$\pm$	0.05	&	0.483	$\pm$	0.003	&	50.37	&	46	&	1.149	&	49	\\
 BD+27 2139		 	&	562	&	5056	$\pm$	205	&	0.70	&	-4.260	$\pm$	0.029	&	4.02	$\pm$	0.14	&	0.810	$\pm$	0.001	&	0.78	$\pm$	0.04	&	0.356	$\pm$	0.002	&	56.80	&	59	&	1.128	&	59	\\
 HIP 61205		 	&	562	&	5854	$\pm$	130	&	0.40	&				&	4.66	$\pm$	0.41	&	0.960	$\pm$	0.001	&	0.91	$\pm$	0.04	&	0.867	$\pm$	0.005	&	36.42	&	29	&	1.026	&	86	\\
 BD+23 2472		 	&	562	&	5408	$\pm$	128	&	0.55	&				&	3.72	$\pm$	0.16	&	0.862	$\pm$	0.002	&	0.81	$\pm$	0.03	&	0.503	$\pm$	0.004	&	49.47	&	50	&	0.451	&	44	\\
HD122862	&	5900	&	5977	$\pm$	93	&	0.36	&				&	2.62	$\pm$	0.04	&	1.285	$\pm$	0.000	&	1.58	$\pm$	0.04	&	2.866	$\pm$	0.005	&	-1.00	&	52	&	9.111	&	216	\\
TYC6779-305-1	&	10	&	5013	$\pm$	51	&	0.73	&	-4.007			&	10.92	$\pm$	0.40	&	1.323	$\pm$	0.009	&	2.10	$\pm$	0.04	&	2.506	$\pm$	0.071	&		&	1	&		&	82	\\
 TYC 6191-552-1		 	&	10	&	4259	$\pm$	198	&	1.10	&				&	6.44	$\pm$	0.19	&	0.951	$\pm$	0.002	&	1.34	$\pm$	0.12	&	0.535	$\pm$	0.005	&	198.81	&	61	&	2.335	&	31	\\
 GSC 06204-00812	 	&	10	&	4234	$\pm$	50	&	1.12	&				&	17.55	$\pm$	0.02	&	0.991	$\pm$	0.003	&	1.47	$\pm$	0.03	&	0.627	$\pm$	0.006	&	180.75	&	11	&	0.980	&	32	\\
 V866			 	&	10	&				&	1.12	&				&	13.42	$\pm$	0.63	&				&				&				&		&	6	&	0.159	&	39	\\
 HD 167389		 	&	414	&	5882	$\pm$	44	&	0.39	&				&	3.09	$\pm$	0.01	&	1.028	$\pm$	0.001	&	1.03	$\pm$	0.01	&	1.146	$\pm$	0.001	&	28.58	&	37	&	1.437	&	168	\\
HIP92680	&	24	&	5338	$\pm$	210	&	0.58	&	-3.906	$\pm$	0.048	&	69.08	$\pm$	3.13	&	1.027	$\pm$	0.001	&	1.17	$\pm$	0.10	&	0.993	$\pm$	0.005	&	46.03	&	47	&	12.519	&	108	\\
 TYC 0486-4943-1	 	&	149	&	4729	$\pm$	79	&	0.87	&				&	8.27	$\pm$	0.41	&	0.725	$\pm$	0.001	&	0.68	$\pm$	0.02	&	0.208	$\pm$	0.001	&	67.80	&	22	&	1.379	&	53	\\
HD186408	&	5800	&	5793	$\pm$	52	&	0.41	&				&	2.66	$\pm$	0.37	&	1.121	$\pm$	0.000	&	1.25	$\pm$	0.02	&	1.588	$\pm$	0.002	&	21.08	&	11	&	0.005	&	207	\\
HD186427	&	8000	&	5777	$\pm$	113	&	0.42	&				&	2.45	$\pm$	0.13	&	1.059	$\pm$	0.000	&	1.12	$\pm$	0.03	&	1.259	$\pm$	0.002	&	26.89	&	13	&	0.159	&	200	\\
 HD 235088		 	&	600	&	5067	$\pm$	46	&	0.69	&	-4.242	$\pm$	0.016	&	4.77	$\pm$	0.40	&	0.814	$\pm$	0.000	&	0.78	$\pm$	0.02	&	0.365	$\pm$	0.001	&	56.36	&	94	&	1.319	&	85	\\
HD191408	&	15000	&	4998	$\pm$	17	&	0.74	&	-4.974	$\pm$	0.014	&	3.02	$\pm$	0.61	&	0.777	$\pm$	0.001	&	0.73	$\pm$	0.01	&	0.297	$\pm$	0.001	&		&	67	&	4.716	&	149	\\
HD196378	&	5300	&	6101	$\pm$	101	&	0.31	&				&	4.44	$\pm$	0.21	&	1.451	$\pm$	0.002	&	1.96	$\pm$	0.07	&	4.817	$\pm$	0.025	&		&	60	&	5.908	&	152	\\
 TYC 1090-543-1		 	&	149	&	4555	$\pm$	176	&	0.94	&				&	16.93	$\pm$	3.28	&	0.693	$\pm$	0.001	&	0.65	$\pm$	0.04	&	0.165	$\pm$	0.001	&	72.37	&	16	&	2.510	&	40	\\
HIP105388	&	30	&	5506	$\pm$	73	&	0.52	&	-4.020	$\pm$	0.040	&	15.48	$\pm$	0.03	&	0.921	$\pm$	0.001	&	0.90	$\pm$	0.03	&	0.672	$\pm$	0.002	&	61.83	&	51	&	14.277	&	81	\\
HD210918	&	8500	&	5737	$\pm$	74	&	0.43	&				&	2.42	$\pm$	0.16	&				&				&				&		&	11	&	1.892	&	415	\\
HIP116748	&	30	&				&	0.76	&	-4.189	$\pm$	0.018	&	16.63	$\pm$	0.98	&				&				&				&		&	14	&	14.116	&	88	\\
 TYC 4610-1318-1	 	&	10	&	5313	$\pm$	150	&	0.60	&	-3.960	$\pm$	0.020	&	9.52	$\pm$	0.49	&	1.038	$\pm$	0.002	&	1.20	$\pm$	0.07	&	1.029	$\pm$	0.008	&	166.49	&	38	&	2.523	&	55	\\
\hline\noalign{\smallskip}
\end{longtable}
\end{landscape}
}


\longtab[3]{
\begin{longtable}{lccc}
\caption{
 { Emission excess.}
  }\label{longtable_fluxes}\\
\hline
 Star & $\log F$(Ca~{\sc ii} H)        & $\log F$(Ca~{\sc ii} K)        & $\log F$(H$\alpha$) \\
      & [ergs$^{\rm -1}$cm$^{\rm -2}$] & [ergs$^{\rm -1}$cm$^{\rm -2}$] & [ergs$^{\rm -1}$cm$^{\rm -2}$] \\
\hline
\endfirsthead
\caption{Continued.} \\
\hline
  Star & $\log F$(Ca~{\sc ii} H)        & $\log F$(Ca~{\sc ii} K)        & $\log F$(H$\alpha$) \\
        & [ergs$^{\rm -1}$cm$^{\rm -2}$] & [ergs$^{\rm -1}$cm$^{\rm -2}$] & [ergs$^{\rm -1}$cm$^{\rm -2}$] \\
\hline
\endhead
\hline
\endfoot
\hline
\endlastfoot
HIP 490	&	6.73	$\pm$	0.74	&	6.65	$\pm$	0.73	&	6.85	$\pm$	0.75	\\
HIP 1481	&	6.53	$\pm$	0.72	&	6.62	$\pm$	0.73	&	6.87	$\pm$	0.76	\\
 TYC 4500478	 	&	6.54	$\pm$	0.72	&	6.43	$\pm$	0.71	&	7.17	$\pm$	0.79	\\
HIP 8486	&	6.09	$\pm$	0.67	&	6.49	$\pm$	0.71	&	6.82	$\pm$	0.75	\\
NGC 752 48	&				&				&	6.32	$\pm$	0.70	\\
Cl* NGC 752 RV 144	&				&				&	5.35	$\pm$	0.60	\\
NGC 752 185	&				&				&	5.98	$\pm$	0.66	\\
NGC 752 211	&				&				&	6.41	$\pm$	0.70	\\
NGC 752 229	&				&				&	5.66	$\pm$	0.62	\\
NGC 752 236	&				&				&	5.60	$\pm$	0.62	\\
NGC 752 244	&				&				&	6.14	$\pm$	0.68	\\
HIP 15310	&	6.31	$\pm$	0.70	&	6.01	$\pm$	0.67	&	6.22	$\pm$	0.68	\\
HIP 16529	&	6.07	$\pm$	0.67	&	5.80	$\pm$	0.64	&	6.00	$\pm$	0.66	\\
HD 22879	&				&				&	5.17	$\pm$	0.59	\\
  HD 283066           	&	5.37	$\pm$	0.59	&	5.82	$\pm$	0.64	&	5.88	$\pm$	0.65	\\
HIP 18327	&	5.82	$\pm$	0.64	&	6.13	$\pm$	0.68	&	6.21	$\pm$	0.68	\\
 V1298 Tau		 	&	6.57	$\pm$	0.72	&	6.50	$\pm$	0.72	&	7.06	$\pm$	0.78	\\
HIP 19098	&	5.64	$\pm$	0.62	&	5.76	$\pm$	0.63	&	6.20	$\pm$	0.68	\\
HIP 19148	&	6.48	$\pm$	0.72	&	6.39	$\pm$	0.71	&	6.40	$\pm$	0.70	\\
 HD 285507		 	&	5.58	$\pm$	0.61	&	5.76	$\pm$	0.63	&	6.05	$\pm$	0.67	\\
HIP 19786	&	6.75	$\pm$	0.74	&	6.52	$\pm$	0.72	&	6.40	$\pm$	0.70	\\
HIP 19793	&	6.31	$\pm$	0.69	&	6.08	$\pm$	0.67	&	6.38	$\pm$	0.70	\\
HIP 19796	&	6.40	$\pm$	0.72	&	6.09	$\pm$	0.72	&				\\
HIP 19859	&	6.60	$\pm$	0.73	&	6.46	$\pm$	0.72	&	6.65	$\pm$	0.73	\\
 TAP 26			 	&	6.29	$\pm$	0.69	&	6.28	$\pm$	0.69	&	6.83	$\pm$	0.75	\\
HIP 20237	&	6.54	$\pm$	0.72	&	6.26	$\pm$	0.69	&	6.63	$\pm$	0.73	\\
  V* V988 Tau         	&	5.89	$\pm$	0.65	&	6.37	$\pm$	0.70	&	6.27	$\pm$	0.69	\\
  HD27771             	&	6.13	$\pm$	0.67	&	5.74	$\pm$	0.63	&	6.12	$\pm$	0.67	\\
HIP 20741	&	6.43	$\pm$	0.71	&	6.65	$\pm$	0.73	&	6.47	$\pm$	0.71	\\
HIP 20815	&	6.59	$\pm$	0.73	&	6.11	$\pm$	0.70	&	6.55	$\pm$	0.72	\\
HIP 20826	&	6.25	$\pm$	0.70	&	6.97	$\pm$	0.77	&	6.50	$\pm$	0.71	\\
HIP 20899	&	6.06	$\pm$	0.70	&	6.73	$\pm$	0.74	&	6.43	$\pm$	0.71	\\
HIP 20978	&				&				&	6.30	$\pm$	0.69	\\
  HD 28462             	&	6.08	$\pm$	0.67	&	6.07	$\pm$	0.67	&	6.14	$\pm$	0.68	\\
 V1202 Tau		 	&	6.56	$\pm$	0.72	&	6.50	$\pm$	0.71	&	6.97	$\pm$	0.77	\\
HIP 21112	&	6.27	$\pm$	0.82	&	6.10	$\pm$	0.77	&	6.25	$\pm$	0.69	\\
HIP 21317	&	6.38	$\pm$	0.70	&	5.40	$\pm$	0.65	&	6.39	$\pm$	0.70	\\
HIP 21654	&	6.92	$\pm$	0.76	&	5.53	$\pm$	0.71	&	6.55	$\pm$	0.72	\\
HIP 22203	&	5.39	$\pm$	0.62	&	6.52	$\pm$	0.72	&	6.31	$\pm$	0.69	\\
  HD 284787            	&	6.04	$\pm$	0.66	&	6.10	$\pm$	0.67	&	6.38	$\pm$	0.70	\\
HIP 22422	&	6.51	$\pm$	0.79	&	5.50	$\pm$	1.32	&				\\
 TYC 5909-319		 	&	6.57	$\pm$	0.72	&	6.52	$\pm$	0.72	&	7.03	$\pm$	0.77	\\
HIP 25486	&	7.00	$\pm$	0.77	&	6.80	$\pm$	0.75	&	7.03	$\pm$	0.77	\\
HD 38283	&				&				&	7.46	$\pm$	0.82	\\
 HD 59747		 	&	6.12	$\pm$	0.67	&	6.18	$\pm$	0.68	&	6.52	$\pm$	0.72	\\
 HD 63433		 	&	6.55	$\pm$	0.72	&	6.51	$\pm$	0.72	&	6.60	$\pm$	0.73	\\
 HD 70573		 	&	6.80	$\pm$	0.75	&	6.71	$\pm$	0.74	&	6.86	$\pm$	0.75	\\
  Cl* NGC 2632 JC 61  	&	6.37	$\pm$	0.70	&	4.54	$\pm$	1.49	&	6.40	$\pm$	0.70	\\
  Cl* NGC 2632 JS 133 	&				&	6.07	$\pm$	0.67	&	6.42	$\pm$	0.71	\\
  Cl* NGC 2632 JS 143 	&				&				&	6.48	$\pm$	0.71	\\
Cl* NGC 2632 JS 247	&				&				&	5.83	$\pm$	0.64	\\
Cl* NGC 2632 JC 141	&	5.82	$\pm$	0.64	&	6.27	$\pm$	0.69	&	6.39	$\pm$	0.70	\\
Cl* NGC 2632 JC 158	&				&	6.11	$\pm$	0.67	&	6.23	$\pm$	0.69	\\
Cl* NGC 2632 JC 172 	&	6.06	$\pm$	0.67	&				&	6.34	$\pm$	0.70	\\
  Cl* NGC 2632 JC 221	&				&				&	6.51	$\pm$	0.72	\\
Cl* NGC 2632 JS 404 	&	6.34	$\pm$	0.70	&	6.41	$\pm$	0.70	&	6.20	$\pm$	0.68	\\
Cl* NGC 2632 JC 234 	&	4.52	$\pm$	1.04	&	6.42	$\pm$	0.71	&	6.49	$\pm$	0.71	\\
K2 101	&				&				&	6.66	$\pm$	0.73	\\
BD+20 2184	&	6.65	$\pm$	0.73	&	6.36	$\pm$	0.70	&	6.32	$\pm$	0.69	\\
Cl* NGC 2632 JC 278 	&	6.28	$\pm$	0.69	&	6.30	$\pm$	0.69	&	6.35	$\pm$	0.70	\\
  Cl* NGC 2632 KW 551 	&				&				&	6.49	$\pm$	0.71	\\
Cl* NGC 2632 JS 563	&				&				&	6.51	$\pm$	0.72	\\
 BD+27 2139		 	&	6.04	$\pm$	0.67	&	6.03	$\pm$	0.66	&	6.40	$\pm$	0.70	\\
 HIP 61205		 	&	6.64	$\pm$	0.73	&	6.65	$\pm$	0.73	&	6.67	$\pm$	0.73	\\
TYC 6779-305	&	6.63	$\pm$	0.73	&	6.21	$\pm$	0.68	&	8.08	$\pm$	0.89	\\
 TYC 6191-552		 	&	6.03	$\pm$	0.66	&	5.79	$\pm$	0.64	&	7.03	$\pm$	0.77	\\
 GSC 06204-00812	 	&	5.09	$\pm$	0.56	&				&	6.06	$\pm$	0.67	\\
 HD 167389		 	&	6.26	$\pm$	0.69	&	6.15	$\pm$	0.68	&	5.56	$\pm$	0.62	\\
 TYC 0486-4943	 	&	6.19	$\pm$	0.68	&	6.01	$\pm$	0.66	&	6.80	$\pm$	0.75	\\
 HD 235088		 	&	6.02	$\pm$	0.66	&	6.03	$\pm$	0.66	&	6.38	$\pm$	0.70	\\
 TYC 1090-543		 	&	6.23	$\pm$	0.69	&	6.21	$\pm$	0.68	&	6.94	$\pm$	0.76	\\
\hline\noalign{\smallskip}
\end{longtable}
}%

\end{appendix}
\end{document}